\documentclass[aps,superscriptaddress,twocolumn,showpacs,preprintnumbers,amsmath,amssymb,showkeys]{revtex4}
\usepackage[cp1251]{inputenc}
\usepackage[english]{babel}
\usepackage{amssymb,amsmath}
\usepackage{amstext,textcomp} 
\usepackage[dvips]{hyperref}
\usepackage[mathscr]{eucal}
\usepackage{longtable}
\usepackage{graphicx}% Include figure files
\begin{document}
\title{Classical and Quantum Electrodynamics  Concept Based on Maxwell Equations' Symmetry}% Force line breaks with \\ 
\author{Dmitri Yerchuck (a),  Alla  Dovlatova (b), Andrey Alexandrov (b)\\
\textit{(a) Minsk State Higher Aviation College, Uborevich Str.77, Minsk, 220096, yearchuck@gmail.com,\\ 
(b) M.V.Lomonosov Moscow State University, Moscow, 119899}}
\date{\today}% 
\begin{abstract}The symmetry studies of Maxwell equations gave new insight on  the nature of electromagnetic (EM) field. It  has in general case quaternion single structure,
consisting of four independent field constituents, which differ with each other by the parities under space inversion
and time reversal. It has also been shown, that for any complex relativistic field the gauge invariant
conserving quantity is two-component scalar or pseudoscalar value - complex charge. Generalized
Maxwell equations for quaternion four-component EM-field are obtained. Invariants for EM-field,
consisting of dually symmetric parts are found. It is shown, that there exists physical conserving
quantity, which is simultaneously invariant under  both Rainich dual and additional hyperbolic dual symmetry transformation of Maxwell
equations. It is spin in general case and spirality in the geometry, when electrical and magnetic vectors $\vec{E}$, $\vec{H}$ are directed
along coordinate axes in ($\vec{E}$, $\vec{H}$) functional space. It is additional proof for quaternion four component 
structure of EM-field to be a single whole. Canonical Dirac quantization method is developed in
two aspects. The first aspect is its application the only to observable quantities. The second aspect
is the realization along with well known time-local quantization of space-local quantization and
space-time-local quantization. It is also shown, that Coulomb field can be quantized in 1D and 2D
systems. New model of photons is proposed. The photons in quantized EM-field are main excitations  in oscillator structure of EM-field, which is equivalent to spin S = 1 "boson-atomic" structure, like  matematically to well known spin S = 1 boson matter  structure - carbon atomic backbone chain structure in many conjugated polymers. They have two kind nature.  The photons of the first kind and  the second kind represent themselves respectively neutral chargeless EM-solitons and charged spinless EM-solitons of  Su-Schrieffer-Heeger family. 
\end{abstract}  
% the Physics and Astronomy
                             % Classification Scheme

\maketitle                         
\section{Introduction}
Electromagnetic (EM) field is well studied, however some new theoretical and experimental results, discussed further, indicate on existence of earlier unknown new general properties, which extend the notion,  concerning EM-field nature itself. Basis equations in electrodynamics (ED) are well known Maxwell equations, which are unique in the sence, that they have the most rich symmetry among all fundamental equations of theoretical physics. The symmetry studies of Maxwell equations have started in 1892, when Heaviside \cite{Heaviside} has paid attention to the symmetry between electrical and magnetic quantities in given equations. So, the symmetry studies of EM-field have a long history, which was starting already in 19-th century and what is interesting, it is continuing hitherto. Let us give very briefly some moments in given studies.

 Mathematical formulation of given symmetry, consisting in  invariance of Maxwell equations for free EM-field under the duality transformations
\begin{equation} 
\label{eq1a}
\vec {E} \rightarrow \pm\vec {H}, \vec {H} \rightarrow \mp\vec {E},
\end{equation}
gave Larmor \cite{Larmor}. Duality transformations (\ref{eq1a}) are particular case of the more general dual transformations, established by Rainich \cite{Rainich}.  Dual transformations produce oneparametric abelian  group $U_1$, which is subgroup  of the group of chiral transformations.  Dual transformations correspond to irreducible representation of  the group of chiral transformations in particular case of quantum number $j = 1$ \cite{Tomilchick} and they are 
\begin{equation} 
\label{eq1b}
\begin{split}
\raisetag{40pt}                                                    
\vec {E'} \rightarrow \vec {E} \cos\theta + \vec {H} \sin\theta\\
\vec {H'} \rightarrow \vec {H} \cos\theta - \vec {E} \sin\theta,
\end{split}
\end{equation}
where parameter $\theta$ is arbitrary continuous variable,
$\theta \in [0,2\pi]$. 
Then it has been found, that maximal local symmetry group of Maxwell equations with sources is fifteen-parametric conform group $C(1,3)$ \cite{Bateman}.  $C(1,3)$ group includes linear 10-parametric Poincare group and linear scale transformations, which produce together the maximal local group of linear  transformations of coordinates and time, under which Maxwell equations with sources are invariant. It includes also local nonlinear conform transformations according to mapping
\begin{equation}
\label{eq1z}
 x_{\mu}: x_{\mu} \to x'_{\mu} = \frac{x_{\mu} - b_{\mu} x_{\nu} x^{\nu}}{1 - 2b_{\nu}x^{\nu} +  x_{\rho}x^{\rho}b_{\tau}b^{\tau}}, 
\end{equation}
where $b_{\mu}$ are parameters, which belongs to the set of real numbers, $b_{\mu} \in R$. 
At the same time the maximal local symmetry group of Maxwell equations without sources is 16-parametric group, which is direct product of $C(1,3)$ group and one-parametric group $U_1$ of Rainich dual transformations, that is $C(1,3) \otimes U_1$. It was proved relatively recently (1967) \cite{Ibragimov}. Some time later (1979) it was established the existance of nonlocal symmetry of Maxwell equations under transformations of 23-dimensional Lee algebra, which is direct product of $C(1,3)$ group and $A_8$-algebra, that is $C(1,3) \otimes A_8$ \cite{Fushchich}. Lee algebra $A_8$ is isomorphous to Lee algebra $U(2) \otimes U(2)$.

In present paper, the role of symmetry of Maxwell equations the only under one-parametric group $U_1$ of Rainich dual transformations will be studied.
 It will be shown, that dual symmetry of Maxwell equations leads to conclusion on compound character of EM-field, consisting in that,  that vector quantities, which characterize EM-field are in general case the four vector-component objects with different t- and  P-parities. We have to say,  that the idea, that vector-functions $\vec {E}(\vec{r},t)$, $\vec {H}(\vec{r},t)$, $\vec {D}(\vec{r},t)$, $\vec {B}(\vec{r},t)$, which characterize EM-field, are compound quantities and inlude both gradient and solenoidal parts, that is uneven and even parts relatively space inversion  transformation, was put forward earlier in \cite{Tomilchick}. There is also theoretical assumption in \cite{Berezin}, where along with usual choice, that is, that electric field  $\vec {E}$ is polar vector, magnetic field  $\vec {H}$ is axial vector, the alternative choice is provided. 
 At the same time given ideas were unknown (and they are almost unknown at present) to general circle  of the researchers in the field. It was also true for us. To conclusion on compound character of EM-field, in particular, to  conclusion, that along with part with polar electrical vector symmetry there is part with the axial symmetry of electrical vector characteristics, which are responsible for optical resonance absorption, Raman scattering,  we came from analysis of experimental results independently  \cite{Yearchuck_Doklady}, \cite{Yearchuck_Yerchak}, \cite{Yearchuck_PL} and  \cite{D_Yearchuck_A_Dovlatova}. In above cited works is presented and theoretically analysed a sufficiently large body of experimental data with the experimental confirmation of the phenomenon of dual symmetry of EM-field vector-functions under space inversion (see for some details Section II). Moreover, the conclusion was arised, that the exhibition of polar or axial properties of EM-field vector-functions depends on the experimental situation.  Given conclusion was to some extent unexpected, since it is considered always in voluminous literature, that electric field vector has to be the only polar vector and magnetic field vector has to be the only axial vector. Given possibility corresponds to known field theory consideration, in which is concluded, that 4-vector of EM-field potential has to be polar and t-even 4-vector \cite{Akhiezer} and, consequently, three-dimensional electric vector quantities have to be polar vectors. It has to remark, that uneven space inversion parity of electric vector quantities has very good experimental confirmation too. For instance, all electric charge transport phenomena can be well described in the suggestion, that electric field vector is polar vector. 
 At the same time given idea in the equations of dual electrodynamics in \cite{Tomilchick}  was presented the only in implicit form.

The conclusions, which follow from experimental data and above cited theoretical ideas and assumptions  have to be mathematically proved more carefully, that will be realized in given paper.  So, the first task of presented paper is to  consider  the dual symmetry  of EM-field, leading to existence of electrical and magnetic vector quantities with both even and uneven parities under improper rotations in more details.

The second task is concerned of more detailed theoretical substantiation for the existence of  two type of scalar quantities - electric and magnetic charges of EM-field on the one hand and description of the EM-field in condensed matter containing along with the particles with electric charge the particles with magnetic charge, including dually charged particles,  on the other hand.
So, the aim of presented paper is the field theory proof for given general properties of  EM-field.

It represents also the interest the  quantum description of EM-field. The well known now
simple formula $E = h \nu$, $h = const > 0$, proposed right at the beginning of the 20th century by Planck \cite{Planck} and  by Einstein
\cite{Einstein} became epoch-making and
 a real symbol of the substantial progress in the science. The interpretation, given by    Einstein indicates
straight on real existence of light quanta of frequency $\nu$  with the total energy $E$, which in its turn has led to 
 a new understanding of the nature of the electromagnetic field.  In fact it was the indication on oscillator structure of EM-field, that is EM-field is the set  of 
physical objects,  which are strongly connected to some periodic with
time period $T = 1/\nu$ process, being to be intrinsic for given  objects, at that  Lorentz invariant
product $E T$ is equal to $h$.  Similar interpretation of 
De Broglie’s relationship \cite{Broglie}  leads to conclusion on quantum nature of the charges.
Although there are the great  achievements in quantum theory, the great challenge to give  an adequate description of  light quanta, which were called by Lewis photons \cite{Lewis} 
still has not brought satisfactory results. Even  Einstein himself  recognizes \cite{Speziali},
 that ”the whole fifty years of conscious brooding have not brought me nearer to the answer
to the question what are light quanta”, and now, half a century later, theoretical physics still
needs progress to present a satisfactory explanation of the   photon nature. We consider
the corresponding theoretically directed efforts to be  necessary and even urgent in view of requirements of the modern science  and engineering, in particular, in
connection with rash progress in nanotechnogy.

The paper is organized in the following way. In Sec.2, Rainich dual symmetry of Maxwell equations is analysed and its experimental confirmation by comparison with literature data is given. In Sec.3, the algebraic properties of 
EM-field functions are summarized. In Sec.4, theoretical foundation of the existence of dually charged particles or quasiparticles is given. Comparison with experiment is also presented.  In Sec.5, the quaternion structure of EM-field is argued in details. In Sec.6, the cavity classical and quantum electrodynamics is considered by taking into account the dual symmetry of  EM-field, here the connection between gauge invariance of  EM-field and analicity of its  vector-functions is considered. In Sec.7, the effect of spin-charge separation in quantized EM-field structure is described. In Sec.8, the conclusions are formulated.

\section{Rainich dual symmetry oF Maxwell equations and its experimental confirmation}

 It is seen from (\ref{eq1b}), that dual transformations have fundamental sense for the symmetry of vector-functions $\vec {E}, \vec {H}$ relatively improper rotations. Dual symmetry  of Maxwell equations indicates, that both electric and magnetic vector force characteristics $\vec {E}$ and $\vec {H}$ of EM-field are possessing by equal rights.  It means in particular, that they both have to possess  by the same symmetry properties including the symmetry properties under improper rotations. Really  the expression (\ref{eq1b}) will be mathematically correct, if vector-functions $\vec {E}\cos\theta, \vec {H}\sin\theta$ (and correspondingly $\vec {H} \cos\theta,  \vec {E} \sin\theta$) in upper (bottom) line will   have the same symmetry, that is both polar or axial ones. It is not evident in given representation, if parameter $\theta$ is arbitrary variable. Let us clarify given situation. 
In matrix form the transformations (\ref{eq1b}) are
\begin{equation}
\label{eq1bc}
 \left[\begin{array} {*{20}c}  \vec {E'} \\ \vec {H'} \end{array}\right] = \left[\begin{array} {*{20}c} \cos\theta&\sin\theta  \\-\sin\theta&\cos\theta \end{array}\right]\left[\begin{array} {*{20}c}  \vec {E} \\ \vec {H} \end{array}\right].
\end{equation}
 At the same time to any complex number $a + ib$ can be set up in conformity the $[2 \times 2]$-matrix according to biective mapping
\begin{equation}
\label{eq1abcd}
 f : a + ib \to \left[\begin{array} {*{20}c} a&-b  \\ b&a \end{array}\right].
\end{equation}
Bijectivity of mapping (\ref{eq1abcd}) indicates on the existence of inverse mapping, that is to any matrix, which has the structure, given by right side in relation (\ref{eq1abcd}), correponds the complex number, determined by left side. 
Consequently, we have
\begin{equation}
\label{eq1bcd}
 \left[\begin{array} {*{20}c}  \vec {E'} \\ \vec {H'} \end{array}\right] = e^ {-i\theta}\left[\begin{array} {*{20}c}  \vec {E} \\ \vec {H} \end{array}\right],
\end{equation}
that is 
 \begin{equation} 
\label{eq1c}
\begin{split}
\raisetag{40pt}                                                    
&\vec {E'} = \vec {E} \cos\theta - i\vec {E} \sin\theta\\
&\vec {H'} = \vec {H} \cos\theta - i\vec {H} \sin\theta.
\end{split}
\end{equation}
  It means, that to real planes, which are determined  by the vectors  $\vec {E}$ and $\vec {H}$  can be set in conformity the complex plane for the vectors $\vec {E'}$ and $\vec {H'}$.
                    
It is evident, that   both the vectors $\vec {E'}$ and $\vec {H'}$ will consist both,  of even  and uneven components under improper rotations. So one component of for instance $\vec {E'}$ will be even under reflection in the plane situated transversely to absciss-axis, the second component will be uneven. 

Therefore dual transformation symmetry of Maxwell equations, established by Rainich \cite{Rainich}, indicates simultaneously on both complex nature of EM-field in general case, and that both electric and magnetic fields are consisting in general case of the components with various parity under improper rotations. In other words the rotation of  EM-field vectors, determined by (\ref{eq1bcd}) is accompanied by appearance of axial component for starting polar electric field and polar component for starting axial magnetic field. 

It has to be taken into account,  that in the case $\theta = 0$ we have well known electrodynamics with odd parity of electric field and even parity of magnetic field. In given case both the vectors are real vectors and by using in calculations the complex vector quantities, which is considered to be a formal, but  convenient mathematical technique, is always used  adding of corresponding complex conjugate  quantities. It is correct for given case and will be incorrect in the case $\theta \neq 0$, since along with adding of corresponding complex conjugate  quantities, which allows to obtain Re$\vec {E}$ and Re$\vec {H}$, Im$\vec {E}$ and Im$\vec {H}$ have  to be retained too.
It is easily to show, that in the case $\theta \neq 0$ the following relationship is taking place
\begin{equation}
\label{eq1bcde}
 \left[ \vec {E}^2 - \vec {H}^2 + 2i(\vec {E}\vec {H})\right]  e^ {-2i\theta} =inv,
\end{equation}
that is, we have at fixed parameter $\theta \neq 0$  two real EM-field invariants
\begin{equation} 
\label{eq1cde}
\begin{split}
\raisetag{40pt}                                                    
&(\vec {E}^2 - \vec {H}^2) \cos 2\theta + 2(\vec {E}\vec {H}) \sin 2\theta = I'_1 = inv \\
&2(\vec {E}\vec {H}) \cos 2\theta  - (\vec {E}^2 - \vec {H}^2) \sin 2\theta = I'_2 = inv.
\end{split}
\end{equation}
  It follows  from  relation (\ref{eq1cde}) that, in particular,  at $\theta = 0$ we have well known EM-field invariants
\begin{equation} 
\label{eq1cdef}
\begin{split}
\raisetag{40pt}                                                    
&(\vec {E}^2 - \vec {H}^2)  = I_1 = inv \\
&(\vec {E}\vec {H}) = I_2 = inv.
\end{split}
\end{equation}
  It is interesting, that at $\theta = 45\textdegree$  and at $\theta = 90\textdegree$ EM-field invariants are determined by the same relation (\ref{eq1cdef}) and by arbitrary  $\theta$ we have two linearly independent combinations of given known invariants. At the same time is it evident, that the following relationship takes place
\begin{equation} 
\label{eq1cdeff}
\begin{split}
\raisetag{40pt}                                                    
 {I'_1}^2 + {I'_2}^2 = {I_1}^2 + 4{I_2}^2 = K = inv. 
\end{split}
\end{equation}
It means, that quantity $K$ is not depending on $\theta$, that is $K$ is invariant of dual transformations. The invariance of quantity $K = {I_1}^2 + 4{I_2}^2$ under dual transformations was found earlier \cite{Tomilchick}, where the physical meaning of given invariance was also explained - the  quantity, equaled to $\frac{1}{4} K$, is square of 4-impulse density.

Examples, where the dual symmetry of  EM-field displays experimentally  itself, are well known.  For instance, the equality of magnetic and electric energy values  for free electromagnetic wave means the invariance of total energy and magnetic and electric energy components taken separately relatively Larmor  transformations. Quite similar situation takes place for  EM-field  in the cavity or in LC-tank. 
Subsequent extension of dual symmetry for the EM-field with sources leads also to requirement of the existence of two type of other physical quantities - two type of charges and currents and two type of  intrinsic moments of the particles or the absorbing (dispersive) centers in condensed matter.  Recently some new theoretical and experimental results were obtained, which concern the dual symmetry of EM-field in the matter with two type of charges and intrinsic moments of quasiparticles. Let us review given results in some details. 
So, the operator equation, describing the optical transition dynamics, has been obtained by using of transition operator method \cite{Yearchuck_Doklady}, \cite{D_Yearchuck_A_Dovlatova}. It has been shown, that given equation is operator equivalent to Landau-Lifshitz (L-L) equation \cite{Landau} in its 
difference-differential form, which takes usual differential form in continuum limit. In view of isomorphism of algebras of 
transition operators $\hat {\vec {\sigma }}_k, k \in N $ and components of the spin $S = 1/2$, the symmetry of Bloch vector $\vec {P}$ under improper rotations  and  physical meaning of all its components in optical Bloch equations  have been established. Let us remember, that optical Bloch equations are the essence of gyroscopic model for spectroscopic transitions proposed for the first time formally \cite{Abragam} by
F.Bloch \cite{Bloch}. It was concluded in \cite{Yearchuck_Doklady}, that Bloch vector is electrical dipole moment, however it is axial vector. In particular, Bloch vector is electrical intrinsic moment, which is proportional to spin moment of optical  centers (electrical "spin" moment), which was predicted by Dirac already in 1928 \cite{Dirac1928}.  
The nature of the second vector, that is, $\vec {E}$, entering
optical Bloch equations, was also clarified. It is usual electric field vector, and it is still and all axial vector.  Given conclusions and the model used at all were experimentally confirmed in \cite{Yearchuck_Yerchak} and \cite{Yearchuck_PL}, in which there was reported on the discovery of ferroelectric  and antiferroelectric  spin wave resonances. They were predicted earlier \cite{Yearchuck_Doklady} on the base of  semiclassical models  for the systems EM-field plus the chain of electrical dipole moments, interacting between themselves, correspondingly, by a mechanism like to Heisenberg exchange or antiferroelectrically. Mechanism of interaction of electrical dipole moments like to Heisenberg exchange was realized in carbynoids by means of formation of lattice of spin-Peierls $\pi$-solitons, being to be electrical dipole moment carriers, in $\pi$-subsystem of valence bonds. Simultaneously in $\sigma$-subsystem of valence bonds Su-Schrieffer-Heeger $\sigma$-polarons also produce lattice, consisting of  two Su-Schrieffer-Heeger $\sigma$-soliton sublattices,  resulting in realization of antiferroelectrical interaction between intrinsic electrical moments (electrical "spin" moments) of $\sigma$-polarons. Especially interesting, that in \cite{Yearchuck_PL} was experimentally proved, that really purely imaginary electrical "spin" moments, in full correspondence with Dirac discovery \cite{Dirac1928}, are responsible for the phenomenon of antiferroelectric  spin wave resonance in the samples studied. It seems to be understandable, that in general case electric dipole moment can be represented by the complex vector-function including along with imaginary part, which corresponds to electrical intrinsic moment, the real part, corresponding  to orbital  motion.

\section{Algebraic properties of EM-field}
Let us summarize some useful results from algebra of the complex numbers. The numbers $1$ and $i$ are usually used to be basis of the linear space of complex numbers over the field of real numbers. At the same time to any complex number $a + ib$ can be set up in conformity the $[2 \times 2]$-matrix aforesaid ( see relation (\ref{eq1abcd})).
 The matrices
\begin{equation}
\label{eq61}
\left[\begin{array} {*{20}c} 1&0  \\ 0&1 \end{array}\right], 
\left[\begin{array} {*{20}c} 0&-i  \\ i&0 \end{array}\right]
\end{equation}
produce basis for complex numbers $\{a + ib\}$, $a,b \in R$  in the linear space  of $[2 \times 2]$-matrices, defined over the field of real numbers. It is convenient often to define the space of complex numbers over the group of real positive numbers, then the dimensionality of the matrices and basis has to be duplicated, since to two unities - positive $1$ and negative $-1$ 
can be set up in conformity the $[2 \times 2]$-matrices according to biective mapping
\begin{equation}
\label{eq62}
 \xi : 1 \to \left[\begin{array} {*{20}c} 1&0  \\ 0&1  \end{array}\right], -1 \to \left[\begin{array} {*{20}c} 0&1  \\ 1&0 \end{array}\right],
\end{equation}
 which allow to recreate the  operations with negative numbers without recourse of negative numbers themselves. Consequently, the following $[4 \times 4]$-matrices, so called [0,1]-matrices, can be basis of complex numbers
\begin{equation}
\label{eq63}
\begin{split}
\raisetag{40pt}
&\zeta : 1 \to [e_1]=\left[\begin{array} {*{20}c}1&0&0&0 \\  0&1&0&0  \\ 0&0&1&0 \\  0&0&0&1 \end{array}\right], \\
&i \to  [e_2]=\left[\begin{array} {*{20}c}  0&1&0&0  \\ 0&0&1&0 \\0&0&0&1\\ 1&0&0&0 \end{array}\right],\\
&-1 \to [e_3]=\left[\begin{array} {*{20}c} 0&0&1&0 \\0&0&0&1\\ 1&0&0&0\\0&1&0&0  \end{array}\right],\\
&-i \to  [e_4]=\left[\begin{array} {*{20}c}  0&0&0&1\\ 1&0&0&0\\0&1&0&0  \\ 0&0&1&0 \end{array}\right].
\end{split}
\end{equation}
The choise of basis is ambiguous. Any four  $[4 \times 4]$  [0,1]-matrices, which satisfy the rules of cyclic recurrence
\begin{equation}
\label{eq64}
i^1 = i, i^2 = -1, i^3= -i, i^4 = 1
\end{equation}
 can be basis of complex numbers.
 In particular, the following 
$[4 \times 4]$  [0,1]-matrices
\begin{equation}
\label{eq65}
\begin{split}
\raisetag{40pt}
&[e'_1]=\left[\begin{array} {*{20}c}1&0&0&0 \\  0&1&0&0  \\ 0&0&1&0 \\  0&0&0&1\end{array}\right], [e'_2]=\left[\begin{array} {*{20}c}  0&0&1&0  \\ 0&0&0&1 \\0&1&0&0\\ 1&0&0&0 \end{array}\right],\\
&[e'_3]=\left[\begin{array} {*{20}c} 0&1&0&0 \\ 1&0&0&0 \\0&0&0&1 \\ 0&0&1&0 \end{array}\right],
[e'_4]=\left[\begin{array} {*{20}c} 0&0&0&1 \\ 0&0&1&0 \\ 1&0&0&0 \\ 0&1&0&0
\end{array}\right]
\end{split}
\end{equation}
can also be basis of complex numbers. Naturally, the set of [0,1]-matrices, given by (\ref{eq65}) is isomorphous to the set, which is given by (\ref{eq63}).
It is evident, that the system of complex numbers can be constructed by infinite number of the ways, at that cyclic basis can consist of $m$ units, $ m \in N$, starting from three. It is remarkable, that the conformity between complex numbers and matrices is realized by biective mappings. It means, that there is also existing the inverse mapping, by means of which to any   squarte matrix, belonging to the linear space with a basis given by (\ref{eq63}), or (\ref{eq65}), or any other, satisfuing the rules of cyclic recurrence like to (\ref{eq64}) can be set up in conformity the complex number. In particular to any Hermitian  matrix $H$ can be set up in conformity the following complex number
\begin{equation}
\label{eq66}
  \zeta: H \to  S + iA = \left[\begin{array} {*{20}c} S&-A  \\ A&S  \end{array}\right], 
\end{equation}
where $S$  and $A$  are symmetric and antisymmetric parts of Hermitian  matrix. 
Given short consideration allows to formulate the following statements. 

1.\textit{Quantized free EM-field is complex field in general case}.

Proof is evident and it is based on  (\ref{eq66}), if to take into account, that quantized free EM-field  can be  determined by Hermitian operators $\hat{\vec{E}}(\vec{r},t)$ and $\hat{\vec{H}}(\vec{r},t)$, representing themselves the full set of quantized free EM-field operator vector-functions, that is, they can serve for basis in corresponding operator vector-functional space (see Sec.IV). Given statement can be generalized.

2.\textit{Any quantumphysical quantity is complex quantity in general case}.

 Proof is evident and it is based on the same relationship, since any quantumphysical quantity is determined by Hermitian operator. Therefore, two sets of observables, which are determined by real functions, correspond to any quantumphysical operator quantity.

Let us consider now the general algebraic properties of EM-field to clarify the symmetry properties under improper rotations first of all of the quantities $\vec {P}$, $\vec {E}$, entering optical Bloch equations, that is, to confirm  the conlusions of Sec.1 from algebraic point of view. It is well known, that real
EM-field can be characterized by both contravariant tensor 
$F^{\mu \nu }$ (or covariant $F_{\mu \nu })$ and contravariant pseudotensor 
$\tilde {F}^{\mu \nu }$, which is dual to $F_{\mu \nu }$ (or 
covariant $\tilde {F}_{\mu \nu }$, which is dual to $F^{\mu \nu })$. For example, $\tilde {F}^{\mu \nu }$ is determined by the relationship $\tilde {F}^{\mu \nu }=\frac{1}{2}e^{\mu \nu \alpha \beta }F_{\alpha \beta }$, where $e^{\mu \nu \alpha \beta}$ is Levi-Civita 4-tensor. It seems to be understandable, that field tensors and pseudotensors  are peer and independent characteristics of EM-field. It follows immediately from general consideration 
of the geometry of Minkowski space $^{1}R_4$. Really, the geometric structure of  pseudo-Euclidean abstract space of index 1, to which Minkowski space is 
isomorphic, determines unambiguously 3 kinds of peer, independent sets of linear centereuclidian
geometrical objects - tensors, pseudotensors and spinors (spin-tensors) \cite{Rashevskii}. The simplest example of practical usage of independency of EM-field tensors and pseudotensors is  
 obtaining  of 
field invariants, see for instance \cite{Landau_Lifshitz_Field_Theory}, \cite {Stephani}, where the independence of EM-field tensors and  pseudotensors is considered to be  going without saying. Algebraic properties of union of two sets of EM-field tensor  and pseudotensor functions of 4-radius-vector $x$
 $\left\{{F}^{\mu \nu }(x)\right\}$ and   $\left\{\tilde {F}^{\mu \nu }(x)\right\}$ respectively  
are summarized in the following statements.

3.\textit{Union of contravariant (or covariant) EM-field tensors and pseudotensors} [\textit{tensor functions of 4-radius-vector} $x$, \textit{determined on some set} $S$ $\subseteq {^{1}R_4}$ \textit{in general case}] \textit{produces linear space} $\left\langle F,+,\cdot \right\rangle$ \textit{over a field of scalars} $P$, \textit{consisting of two invariant subspaces (correspondingly, tensor  and pseudotensor ones).}

4.\textit{Union of contravariant (or covariant) EM-field tensors and pseudotensors} [\textit{tensor functions of 4-radius-vector} $x$, \textit{determined on some set} $S$ $\subseteq {^{1}R_4}$ \textit{in general case}] \textit{produces linear algebra} $\mathfrak F$ = $\left\langle \mathfrak F,+,\cdot, \ast \right\rangle$ \textit{with algebraic operations of proceeding to dual elements by means of convolution with Levi-Civita 4-tensor, composition (addition) of "vectors" and multiplication of "vectors"  by scalar}.

Term "vector"   means  an element of  tensor-pseudotensor union. The proofs of statements 3 and 4 are simple, and they do not produce here.
  Let us  define on the space $\left\langle F,+,\cdot \right\rangle$ the functional $\Phi(x)$ by the following relationships
\begin{equation}
\label{eq4aa}
 \Phi[{F}^{\mu \nu}(x)] \equiv \left\langle {F}^{\mu \nu}(x) | \Phi\right\rangle = F^{\mu \nu}(x)  F_{\mu \nu}(x^*), \end{equation} and 
\begin{equation}
\label{eq4bb}
 \Phi [\tilde {F}^{\mu \nu}(x)] \equiv \left\langle \tilde {F}^{\mu \nu}(x) | \Phi \right\rangle = \tilde {F}^{\mu \nu}(x) \tilde{F}_{\mu \nu }(x^*). 
\end{equation}
We can also define on the space $\left\langle F,+,\cdot \right\rangle$ the functional $\tilde{\Phi}(x)$ by the following relationships
 \begin{equation}
\label{eq4cc}
 \tilde{\Phi}[{F}^{\mu \nu}(x)] \equiv \left\langle F^{\mu \nu}(x) | \Phi\right\rangle = F^{\mu \nu}(x) \tilde {F}_{\mu \nu}(x^*) \end{equation} and 
\begin{equation}
\label{eq4dd}
 \tilde{\Phi}[ \tilde {F}^{\mu \nu}(x)] \equiv \left\langle \tilde {F}^{\mu \nu}(x) | \Phi \right\rangle = \tilde {F}^{\mu \nu}(x) F_{\mu \nu}(x^*). 
\end{equation} 
 In Eq.(\ref{eq4aa} to \ref{eq4dd}) $x^*$ is fixed value of 4-radius-vector $x$.
It is clear, that $\Phi$ is antilinear functional on the space $F$, if  field of scalars is  field of complex numbers C. In  particular, if  field of scalars is  field of real numbers R, $\Phi$ is linear functional. Then the statement 5 takes place.

5.\textit{The set of  functionals} $\left\{\Phi [\tilde {F}^{\mu \nu}(x)]\right\}$, $\left\{\tilde{\Phi}[\tilde {F}^{\mu \nu}(x)]\right\}$, \textit{preassigned on the space} $\left\langle F,+,\cdot \right\rangle$, \textit{produces  linear space} $\left\langle \Phi',+,\cdot \right\rangle$ \textit{over a field of scalars} $P$, \textit{which is dual to the space} $\left\langle F,+,\cdot \right\rangle$, \textit{however it is nonselfdual}.

Here $\Phi'$ is union of $\Phi$ and $\tilde{\Phi}$. The proof of statement is simple, and it does not produce.
Quite analogous statement can be formulated, if instead of a field of scalars $P$ some set of pseudoscalars $\tilde P$ will be taken into consideration.  
Statement 5 (for both the cases) can be expressed shortly by the relationship 
\begin{equation}
\label{eq5a}
 \left\langle \Phi',+,\cdot \right\rangle = \left\langle {F}^{\times},+,\cdot \right\rangle.
 \end{equation} 
 The nonselfduality of  ${F}^{\times}$ seems to be substantional, and it determines the practical significance of given statement. Really, from the statement 5 follows the
necessity to take into consideration always both the spaces, that is   $\left\langle F,+,\cdot \right\rangle$ and  $\left\langle \Phi' + \cdot \right\rangle$, by the study of any physical process with participation of EM-field. More strictly, known Gelfand triple, which includes together with spaces $\left\langle F + \cdot \right\rangle$ and  $\left\langle \Phi' + \cdot \right\rangle$ the Hilbert space with topology, determined in the proper way, has to be taken into consideration, see, for example \cite{Bohm}. In other words, for full physical description of dynamical systems, interacting with EM-field, and for description of any physical phenomena at all, where EM-interaction presents, it 
is necessary to study the response to two Gelfand triples, determined  correspondingly over the scalar $P$ field and over pseudoscalar $\tilde P$ set. 

It is advisable to indicate, that pseudoscalars' set $\tilde P$ does not produce a field, although  given set  
produces an additive group. It is evident, that the union of sets of scalars $P$ and pseudoscalars $\tilde P$ produces the ring ${P^{'}}$ without unit.  It leads to union of linear space $\left\langle F,+,\cdot \right\rangle$ over a field of scalars  $P$ and linear space $\left\langle F,+,\cdot \right\rangle$ over a group of pseudoscalars $\tilde P$, if we define both the tensor functions ${F}^{\mu\nu}$ and $\tilde {F}^{\mu \nu}$ over a ring of scalar and pseudoscalar union $\tilde {P^{'}}$. Let us designate  linear space obtained $\left\langle \mathcal F,+,\cdot \right\rangle$.  The union of sets of scalars $P$ and pseudoscalars $\tilde P$  leads also to the union of linear algebras $\mathfrak F$ = $\left\langle \mathfrak F,+,\cdot, \ast \right\rangle$ prescribed over scalar $P$ field and pseudoscalar $\tilde P$ set   by means of  their definition over a ring of scalar and pseudoscalars $\tilde {P^{'}}$. It is clear, that convolution of algebra elements with Levi-Civita 
4-tensor, that is   proceeding to dual elements, realizes automorphism. 

It is easily to show, that the space $\left\langle \mathcal F,+,\cdot \right\rangle$ is selfdual. Then, foregoing practical remark can be reformulated - the solution of one or another task with EM-field participation has to be performed in the  space $\left\langle \mathcal F,+,\cdot \right\rangle$ over a ring of scalars and pseudoscalars $\tilde {P^{'}}$.  It is also evident, that partition of given space $\left\langle \mathcal F,+,\cdot \right\rangle$ into  four invariant subspaces $\left\langle\mathcal F^{(i)},+,\cdot \right\rangle$, $i = \overline {1,4}$,  can be realized. Elements of the first subspace $\left\langle \mathcal F^{(1)},+,\cdot \right\rangle$ are genuine EM-field tensors (tensor function in general case), determined over a scalar field $P$. Elements of the second subspace $\left\langle \mathcal F^{(2)},+,\cdot \right\rangle$ are also genuine EM-field tensors (tensor function in general case), determined however over a pseudoscalar $\tilde P$ additive group. Elements of the third subspace $\left\langle \mathcal F^{(3)},+,\cdot \right\rangle$ are EM-field pseudotensors (pseudotensor function in general case), determined over a scalar field $P$. Elements of the fourth subspace $\left\langle \mathcal F^{(4)},+,\cdot \right\rangle$ are EM-field pseudotensors (pseudotensor function in general case), determined now over a pseudoscalar $\tilde P$ additive group. Let us characterize the symmetry kind under improper rotations of the components of tensor elements for each subspace, that is, let us thereby establish the symmetry kind of vectors of electric field $\vec {E}$ and magnetic field $\vec {H}$. In subspaces  $\left\langle \mathcal F^{(1)},+,\cdot \right\rangle$,  $\left\langle \mathcal F^{(2)},+,\cdot \right\rangle$  vector $\vec {E}$ is polar vector and vector $\vec {H}$ is axial. Given conclusion is evident for the first subspace. Vector $\vec {E}$ in the second subspace is dual vector to antisymmetric 3-\textit{pseudo}tensor, that determines its polar character and vector $\vec {H}$, respectively, is axial. At the same time, in contrast to the case 1, the components of vector $\vec {E}$ correspond now to pure space components of EM-field  4-\textit{pseudo}tensor ${\tilde {F}^{\mu \nu}}$ in given case. The components of vector $\vec {H}$ correspond for the case 2 to time-space mixed components of given  4-\textit{pseudo}tensor, that determines  the axial symmetry of vector $\vec {H}$. Arbitrary element of subspace  $\left\langle \mathcal F^{(3)},+,\cdot \right\rangle$ can be represented in the form of
 $ \alpha {F}^{\mu \nu }(x_1) + \beta {F}^{\mu \nu }(x_2)$, where
 $\alpha, \beta \in \tilde {P}$ and $x_1, x_2 \in S \subseteq {^{1}R_4} $. It is 4-\textit{pseudo}tensor, since $\alpha, \beta \in \tilde {P}$. Its space components are in fact the components of antisymmetric 3-\textit{pseudo}tensor, which determine dual to given tensor polar magnetic field vector $\vec {H}$, while time-space mixed components are the components of axial 3-vector $\vec {E}$ of electric field.   
Therefore, the symmetry properties of the components of vectors $\vec {E}$ and $\vec {H}$ under  improper rotations in the case 3 will be opposite to the case 1. It is evident, that in the 4-th case the symmetry properties of the components of $\vec {E}$ and $\vec {H}$ under  improper rotations will be opposite to the case 2. 
 
 Given consideration has clear mathematical and physical meaning. For instance,  case 3 means mathematically,  that if,  in particular, in fixed point of 3D-space the vectors $\vec {E_0}$ and $\vec {H_0}$  are polar and axial vectors correspondingly, the fields (in vector analysis meaning) of these vectors, that is, the vector-functions, corresponding to given vectors, can have other symmetry properties. For instance, vector-functions $\vec {E}(x) = \vec {E_0} \sin x$ and $\vec {H}(x) = \vec {H_0} \sin x$ will have opposite symmetry properties in comparison with $\vec {E_0}$ and $\vec {H_0}$ under inversion of $x$-coordinate. Physically it means, that interactions of EM-field with the centers, which are 1D-, 2D- or 3D-extended in 3D-space can be quite different  in comparison with interaction of EM-field with point centers like to nuclei.

It is also understandable, that, if electric field components correspond to above considered case 3 or 4 
(and consequently electric dipole moments are also pseudovectors), the equation of dynamics of optical transitions will have the structure, which is
mathematically equivalent to the structure of the equation for 
dynamics of magnetic resonance transitions (by which magnetic field components correspond to the case 1 or 2). In other words, mathematical abstractions in optical Bloch equation
 become, in agreement with results \cite{Yearchuck_Doklady}, \cite{Yearchuck_Yerchak}, real physical meaning, that is really $\vec{E}$ is the part of intracrystalline and 
external electric field, which has axial vector symmetry, $\vec{P}$ is electrical moment, which seems to be built like 
to magnetic moment. 

Let us compare foregoing general algebraic  properties of EM-field tensors  with general algebraic properties of EM-field potentials in the matter. The components of 4-vector $A_\mu$ of EM-field potentials in the matter (like to any 4-vector) have to transform under Lorentz group representations, at that, under general Lorentz group in general case. Let the components of 4-vector $A_\mu$ transform   according to the representation $D({L^{(+)}_+})$ of proper Lorentz group ${L^{(+)}_+}$, corresponding to proper orthochronous transformations, then in the case of improper orthochronous transformations ${L^{(+)}_-}$, proper nonorthochronous ${L^{(-)}_+}$, improper nonorthochronous ${L^{(-)}_-}$ transformations of general Lorentz group $L$ \cite{Fedorov} 4-vector $A_\mu$ will transform under direct product of the following representations 
\begin{equation} 
\label{eq5} 
\begin{split}
\raisetag{40pt}
D({L^{(+)}_+}) \otimes {D(P)}, \\ D({L^{(+)}_+}) \otimes {D(P^{'})}, \\ D({L^{(+)}_+}) \otimes {D(P)} \otimes {D(P^{'})}, 
\end{split}
\end{equation} 
correspondingly, where $D(P)$ is  some representation of space inversion group $P$, $D(P^{'})$ is some representation of time reversal group $P^{'}$, where both ones are subgroups of general Lorentz group $L$. Therefore, 4-vector $A_\mu$ of EM-field potential along with 
known possibility to be polar t-even 4-vector can also be axial t-even, polar t-uneven and axial t-uneven 4-vector, if to choose uneven representations of space inversion and time reversal groups. Hence we will also have 4 possibilities for symmetries of EM-field tensor and its components under improper rotations. It corresponds,   in other words, to partition of
linear space $\left\langle\mathcal F,+,\cdot \right\rangle$ over the ring of  scalars and pseudoscalars, the vectors in which are sets of contravariant (or covariant) EM-field tensors and pseudotensors $\left\{{F}^{\mu\nu}\right\}$, $\left\{\tilde{F}^{\mu\nu}\right\}$, (or $\left\{{F}_{\mu\nu}\right\}$, $\left\{\tilde{F}_{\mu\nu}\right\}$) into 4 subspaces. However from given simple group-theory analysis does not follow the necessity to take all the field symmetry variants into consideration by  interaction of starting free EM-field with matter in contrast to foregoing more general algebraic consideration.

The result obtained allows to suggest, that free EM-field is 4-fold degenerated under improper rotations. The interaction of EM-field with device (or, generally, with some substance) can relieve degeneracy and can lead, especially by interaction with extended centers, to unusual, discussed in Sec.1, symmetry of field vector-functions. On the other hand, there is CPT-invariance requirement, and optical resonance system has to satisfy to it too. However, given contradiction seems to be illusory. To resolve it, we have to take into account, that spectroscopic transitions are not instantaneous, that is, it seems to be taking place the formation of resonance state {field + matter} or, in particular, {field + device}, which can have very long life time in comparison with field mode period, determined for instance by the time of Rabi-oscillation amplitude damping. Given conclusion becomes especially actual, if to take into account recent theoretical discovery of new quantum optics phenomenon - Rabi waves  formation \cite{Slepyan_Yerchak} for the systems with strong electron-photon interaction. By Rabi waves  formation the interaction process of the photons with matter is so long, that it can be visible by the detection by usual stationary spectroscopy methods by means of appearance of additional spectral lines, that has to be taken into consideration by interpretation of spectroscopic results. 

Realization of concrete EM-field symmetry state under improper rotations (one of 4 possible) will, evidently, be determined by symmetry characteristics of the states of centers in the matter, interacting with EM-field by taking into account the requirement of CPT-invariance for resonance state {field + matter}. In other words, suitable symmetry variant of EM-field vectors will be quasi "choosed" by centers in the matter themselves. They determine in fact the symmetry kind of EM-field, propagating in the matter, that is, in other words, the symmetry of EM-field vector-functions. 

We suggest also, that by interaction of EM-field with the matter the elementary charge carrier size, that is electron size, can be taking into consideration for the determination of the concept of lengthy centers. Electron size does not exceed the aforesaid (Sec.I) value $~10^{-16}$ cm. Let us remember, that given evaluation follows from the conclusion on applicability of quantum electrodynamics theory up to distances $~10^{-16}$ cm. Therefore, it seems to be reasonable to suggest, that any individual atom can be considered relatively the electron size to be extended  3D-center. It means, that even in atomic spectroscopy for dipole moments and electric field strengths the space coordinate dependent vector-functions have to be used instead constant (that is independent on space coordinate) vectors and first of all the symmetry of given vector-functions under improper rotations has to be taken into account. 

Therefore foregoing simple algebraic consideration of the properties of EM-field vector-functions leads in fact to conclusion, that free EM-field is complicated dynamical system and consist of two independent fields with various P-parity, which in turn  consist of two independent fields with various  t-parity. Given conclusion has to be taken into account by interpretation of experimental data.

It seems to be reasonable along with considered symmetry of EM-field under improper rotation, by taking also into account the suggested role of elementary electric charge size to be a space scaling factor, to consider in more details the gauge symmetry of EM-field, which is concerned of charges immediately.

\section{Dually charged quasiparticles - theory and its experimental confirmation}
It is remarkable, that the 
 concept to take into consideration the complex charge in EM-field theory was proposed  already in 1981 in \cite{Tolkachev} by the description of electrodynamics of dually charged particles (then, naturally, hypothetical ones), and it was done  for the first time (to our knowledge).  At the same time the point of view,  that EM-field is vector real field only and that  it cannot be characterized by any charges at all, is dominating upto now, and we understand well, that  the properties foregoing have to be theoretically argued in more detail.

\subsection{Additional gauge invariance of complex relativistic fields}

We will argue in the next Section, that EM-field in the matter has to be considered in many practical cases to be complex field, that is for correct description of the system  (EM-field + matter) two sets of real EM-field vector-functions have to be taken into consideration. The simplest example is description of optical absorption, transmission, reflection, scattering or luminescece experiments in additional external electric or magnetic field. In given Section we will prove the idea, that for any complex field the conserved quantity, corresponding to its gauge symmetry, that is charge, can be in general case also complex.

 Let $u(x)$ = $\left\{ {\,u_{i}(x) \,} \right\}$, $i = \overline{1,n}$, the set of the functions of some complex relativistic field, that is, scalar, vector or spinor field, given in some space of Lorentz group representations. It is well known, that Lagrange equations for any complex relativistic field can be represented in the form of one matrix relativistic differential equation of the first order in partial derivatives, that is in the form of so called generalized relativistic equation, and analogous equation for the field with Hermitian conjugated (complex conjugated in the case of scalar fields) functions $u^{+}(x)$ = $\left\{ {\,u_{i}^{+}(x) \,} \right\}$. The equation for the set $u(x)$ of field functions is
\begin{equation}
\label{eq4a}
(\alpha_{\mu} \partial_{\mu} + \kappa \alpha_{0}) u(x) = 0.
\end{equation}
Similar equation for the field with Hermitian conjugated (complex conjugated in the case of scalar fields) functions, that is for the functions
  $u^{+}(x) = \left\{ {\,u_{i}^{+}(x) \,} \right\}$,   $i = \overline{1,n}$, is
\begin{equation}
\label{eq4b}
\partial_{\mu} u^{+}(x) \alpha_{\mu} + \kappa u^{+}(x) \alpha_{0} = 0.
\end{equation}

In equations (\ref{eq4a}, \ref{eq4b}) $\alpha_{\mu}, \alpha_{0}$ are matrices with constant numerical elements. They have an order, which coincides with dimension of corresponding space of Lorentz group representation, realized by $\left\{ {\,u_{i}(x) \,} \right\}$, $i = \overline{1,n}$. In particular, they are $[n\times n]$- matrices, if $\left\{ {\,u_{i}(x) \,} \right\}$, $i = \overline{1,n}$ are scalar functions. It is evident, that the transformation
\begin{equation}
\label{eq10}
u'(x) = \beta exp(i \alpha) u(x),
\end{equation}
where $\alpha,\beta\in R$,  and analogous transformation for Hermitian conjugate functions (or complex conjugate functions in the case of scalar fields) 
\begin{equation}
\label{eq11}
u'^{+}(x) = \beta exp(-i \alpha) u^{+}(x)
\end{equation}
keep Lagrange equations $(\ref{eq4a}, \ref{eq4b})$ to be invariant. It is understandable that transformation of field functions by relationships  $(\ref{eq10}), (\ref{eq11})$ is equivalent to multiplication of field functions  by arbitrary complex number. It is well known, that given linear transformation  is the simplest example of isomorphism of corresponding linear space, which is given over the field of complex numbers, onto itself, that is, in the case considered  the relationships  $(\ref{eq10}, \ref{eq11})$ give  automorphism of the space of  field functions.  Automorphism of any linear  space leads to some useful properties of the objects, which belong to given space. For instance, if to set up in a correspondence to the space of field function the affine space, then conservation laws of collinearity of the points and of simple relation of the triple of collinear points will be fulfilled by automorphism in given affine space. Consequently,  we have to expect the physical consequences of given algebraic property in the case of physical spaces. Conformably to the case considered
 we have in fact gauge transformation of field functions, which is more general in comparison with usually used. The set $(\beta exp(-i \alpha)$ for all possible $\alpha$, $\beta \in R$ produces the group $\Gamma$, which is direct product of known symmetry group $U_1$, and multiplicative group $\mathfrak R$ of all real numbers (without zero). Therefore, in the case considered the symmetry group of given complex field acquires additional parameter. So, we will have
\begin{equation}
\label{eq12}
\Gamma(\alpha, \beta) = U_1(\alpha) \otimes \mathfrak R (\beta)
\end{equation}
Let us find the irreducible representations of the group $\mathfrak R(\beta)$. It has to be taken into account, that the group $\mathfrak R(\beta)$ is abelian group and its irreducible representations $T(\mathfrak R)$ are onedimensional. So, the mapping
\begin{equation}
\label{eq13}
T: \mathfrak R \rightarrow T(\mathfrak R)
\end{equation}
is isomorphism, where
$T(1) = 1$.
Therefore, for $\forall (\beta, \gamma)$ of pair of elements of group $\mathfrak R (\beta)$ the following relationship takes place
\begin{equation}
\label{eq14}
T(\beta, \gamma) = T(\beta) T(\gamma).
\end{equation}
Then, it is easy to show,
that 
\begin{equation}
\label{eq16}
T(\beta) = \beta^ {\frac{\partial{T}}{\partial\gamma}{(1)}}.
\end{equation}
The value ${\frac{\partial{T}}{\partial\gamma}{(1)}}$ can be obtained from the condition
\begin{equation}
\label{eq17}
T(-\beta)= -T(\beta).
\end{equation}
Consequently, we have
\begin{equation}
\label{eq18a}
T(\beta) = \beta^{2k+1}= exp[{(2k+1) ln\beta}],
\end{equation}
where $k \in N$. Then irreducible representations of the group $\Gamma(\alpha, \beta)$ represent
 direct product of irreducible representations of the groups $U_1(\alpha)$ and $\mathfrak R(\beta)$ 
\begin{equation}
\label{eq18b}
T(U_1(\alpha)) \otimes T(\mathfrak R(\beta)) = exp(-i m \alpha) exp[{(2k+1) ln\beta}],
\end{equation}
where
$m, k = 0, \pm1, \pm2, ...$ .

It is clear, that some conserved quantity has to correspond to gauge symmetry of the field, which is determined by the group $\mathfrak R(\beta)$. Thus we arrive at a formulation of the following statement.

6.\textit{Conserving quantity - complex charge, which is invariant under total gauge transformations, corresponds to any complex relativistic field (scalar, vector, spinor).}
 
\textbf{Proof}.
 Really, since generalized relativistic equations are invariant under transfomations ($\ref{eq10}, \ref{eq11}$) and variation of action integral with starting Lagrangian is equal to zero, then variation of action integral with transformed Lagrangian in accordance with ($\ref{eq10}, \ref{eq11}$) will also be zero. Consequently, all the conditions of applicability of N\"{o}ther theorem, by proof of which the only invariance under Lagrange equations is sufficient, \cite{Noether}, are held true. According to N\"{o}ther theorem, the conserved quantity, corresponding to $\nu -th$ parameter ($\nu = \overline{1,k}$) by invariance of field under some $\textit{k}$-parametric symmetry group, is
\begin{equation}
\label{eq19a}
Q_\nu(\sigma) = \int\limits_{(\sigma)}\theta_{\mu \nu}d\sigma_\mu = const,
\end{equation}
where
\begin{equation}
\label{eq19b}
\theta_{\mu \nu} = \frac{\partial{L}}{\partial(\partial_{\mu}u_i)} [\partial_{\rho}u_i X_{\rho \nu} - Y_{i \nu}] - L X_{\mu\nu},
\end{equation}
L is field Lagrangian, $\sigma$ is any spacelike hypersurface, $\sigma$ $\subset {^{1}R_4}$.  We have to pay attention to typical mistake,  which is abundant in the literature, consisting in that, that for applicability of N\"{o}ther theorem the Lagrangian invariance under corresponding symmetry transformations is required. At the same time the only invariance of Lagrange equations under corresponding symmetry transformations, which certainly takes place in given case, is necessary (see proof of N\"{o}ther theorem). The matrices $X_{\rho \nu}, \text{Y}_{i \nu}$ are determined by matrix representations $\left\|(I_{\nu})_{\mu\ \nu}\right\|$ and $\left\|(J_{\nu})_{i k}\right\|$ of infinitesimal operators of symmetry group in coordinate space and in the space of field functions respectively in accordance with the following relationships
\begin{equation}
\label{eq19c}
X_{\rho \nu} = (I_{\nu})_{\mu \alpha} x_\alpha,
Y_{i \nu} = (J_{\nu})_{i k} u_{k}.
\end{equation}
So, using N\"{o}ther theorem, we obtain for 4-vector $\theta_\mu$ the following expression
\begin{equation}
\label{eq19}
\theta_{\mu} = -\frac{\partial{L}}{\partial(\partial_{\mu}u_i)} u_{i} -\frac{\partial{L}}{\partial(\partial_{\mu}u_i^{*})} u_{i}^*,
\end{equation}
Components of 4-vector $\theta_\mu$ satisfy to continuity equation
\begin{equation}
\label{eq20}
\partial_{\mu}{\theta_{\mu}} = 0.
\end{equation}
Conserving quantity, corresponding to (\ref {eq19}), is also charge, which is equal to
\begin{equation}
\label{eq21}
Q^{'}_{2} = iQ_{2} = -i \int\theta_{4}d^{3}x.
\end{equation}
So $iQ_{2}$ is determined by relationship 
\begin{equation}
\label{eq22}
iQ_{2} = i\int[\frac{\partial{L}}{\partial(\partial_{4}u_i)} u_{i} + \frac{\partial{L}}{\partial(\partial_{4}u_i^{*})} u_{i}^*]d^{3}x.
\end{equation}

It is seen from relationship (\ref{eq22}), that obtained additional charge is purely imaginary quantity. At the same time known conserved quantity for any complex field, for instance, for Dirac field, is well known real guantity - charge $Q_{1}$, which is consequence of gauge symmetry, consisting in the invariance of Lagrange equations under the transformations
\begin{equation}
\label{eq22a}
u'(x) = exp(i \alpha) u(x) 
\end{equation}
and
\begin{equation}
\label{eq22b}
 u'^{+}(x) = exp(-i \alpha) u^{+}(x). 
\end{equation}
In general case $Q_{1}$, see, for instance, \cite{Bogush}, is 
\begin{equation}
\label{eq23}
Q_{1} = -\int[\frac{\partial{L}}{\partial(\partial_{4}u_i)} u_{i} - \frac{\partial{L}}{\partial(\partial_{4}u_i)^{*}} u_{i}^*]d^{3}x. 
\end{equation}
 
 Therefore
 any relativistic complex field can be characterized by complex conserving quantity $Q$, that is by complex charge, which can be represented in the form
\begin{equation} 
\label{eq23a} 
Q = Q_{1} + iQ_{2}. 
\end{equation} 
The  statement is proved.

It is remakable, that, like to mechanics, a number of conservation laws, which can have  EM-field, are optional in their simultaneous fulfilment. In particular, it is evident, that by automorphic transformation of the space of EM-field functions by relationship (\ref{eq10}) the conservation law for charge will  always take place. At the same time the energy conservation law and the conservation of Poynting vector will be fullfilled, if given transformation is applied to EM-field potentials.  The force characteristics, that is $\vec{E}$-, $\vec{H}$-vector functions can be used to be basis for free EM-field description, since they will represent the full set in free  EM-field case. However the energy conservation law and the conservation of Poynting vector, that is mathematical construction, to which enter $\vec{E}$-, $\vec{H}$-vector functions, will not  be fullfilled by transformation (\ref{eq10}) at arbitrary $\beta$.  We see, that the charge conservation law for EM-field is fullfilled  even through the energy conservation law does not take place.  Therefore, the charge conservation law can be considered in given meaning to be more fundamental.

There seems to be essential, that the existence of dually charged particles in condensed matter was  experimentally confirmed. Given conclusion was done  from the comparison of experimental results, reported in \cite{Ertchak_J_Physics_Condensed_Matter}, \cite{Ertchak_Carbyne_and_Carbynoid_Structures} and in \cite{Yearchuck_Yerchak}. Let us give some details, concerning given experimental confirmation (which was done to our knowledge for the first time). Carbynoid samples, studied in cited works, were active both in magnetic resonance and optical spectroscopy [infrarot (IR) absorption, reflection and Raman scattering (RS) spectroscopy]. Carbynoid samples, representing themselves the systems of carbon chains, can be considered to be the most simple modelling systems for verification of theoretically predicted effects for quasionedimensional structures with very many practical applications. The samples (designated "samples A" in \cite{Yearchuck_Yerchak}), in which the ferromagnetic and ferroelectric spin wave resonances (FMSWR and FESWR) were observed on the same chain structures, have been used. Both the resonances can be described by in fact the same equations, obtained in \cite{Yearchuck_Doklady} and modified by taking into consideration 
the relaxation processes in  \cite{Yearchuck_Yerchak}. So, the equation for the description of optical transition dynamics in a chain  
is 
\begin{equation}
\label{eq1}
\begin{split}
\raisetag{40pt}
&\frac{\partial \vec {S}(z)}{\partial t} = \left[ {\vec {S}(z)\times \gamma_{E} 
\vec {E}} \right] -\\
&\frac{4a^2J_{E}}{\hbar ^2}\left[ {\vec {S}(z)\times \nabla 
^2\vec {S}(z)} \right] + \frac{\vec {S}(z)-\vec {S}_0 (z)}{\tau },
\end{split}
\end{equation}
where $\vec {S}(z)$ is electric analogue of intrinsic magnetic moment, $\gamma_{E}$, $J_{E}$ are optical analogues of gyromagnetic ratio and exchange 
interaction constant respectively, $\hbar$ is Planck's constant, $a$ is lattice 
spacing, $\vec {E}$ is electric field, 
vector-function $\vec {S}_0 (z)$ is equilibrium value of axially symmetric electrical dipole moment (in particular, electrical "spin" moment), $\tau$ is relaxation time. 
Vector-functions $\vec {S}(z)$, $\vec {S}_0 (z)$ acquire in the case of FMSWR the meaning of magnetic  moment vector-functions, $E_{1} $ is replaced in FMSWR-case by $H_{1}$, that is, by amplitude of magnetic component of external oscillating EM-field, $J_{E}$ is replaced by the exchange interaction constant 
$J_{H}$, $\gamma_{E}$ by $\gamma_{H}$. For the agreement with experiments, in which the values of oscillating external electric field 
components $E^x, E^y$ in $(\ref{eq1})$ are greatly less in 
comparison with the value of intracrystalline electric field component 
$E^z$, and consequently under analogous relationships between corresponding  components of total 
electrical dipole moment (since x-, y-dipole components are induced by weak external electric field)
 the linearized equation in \cite{Yearchuck_Yerchak} was obtained. It was solved under  additional assumption, that equilibrium distribution of $\vec {S}_0 (z)$ along the 
chain is homogeneous. All the assumptions were entirely correct for IR measurements. The solution of linearized equation gives 
 the relationships for a shape and 
amplitudes of resonance modes and for dispersion law. They are
\begin{equation}
\label{eq2}
a_{n} = \left\{ {{\begin{array}{*{40}c}
 {-\frac{i \gamma_{E} S \tau^2 E_{1}}{\pi n} \frac{\left[{(\omega_n - \omega
) - \frac{i}{\tau}} \right]}{\left[ {1 + (\omega_{n} - \omega)^2 \tau^2} 
\right]},\,\,n = 1,\,3,\,5,... \hfill} \\
{\,0,\,\,\,\,\,\,\,\,\,\,\,\,\,\,\,\,\,\,\,\,\,\,\,\,\,\,\,\,\,\,\,\,\,\,\,\,\,n = 2,\,4,\,6,...} 
\hfill, \\
\end{array} }} \right. 
\end{equation}
\begin{equation}
\label{eq3}
\nu _n =\nu _0 - \mathfrak A n^2,
\end{equation}
where $n\in N$ including zero, $\nu_{n}$ is a frequency of \textit{n-th} mode, $\mathfrak A$ is 
a material parameter ($\mathfrak A = \frac{2 \pi {a}^{2} S \left|J_E\right|}{\hbar^2 L^{2}}  > 0, L$ is chain length).
Like to magnetic resonance measurements, $Re\,a_n$ is proportional to absorption signal, $Im\,a_n$ is proportional to dispersion signal. 
It was found the following.

1.Dispersion law (\ref{eq3})  holds true both by IR- and RS-detection of FESWR. 

2.The excitation of the only uneven modes by IR-detection (by which the experimental conditions were corresponding to applicability of linearized equation) takes place in accordance with (\ref{eq2}). 

3.Inversely proportional dependence  of the amplitudes of modes (at resonance)  on mode number $n$ holds also true in accordance with (\ref{eq2}), however also the only by IR-detection of FESWR. 

4.Splitting of Raman active vibration modes is characterized by the value of parameter $\mathfrak{A}$, being greater
approximately by factor 2, than parameter $\mathfrak{A}$, which characterizes IR FESWR spectra 
(by the frequencies of zero modes reduced by means of linear approximation 
procedure to the same value),  that  confirms well the theoretical prediction in \cite{Yearchuck_Doklady}.

 Moreover, the value of spin $S$, equal for optically active local centers studied to 1/2, has been obtained   for the first time   from pure optical experiment at all, demonstrating the advantages of transition operator method in comparison with density matrix formalism in description of dynamics of spectroscopic transitions, which was discussed in detail in  \cite{D_Yearchuck_A_Dovlatova}. 
The agreement between theory and experiment obtained denotes unambiguously  the reliability of identification of  FESWR phenomenon  on the one hand and  the validity of the conclusion on existence of axial symmetry of electric vector-functions in spectroscopic transition phenomena on the other hand.  Analogous conclusions on coincidence between  theory and experiment  take place for AFESWR phenomenon. We wish to pay attention, that both the phenomena are rather general and there are many experimental works, where FESWR and especially AFESWR were registered, however they were unidentified. For instance, we can point out  the work \cite{Winter_Kuzmany} with unidentified FESWR registration. In \cite{Winter_Kuzmany} the characteristic splitting of two lowest (Hg)-modes, corresponding  to  the phenomenon of FESWR, has been observed  by Raman scattering study in a single crystal of metallic potassium fulleride at 80K. Our analysis of given results gives the value of splitting parameter $\sim 1.6 cm^{-1} $ for (Hg(2))-mode, that is substantially smaller in comparison with splitting parameter in $ 62.2 cm^{-1}$, which was found in  carbynoid samples.  It allows therefore to compare quantitatively the effectivity of exchange interaction between electric dipole moments in carbynoids and potassium fulleride. We see, that it differs by more than order of value. It concerns also the first task.

The most substantional for the second task is that, that  the ratio of two exchange constants $J_{E }/J_{H}$  
was obtained from the ratio of  the values of splitting parameters  $\mathfrak{A}^E$ in IR FESWR spectra and $\mathfrak{A}^H$ in the spectra of FMSWR in A-sample reported earlier in \cite{Ertchak_J_Physics_Condensed_Matter}, \cite{Ertchak_Carbyne_and_Carbynoid_Structures}. The range of given ratio is $(1.2 - 1.6)10^{4}$. The 
appearance of two exchange constants, which differ  each other more 
than 4 order for the same chain structures is in fact the direct indication, that EM-field in the matter has along with complex vector characteristics the complex scalar characteristic - charge, which can be proposed to be consisting of elecric and magnetic components,  corresponding to real and imaginary parts correspondingly. Moreover, the values of  the ratio $J_{E }/J_{H}$ of exchange constants, observed in the same sample on the same carbon chains, allows  to evaluate the ratio of electric and magnetic $e_{H} \equiv g$ to $e_{E}\equiv e$  components of complex charge of simultaneously optically and magnetically active centers - spin-Peierls solitons. It is sufficient to take into account the relation  $\frac{g}{e} \sim \sqrt{\frac{J_{E}}{J_{H}}} $, which is valid in the suggestion, that the change of the space distribution of the wave function  of given local quasiparticles  (characteristic size of spin-Peierls solitons is evaluated in 20 interatomic units) in magnetic field and by its absence is vanishing. Hence $\frac{g}{e}\approx (1.1 - 1.3)10^{2}$.    It is obviously, that given numerical evaluation is agreeing well with Dirac charge quantization theory \cite{Dirac P.A M}, that is, with relationship $\frac{g}{e} = 68.5 e n$, where $n = ±1, ±2,...,$ at n = 2. Some deviation is quite acceptably, since the guess is rather crude. Given result seems to be strong  experimental proof of existence of both types of charges (electric and magnetic ones) in condensed matter. It leads to conclusion, that EM-field with the sources is also  dually symmetric field in general case. Therefore, given comparison allows to conclude on experimental discovery of dually charged particles, the existence of which and their properties were intensively theoretically discussed in a number of publications of Belarusian theoretical physics branch of V.Fock schools of thought, see for instance, monographs \cite{Tomilchick}, \cite{Berezin} and many references therein.

Although spin-Peierls solitons are quasiparticles, which are relatively strong localized   in comparison with chain length, their characteritic localization length (and, consequently, upper boundary for characteritic localization length of magnetic charge) is exceeding in more than eight order the very strong  localization degree of electric charge, realized  by electrons. Electron size does not exceed the value $~10^{-16}$ cm \cite {Ph.Enc.}. At the same time  experimental data discussed, indicating on the existence of magnetic charge, do not indicate lower space localization boundary for given two magnetic charge quanta. It means, that  the existence of Dirac monopoles, that is, the particles with strong localization degree of magnetic charge, being to be comparable with the localization degree of electric charge in electrons remains to be open. Nevertheless, it seems to be interesting, that there is the correlation between the ratio of  upper boundary for localization degree of magnetic charge of spin-Peierls solitons to  localization degree of electric charge in electrons and the ratio $J_{E }/J_{H} = (1.2 - 1.6)10^{4}$ of exchange 
constants. Square root of the ratio of characteritic localization length of spin-Peierls solitons in carbynoids to localization length of electric charge in electrons is evaluated in $\sim 5 10^{4}$, that is, both the ratios are  of the same order of values. It can additionally mean, that both the scalar characteristics, that is,  electric and magnetic charges can be quantized, at that the ratio of both the quanta has to be the conserving quantity.  

Therefore from analysis of known experimental and theoretical works we have two reasonable  conclusions of  the real existence of  new general peculiatities of  EM-field. They are 1) EM-field has compound character, that is, it represents itself the superposition of components with various time inversion (t) and space inversion (P) parities.  On the other hand above considered analysis shows, that dual transformations, determined by (\ref{eq1bcd}) convert real EM-field into complex (and vice versa). It means, that  2) EM-field itself has to possess in general case by the charge, which seems  to be also complex in general case.

\section{Quaternion Nature  of EM-Field}

The existing interpretation of Maxwell equations with sources have some dissatisfaction of a following nature. In fact, every equation in mathematical form A = B means that A and B are different notations for the
same quantity. In  existing interpretation of Maxwell equations are absent the indications concerning the physical quantity, which has to characterise simultaneously the local properties of the field and  local properties  of the charged particles,
that is, which has to be  represented simultaneously through the field derivatives, and through the characteristics of
the charged particles.  If the EM-field, considered to be a physical
object, cannot carry  charge, then from
physical point of view both  two quantities have
different physical nature and cannot be equivalent. 
The presence of complex charge to be characteristic of EM-field allows to remove given contradiction. It means that 4-vector of current $j_\mu$ for any complex field is complex vector. In its turn, it means, that independently on starting origin of the charges and currents in the matter [they can be result of presence of Dirac field or another complex field] all the characteristics of EM-field in the matter have also to be complex-valued. Given conclusion follows immediately from Maxwell equations, since complex current $j_\mu$ enters explicitly Maxwell equations. It is also substantial, that Maxwell equations are invariant under the transformation of EM-field functions by relationship (\ref{eq10}). 

Now we will show, that generalization of Maxwell equations by means of their representation in quaternion form which represent itself four independent constituents of EM-field is direct consequence of dual invariance symmetry. 

\subsection{Generalized  Maxwell Equations}
Symmetry studies of electromagnetic (EM) field have a long history, which was starting already in 19-th century from the work of Heaviside \cite{Heaviside}, where the existence of the symmetry of Maxwell equations under  electrical and magnetic quantities was  remarked for the first time. Mathematical formulation of given symmetry gave Larmor \cite{Larmor}. It is consisting in  invariance of Maxwell equations for free EM-field under the  transformations
\begin{equation} 
\label{eq1aq}
\vec {E} \rightarrow \pm\vec {H}, \vec {H} \rightarrow \mp\vec {E},
\end{equation}
The  transformations (\ref{eq1aq}) are called duality transformations, or  Larmor  transformations.
 Larmor transformations (\ref{eq1aq}) are particular case of the more general dual transformations, established by Rainich \cite{Rainich}. Dual transformations produce oneparametric abelian  group $U_1$, which is subgroup  of the group of chiral transformations of massless fields.  Dual transformations correspond to irreducible representation of  the group of chiral transformations of massless fields in particular case of quantum number $j = 1$ \cite{Tomilchick} and they are 
\begin{equation} 
\label{eq1bq}
\begin{split}
\raisetag{40pt}                                                    
\vec {E'} \rightarrow \vec {E} \cos\theta + \vec {H} \sin\theta\\
\vec {H'} \rightarrow \vec {H} \cos\theta - \vec {E} \sin\theta,
\end{split}
\end{equation}
where parameter $\theta$ is arbitrary continuous variable,
$\theta \in [0,2\pi]$. In fact the expression (\ref{eq1bq}) is indication in implicit form on compound character of EM-field. Really at fixed $\theta$
the expression (\ref{eq1bq}) will be mathematically correct, if vector-functions $\vec {E}$, $\vec {H}$  will   have the same symmetry under improper rotations, that is concerning the parity $P$ under space inversion, both be polar or axial ones, or be both consisting of polar and axial components simultaneously. Analogous conclusion takes place regarding the parity $t$ under time reversal. The possibility to have the same symmetry, that is, the situation, when both the vector-functions $\vec {E}$, $\vec {H}$  are pure polar (axial) vector-functions, or both ones t-even (t-uneven) simultaneously contradicts to experiment. Consequently it remains the variant, that vector-functions $ \vec {E'}$, $\vec {H'}$ in the expression (\ref{eq1bq}) are compound and consits of the components with even and uneven parities under improper rotations. It  is in agreement with overview on compound symmetry structure of EM-field vector-functions $\vec {E}(\vec{r},t)$, $\vec {H}(\vec{r},t)$, $\vec {D}(\vec{r},t)$, $\vec {B}(\vec{r},t)$, consisting of both the gradient and solenoidal parts, that is uneven and even parts under space inversion in \cite{Tomilchick}, where compound symmetry structure of EM-field vector-functions is represented to be self-evident. It corresponds also to theoretical assumption in \cite{Berezin}, where along with usual choice, that is, that electric field  $\vec {E}$ is polar vector, magnetic field  $\vec {H}$ is axial vector, the alternative choice is provided. The conclusion can be easily proved, if to represent relation (\ref{eq1bq}) in matrix form  

\begin{equation}
\label{eq1bcq}
 \left[\begin{array} {*{20}c}  \vec {E'} \\ \vec {H'} \end{array}\right] = \left[\begin{array} {*{20}c} \cos\theta&\sin\theta  \\-\sin\theta&\cos\theta \end{array}\right]\left[\begin{array} {*{20}c}  \vec {E} \\ \vec {H} \end{array}\right].
\end{equation}
We see, that given matrix has the form, which allows set up in conformity to it the comlex number  according to biective mapping like to (\ref{eq1abcd}).
Consequently, we have
\begin{equation}
\label{eq1bcdq}
 \left[\begin{array} {*{20}c}  \vec {E'} \\ \vec {H'} \end{array}\right] = e^ {-i\theta}\left[\begin{array} {*{20}c}  \vec {E} \\ \vec {H} \end{array}\right],
\end{equation}
that is 
 \begin{equation} 
\label{eq1cq}
\begin{split}
\raisetag{40pt}                                                    
&\vec {E'} = \vec {E} \cos\theta - i\vec {E} \sin\theta\\
&\vec {H'} = \vec {H} \cos\theta - i\vec {H} \sin\theta.
\end{split}
\end{equation}
  It means, that to real plane, which is determined  by the vectors  $\vec {E}$ and $\vec {H}$  can be set in conformity the complex plane for the vectors $\vec {E'}$ and $\vec {H'}$.
                    
It is evident now, that really  both the vectors $\vec{E'}$ and $\vec{H'}$ are consisting of  both,  $P$-even  and $P$-uneven components. So the first component of, for instance, $\vec {E'}$ will be $P$-even under reflection in the plane situated transversely to absciss-axis, the second component will be $P$-uneven. 

Therefore, dual transformation symmetry of Maxwell equations, established by Rainich \cite{Rainich}, indicates simultaneously on both complex nature of EM-field in general case, and that both electric and magnetic fields are consisting in general case of the components with various parity under improper rotations. 

Let us find the invariants of dually transformed EM-field. It is easily to show, that  the following relationship is taking place
\begin{equation}
\label{eq1bcdeq}
 \left[ \vec {E}^2 - \vec {H}^2 + 2i(\vec {E}\vec {H})\right]  e^ {-2i\theta} =inv,
\end{equation}
that is, we have at fixed parameter $\theta \neq 0$  two real EM-field invariants
\begin{equation} 
\label{eq1cdeq}
\begin{split}
\raisetag{40pt}                                                    
&(\vec {E}^2 - \vec {H}^2) \cos 2\theta + 2(\vec {E}\vec {H}) \sin 2\theta = inv \\
&2(\vec {E}\vec {H}) \cos 2\theta  - (\vec {E}^2 - \vec {H}^2) \sin 2\theta = inv.
\end{split}
\end{equation}
  It follows  from  relation (\ref{eq1cdeq}) that, in particular,  at $\theta = 0$ we have well known \cite{Landau_Lifshitz_Field_Theory} EM-field invariants
\begin{equation} 
\label{eq1cdefq}
\begin{split}
\raisetag{40pt}                                                    
&(\vec {E}^2 - \vec {H}^2)  = inv \\
&(\vec {E}\vec {H}) = inv.
\end{split}
\end{equation}
  It is interesting, that at $\theta = 45\textdegree$  and at $\theta = 90\textdegree$ the invariants of dually transformed EM-field  are determined by the same relation (\ref{eq1cdefq}) and by arbitrary  $\theta$ we have two linearly independent combinations of given known invariants.

Subsequent extension of dual symmetry for the EM-field with sources leads also to known requirement of the existence of two type of other physical quantities - two type of charges and two type of  intrinsic moments of the particles or the absorbing (dispersive) centers in condensed matter. They  can be considered to be the components of complex charge, or dual charge (in another equivalent terminology) of so called dually charged particles, which were described theoretically  (see \cite{Tomilchick}). We have to remark, that setting into EM-field theory of two type of charges and two type of  intrinsic moments is actual, since recent discovery of dually charged quasiparticles \cite{Yearchuck_Yerchak} and the particles with pure imaginary electric intrinsic moments \cite{Yearchuck_PL} in condensed matter, - respectively, spin-Peierls $\pi$-solitons and Su-Schrieffer-Heeger $\sigma$-solitons in carbynoids, is clear experimental proof, that  EM-field theory with complex charges and complex intrinsic moments has physical content. EM-field theory with complex charges and complex intrinsic moments ceases consequently to be the only formal  model, which  although is very suitable for  many technical calculations, but was considered upto now to be mathematical abstraction, in which magnetic charges and magnetic currents are fictitious quantities. Similar conclusion concerns
 the conception of complex characteristics of EM-field in the matter.  Given conception agrees well with all practice of electric circuits' calculations. Very fruitfull mathematical method for electric circuits' calculations, which uses all complex electric characteristics, see for example \cite {Angot}, was also considered earlier the only to be  formal, but  convenient mathematical technique. Informality of given technique gets now natural explanation.
 It has to be taken into account however,  that in the case $\theta = 0$ we have well known electrodynamics with odd P-parity of electric field and even P-parity of magnetic field. In given case all the EM-field characteristics   are real  and by using in calculations of the complex  quantities, we have always  add the corresponding complex conjugate  quantities. At the same time in the case $\theta \neq 0$ it will be incorrect, since both the real observable quantities, which corresponds to any complex EM-field characteristics  have  to be retained. Independent conclusion follows  also from gauge invariance above considered (Sec.I).
Really, the presence of complex charge means that 4-vector of current $j_\mu$ for any complex field is complex vector. In its turn, it means, that independently on starting origin of the charges and currents in the matter [they can be result of presence of Dirac field or another complex field] all the characteristics of EM-field in the matter have also to be complex-valued. Given conclusion follows immediately from Maxwell equations, since complex current $j_\mu$ enters explicitly Maxwell equations. It is also substantial, that Maxwell equations are invariant under the transformation of EM-field functions by relationship (\ref{eq10}).

Let us designate the terms in (\ref{eq1cq})
 \begin{equation} 
\label{eq10c}
\begin{split}
\raisetag{40pt}                                                    
& \vec {E} \cos\theta = \vec{E}^{[1]},  \vec {E} \sin\theta = \vec{E}^{[2]} \\
&\vec {H} \cos\theta = \vec{H}^{[1]}, \vec {H} \sin\theta =  \vec{H}^{[2]}.
\end{split}
\end{equation}
The Maxwell equations for the EM-field ($\vec {E'}$, $\vec {H'}$) in the matter in general case of both type of charged particles (that is electrically and magnetically charged), including dually charged particles are
\begin{equation}
\label{eq4abc}
\left[\nabla\times\vec{E'}(\vec{r},t)\right] = -\mu_0 \frac{\partial \vec{H'}(\vec{r},t)}{\partial t} - \vec{j'_g}(\vec{r},t), 
\end{equation}
\begin{equation}
\label{eq2abc}
\left[\nabla\times\vec{H'}(\vec{r},t)\right] = \epsilon_0 \frac{\partial\vec{E'}(\vec{r},t)}{\partial t } + \vec{j'_e}(\vec{r},t), 
\end{equation}
\begin{equation}
\label{eq4abcd}
(\nabla \cdot \vec{E'}(\vec{r},t)) = \rho'_e(\vec{r},t),
\end{equation}
\begin{equation}
\label{eq4abcde}
(\nabla \cdot \vec{H'}(\vec{r},t)) = \rho'_g(\vec{r},t),
\end{equation}
where $\vec{j'_e}(\vec{r},t)$,  $\vec{j'_g}(\vec{r},t)$ are respectively electric  and magnetic current densities,  $\rho^{'}_e(\vec{r},t)$, $\rho^{'}_g(\vec{r},t)$ are respectively electric  and magnetic charge densities.
Taking into account the relation (\ref{eq1c}) and (\ref{eq10c}) the system (\ref{eq4abc}), (\ref{eq2abc}), (\ref{eq4abcd}), (\ref{eq4abcde}) can be rewritten
\begin{equation}
\label{eq5abc}
\begin{split}
\raisetag{40pt}
&\left[\nabla\times(\vec{E}^{[1]}(\vec{r},t) - i\vec{E}^{[2]}(\vec{r},t)) \right] = \\
&- \mu_0 \left[\frac{\partial \vec{H}^{[1]}(\vec{r},t)}{\partial t} - i \frac{\partial \vec{H}^{[2]}(\vec{r},t)}{\partial t}\right]\\ 
&- \vec{j_g}^{[1]}(\vec{r},t) + i \vec{j_g}^{[2]}(\vec{r},t), 
\end{split}
\end{equation}
\begin{equation}
\label{eq6bcd}
\begin{split}
\raisetag{40pt}
&\left[\nabla\times(\vec{H}^{[1]}(\vec{r},t) - i\vec{H}^{[2]}(\vec{r},t)) \right] = \\
&\epsilon_0 \left[\frac{\partial \vec{E}^{[1]}(\vec{r},t)}{\partial t} - i \frac{\partial \vec{E}^{[2]}(\vec{r},t)}{\partial t}\right]\\ 
&+ \vec{j_e}^{[1]}(\vec{r},t) - i \vec{j_e}^{[2]}(\vec{r},t), 
\end{split}
\end{equation}
\begin{equation}
\label{eq7abcd}
(\nabla \cdot (\vec{E}^{[1]}(\vec{r},t) - i\vec{E}^{[2]}(\vec{r},t))) = \rho^{[1]}_e(\vec{r},t) - i\rho^{[2]}_e(\vec{r},t),
\end{equation}
\begin{equation}
\label{eq8abcd}
(\nabla \cdot (\vec{H}^{[1]}(\vec{r},t) - i\vec{H}^{[2]}(\vec{r},t))) = \rho^{[1]}_g(\vec{r},t) - i\rho^{[2]}_g(\vec{r},t),
\end{equation}
where $\vec{j_e}^{[1]}(\vec{r},t)$, $\vec{j_e}^{[2]}(\vec{r},t)$, $\vec{j_g}^{[1]}(\vec{r},t)$,  $\vec{j_g}^{[2]}(\vec{r},t)$ are correspondingly electric  and magnetic current densities, which by dual transformations are obeing to relation like to (\ref{eq1c}) \cite{Tomilchick}, they  are designated like to (\ref{eq10c}),
$\rho^{[1]}_e(\vec{r},t)$, $\rho^{[2]}_e(\vec{r},t)$, $\rho^{[1]}_g(\vec{r},t)$,  $\rho^{[2]}_g(\vec{r},t)$ are correspondingly electric  and magnetic charge densities, which transformed and designated like to field strengths and currents. In fact in the system of equations (\ref{eq5abc}), (\ref{eq6bcd}), (\ref{eq7abcd}), (\ref{eq8abcd}) are integrated Maxwell equations for two kinds of 
EM-fields (photon fields in quantum case), which differ by parities of vector and scalar quantities, entering in equations, under space inversion. 
 So, the components $\vec{E}^{[1]}(\vec{r},t)$, $\vec{H}^{[2]}(\vec{r},t)$, $\vec{j_e}^{[1]}(\vec{r},t)$ have uneven parity, $\vec{E}^{[2]}(\vec{r},t)$, $\vec{H}^{[1]}(\vec{r},t)$,$\vec{j_e}^{[2]}(\vec{r},t)$ have even parity, $\rho^{[1]}_e(\vec{r},t)$, $\rho^{[2]}_g(\vec{r},t)$ are scalars, $\rho^{[2]}_e(\vec{r},t)$, $\rho^{[1]}_g(\vec{r},t)$ are pseudoscalars. In the case, when $\vec{j'_g}(\vec{r},t) = 0$, $\rho'_g(\vec{r},t) = 0$ we obtain the equations of usual singly charge electrodynamics for compound EM-field in mathematically correct form, which allows to separate the components of EM-field with various parities P under space inversion. It is remarkable, that the idea, that vector quantities, which characterize EM-field, are compound quantities and inlude both gradient and solenoidal parts, that is uneven and even parts under space inversion was put forward earlier in \cite{Tomilchick}. At the same time in the equations of dual electrodynamics given idea was presented the only in implicit form. The representation in explicit form by equations (\ref{eq5abc}), (\ref{eq6bcd}), (\ref{eq7abcd}), (\ref{eq8abcd}) seems to be actual, since field vector and scalar functions with various t- and P-parities are mathematically heterogeneous and, for instance, their simple linear combination, for instance, for $P$-uneven and $P$-even  electric field vector-functions $\vec{E}^{[1]}(\vec{r},t)$, $\vec{E}^{[2]}(\vec{r},t)$
\begin{equation}
\label{eq5abcdef}
\alpha_1 \vec{E}^{[1]}(\vec{r},t)  + \alpha_2\vec{E}^{[2]}(\vec{r},t)
\end{equation}
with coefficients $\alpha_1$, $\alpha_2$ from the field of real numbers, which is taking place in some theoretical and experimental works, is collage. Similar situation was discussed in \cite{D_Yearchuck_A_Dovlatova} by analysis of Bloch vector symmetry under improper rotations. 
 Mathematically the objects,
which are like to (\ref{eq5abcdef}) can exist. Actually to the set
of $ \{\vec{E}^{[1]}(\vec{r},t)\}$
and to the set of $\{\vec{E}^{[2]}(\vec{r},t)\}$
can  be put in
correspondence the affine space. However given affine space corresponds
to direct sum of two usual vector spaces, consisting of
different physically objects, that is, it represents in fact also collage. Really,
given direct sum can be represented by direct sum of linear capsule
\begin{equation}
\label{eq3a}
\left\{\alpha^k_1\vec{E}^{[1]}(\vec{r},t)| \alpha^k_1 \in R, k \in N \right\},
\end{equation}
representing itself three-dimensional
vector space of the set $\{\vec{E}^{[1]}(\vec{r},t)\}$
and linear capsule 
\begin{equation}
\label{eq3b}
\left\{\alpha^l_2\vec{E}^{[2]}(\vec{r},t)| \alpha^l_2 \in R, l \in N \right\},
\end{equation}
representing itself three-dimensional
vector space of the set $\{\vec{E}^{[2]}(\vec{r},t)\}$. It is substantial,
that both the  vector
spaces cannot be considered to be subspaces of any three-dimensional or six-dimensional
vector spaces, consisting of uniform objects. Moreover, it is evident, that
the affine space, defined in that way cannot be metrizable, when
considering it to be a single whole. It means in turn, that the set of objects, given by (\ref{eq5abcdef})
are not vectors in usual algebraic meaning. Even Pythagorean theorem,
for instance, cannot be used. 

It can be shown, that the system, analogous to (\ref{eq5abc}), (\ref{eq6bcd}), (\ref{eq7abcd}), (\ref{eq8abcd}) can be obtained for the second pair   of 
EM-fields (photon fields in quantum case), which differ by parities of vector and scalar quantities, entering in equations, under time reversal.  Really it is easily to see, that Maxwell equations along with dual transformation symmetry, established by Rainich, given by relations (\ref{eq1b}) - (\ref{eq1c}), are symmetric relatively the dual transformations of another kind of all the vecor and scalar quantities, characterizing EM-field, which, for instance, for electric and magnetic field strengh vector-functions can be presented in the following matrix form
\begin{equation}
\label{eq1bca}
 \left[\begin{array} {*{20}c}  \vec {E{''}} \\ \vec {H{''}} \end{array}\right] = \left[\begin{array} {*{20}c} \cosh\vartheta& i\sinh\vartheta  \\ -i\sinh\vartheta&\cosh\vartheta \end{array}\right]\left[\begin{array} {*{20}c}  \vec {E} \\ \vec {H} \end{array}\right],
\end{equation}
where  $\vartheta$ is arbitrary continuous parameter,
$\vartheta \in [0,2\pi]$. The relation (\ref{eq1bca}) can be rewritten in the form
\begin{equation}
\label{eq1bcae}
 \left[\begin{array} {*{20}c}  \vec {E{''}} \\ \vec {H{''}} \end{array}\right] = \left[\begin{array} {*{20}c} \cos i\vartheta& \sin i\vartheta  \\ -\sin i\vartheta&\cos i\vartheta \end{array}\right]\left[\begin{array} {*{20}c}  \vec {E} \\ \vec {H} \end{array}\right]. 
\end{equation}
In particular, if $\vartheta$ is polar angle of coordinate system in the plane, determined by $\vec {E}$ and $\vec {H}$, the transformations (\ref{eq1bca}) represent themselves hyperbolic rotations in $(\vec {E}, \vec {H})$-plane. Let us call the transformations (\ref{eq1bca}) by hyperbolic dual transformations. It represents the interest to consider the following particular case of hyperbolic dual transformations. We can define parameter $\vartheta$   according to relation
\begin{equation}
\label{eq2bcae}
\tanh\vartheta = \frac{V}{c} = \beta,
\end{equation}
where $V$ is velocity of the frame of reference, moving along $x$-axis in 3D subspace of ${^1}R_4$  Minkowski space. We can  also to set up in conformity to the plane $(x_2,x_3)$ in  Minkowski space, the plane $ (\vec {E}, \vec {H})$ , in which $\vec {E}, \vec {H}$  are orthogonal and are directed along absciss and ordinate axes correspondingly (or vice versa). Then we obtain
\begin{equation}
\label{eq3bcae}
\begin{split}
\raisetag{40pt}
&|\vec {E{''}}| = \frac {|\vec {E}| + \beta|\vec {H}|}{\sqrt{1-\beta^2}}\\ 
&|\vec {H{''}}| = \frac {|\vec {H}| - \beta|\vec {E}|}{\sqrt{1-\beta^2}}, 
\end{split}
\end{equation}
(or similar relations, in which $\vec {E},  \vec {H}$ are interchanged by places).
In vector form given transformations are 
\begin{equation}
\label{eq4bcae}
\begin{split}
\raisetag{40pt}
&\vec {E{''}} = \frac {\vec {E} + \frac{1}{c}[\vec {H} \times \vec{V} ]}{\sqrt{1-\beta^2}}\\ 
&\vec {H{''}} = \frac {\vec {H} - \frac{1}{c}[\vec {E} \times \vec{V} ]}{\sqrt{1-\beta^2}}, 
\end{split}
\end{equation}
Therefore it is seen, that $\vec {E}$,  $\vec {H}$ are transformed like to $x_0$ and $x_1$ coordinates (or vice versa) of the space ${^1}R_4$. It follows from here, that both the vectors $\vec {E{''}}$,  $\vec {H{''}}$ have $t$-even and $t$-uneven components in general case. We see also that Lorentz-invariance of Maxwell equations is particular case of hyperbolic dual symmetry. It means, that restriction to only Lorentz-invariance in consideration of Maxwell equations' symmetry, which is usually used, constricts the concept on the EM-field itself and it is thereby constricting the possibilities for completeness of its practical usage.
Taking into account 
(\ref{eq1abcd}) we obtain the relations, which are similar to (\ref{eq1bcd}), which can be rewritten in the similar to (\ref{eq1c}) form, that is, we have
 \begin{equation} 
\label{eq1c'}
\begin{split}
\raisetag{40pt}                                                    
&\vec {E{''}} = \vec {E} \cos i\vartheta - i\vec {E} \sin i\vartheta\\
&\vec {H{''}} = \vec {H} \cos i\vartheta - i\vec {H}  \sin i\vartheta.
\end{split}
\end{equation}
It is proof in general case, that each of two independent  Maxwellian field components with even and uneven parities under space inversion is also compound and it consists of two independent   components with even and uneven parities under time reversal. Then imposing designations 
 \begin{equation} 
\label{eq10cc}
\begin{split}
\raisetag{40pt}                                                    
& \vec {E} \cos i\vartheta = \vec{E}^{[3]},  \vec {E} \sin i\vartheta = \vec{E}^{[4]} \\
&\vec {H} \cos i\vartheta = \vec{H}^{[3]}, \vec {H} \sin i\vartheta =  \vec{H}^{[4]}
\end{split}
\end{equation}
and considering the vector-functions $(\vec{E}^{[1]}(\vec{r},t)$, $\vec{E}^{[2]}(\vec{r},t))$ and $(\vec{H}^{[1]}(\vec{r},t)$, $\vec{H}^{[2]}(\vec{r},t))$ to be definitional domain for the vector-functions $ \vec {E{''}}(\vec{r},t)$, $\vec {H{''}}(\vec{r},t)$ correspondingly,
the Maxwell equations for the components of the field $(\vec{E}^{[1]}$, $\vec{H}^{[1]})$ and $(\vec{E}^{[2]}$, $\vec{H}^{[2]})$ have the same form and they are 
\begin{equation}
\label{eq5abcc}
\begin{split}
\raisetag{40pt}
&\left[\nabla\times(\vec{E}^{[3]}(\vec{r},t) - i\vec{E}^{[4]}(\vec{r},t)) \right] = \\
&- \mu_0 \left[\frac{\partial \vec{H}^{[3]}(\vec{r},t)}{\partial t} - i \frac{\partial \vec{H}^{[4]}(\vec{r},t)}{\partial t}\right]\\ 
&- \vec{j_g}^{[3]}(\vec{r},t) + i \vec{j_g}^{[4]}(\vec{r},t), 
\end{split}
\end{equation}
\begin{equation}
\label{eq6bcdd}
\begin{split}
\raisetag{40pt}
&\left[\nabla\times(\vec{H}^{[3]}(\vec{r},t) - i\vec{H}^{[4]}(\vec{r},t)) \right] = \\
&\epsilon_0 \left[\frac{\partial \vec{E}^{[3]}(\vec{r},t)}{\partial t} - i \frac{\partial \vec{E}^{[4]}(\vec{r},t)}{\partial t}\right]\\ 
&+ \vec{j_e}^{[3]}(\vec{r},t) - i \vec{j_e}^{[4]}(\vec{r},t), 
\end{split}
\end{equation}
\begin{equation}
\label{eq7abccd}
(\nabla \cdot (\vec{E}^{[3]}(\vec{r},t) - i\vec{E}^{[4]}(\vec{r},t))) = \rho^{[3]}_e(\vec{r},t) - i\rho^{[4]}_e(\vec{r},t),
\end{equation}
\begin{equation}
\label{eq8abccd}
(\nabla \cdot (\vec{H}^{[3]}(\vec{r},t) - i\vec{H}^{[4]}(\vec{r},t))) = \rho^{[3]}_g(\vec{r},t) - i\rho^{[4]}_g(\vec{r},t),
\end{equation}
where $\vec{j_e}^{31]}(\vec{r},t)$, $\vec{j_e}^{[4]}(\vec{r},t)$, $\vec{j_g}^{[3]}(\vec{r},t)$, $\vec{j_g}^{[4]}(\vec{r},t)$ are, correspondingly, electric  and magnetic current densities,
$\rho^{[3]}_e(\vec{r},t)$, $\rho^{[4]}_e(\vec{r},t)$, $\rho^{[3]}_g(\vec{r},t)$,  $\rho^{[4]}_g(\vec{r},t)$ are, correspondingly, electric  and magnetic charge densities, which transformed and designated like to field strengths and currents. In fact  the system of equations (\ref{eq5abcc}), (\ref{eq6bcdd}), (\ref{eq7abccd}), (\ref{eq8abccd}) represent itself corredtly integrated Maxwell equations for two kinds of 
EM-fields (photon fields in quantum case), which differ by parities of vector and scalar quantities, entering equations, under time reversal. So, the components $\vec{E}^{[3]}(\vec{r},t)$, $\vec{H}^{[4]}(\vec{r},t)$, $\vec{j_e}^{[3]}(\vec{r},t)$ have uneven parity, $\vec{E}^{[4]}(\vec{r},t)$, $\vec{H}^{[3]}(\vec{r},t)$, $\vec{j_e}^{[4]}(\vec{r},t)$ have even parity, $\rho^{[3]}_e(\vec{r},t)$, $\rho^{[4]}_g(\vec{r},t)$ are scalars, $\rho^{[4]}_e(\vec{r},t)$, $\rho^{[3]}_g(\vec{r},t)$ are pseudoscalars. In the case, when $\vec{j{''}_g}(\vec{r},t) = 0$, $\rho{''}_g(\vec{r},t) = 0$ we obtain the equations of usual singly charge electrodynamics for two  components of EM-field with various parities under space inversion, at that either of the two consist also of two  components of EM-field with various parity under time reversal.

It is easily to see,  that invariants
for EM-field, consisting of two  hyperbolic dually symmetric parts, that is at  $\vartheta \neq 0$ have the form, analogous to (\ref{eq1bcde}) and they can be obtained, if parameter $\theta$ to replace by $i\vartheta$. They are \begin{equation}
\label{eq1bcdee}
 \left[ \vec {E}^2 - \vec {H}^2 + 2i(\vec {E}\vec {H})\right]  e^ {2\vartheta} =inv.
\end{equation}
Consequently, two real invariats at  $\vartheta \neq 0$ have the form
\begin{equation} 
\label{eq1cacde}
\begin{split}
\raisetag{40pt}                                                    
&(\vec {E}^2 - \vec {H}^2) e^{2\vartheta} = I_1{''} = inv, \\
&2(\vec {E}\vec {H}) e^{2\vartheta} = I_2{''}  = inv.
\end{split}
\end{equation}
  It follows  from  relation (\ref{eq1cacde}), that in  both the cases, that is  at $\vartheta = 0$ and  at fixed $\vartheta \neq 0$, we obtain in fact well known EM-field invariants, since factor $e^ {2\vartheta}$ at fixed $\vartheta$ seems to be insufficient. At the same time at arbitrary $\vartheta$ the relation
\begin{equation} 
\label{eq1cacdeg}
\begin{split}
\raisetag{40pt}                                                    
 \frac {I_1{''}}{I_2{''}} =                        
 \frac {I_1}{I_2} = W = inv 
\end{split}
\end{equation}
is taking place. It is seen, that the value of $W$ is independent on $\vartheta$. It means physically, that the absolute values of both the  vector-functions $\vec {E}(\vec{r},t)$ and $\vec {H}(\vec{r},t)$ are changed synchronously by hyperbolic dual transformations.

So, the usage of complex number theory
allows to represent  correctly the electrodynamics for two photon fields,  which differs by parities under space inversion or time reversal by  the same single system of generalized  Maxwell equations. At the same time we have two related sets, that is pairs of complex vector and scalar functions, which are ordered in their P- and t-parities. It corresponds to definition   of quaternions. 
Really any quaternion number $x$ can be determined according to relation
\begin{equation} 
\label{eq1cacdegh}
\begin{split}
\raisetag{40pt}
&x = (a_1 + ia_2)e + (a_3 + ia_4)j,
\end{split}
\end{equation}
where $\{a_m\}\in R, m = \overline{1,4}$ and
 $e, i, j, k$ produce basis, elements of which are satisfying the conditions  
\begin{equation} 
\label{eq15cacdegh}
\begin{split}
\raisetag{40pt}\\
&(ij) = k, (ji) = -k, (ki) = j, (ik)= -j,\\
&(ei)= (ie) = i, (ej)= (je) = j, (ek) = (ke) = k.
\end{split}
\end{equation}

Let us designate the quantities
\begin{equation} 
\label{eq1cadegh}
\begin{split}
\raisetag{40pt} 
&(\vec{E}^{[1]}(\vec{r},t) - i\vec{E}^{[2]}(\vec{r},t)) +
(\vec{E}^{[3]}(\vec{r},t) - i\vec{E}^{[4]}(\vec{r},t)) j =\\ &\vec{\mathfrak{E}}(\vec{r},t)\\
&(\vec{H}^{[1]}(\vec{r},t) - i\vec{H}^{[2]}(\vec{r},t)) +
(\vec{H}^{[3]}(\vec{r},t) - i\vec{H}^{[4]}(\vec{r},t)) j =\\ &\vec{\mathfrak{H}}(\vec{r},t)\\
&(\vec{j_e}^{[1]}(\vec{r},t) - i \vec{j_e}^{[2]}(\vec{r},t)) +
(\vec{j_e}^{[3]}(\vec{r},t) - i \vec{j_e}^{[4]}(\vec{r},t)) j =\\ &\vec{\mathfrak{j_e}}(\vec{r},t)\\
&(- \vec{j_g}^{[1]}(\vec{r},t) + i \vec{j_g}^{[2]}(\vec{r},t)) +
(- \vec{j_g}^{[3]}(\vec{r},t) + i \vec{j_g}^{[4]}(\vec{r},t)) j =\\ &\vec{\mathfrak{j_g}}(\vec{r},t)\\
&(\rho^{[1]}_e(\vec{r},t) - i\rho^{[2]}_e(\vec{r},t)) +
(\rho^{[3]}_e(\vec{r},t) - i\rho^{[4]}_e(\vec{r},t)) j =\\ &\mathfrak{\rho_e}(\vec{r},t)\\
&(\rho^{[1]}_g(\vec{r},t) - i\rho^{[2]}_g(\vec{r},t)) +
(\rho^{[3]}_g(\vec{r},t) - i\rho^{[4]}_g(\vec{r},t)) j =\\ &\mathfrak{\rho_g}(\vec{r},t),
\end{split}
\end{equation}
where
 \begin{equation} 
\label{eq31cadegh}
\begin{split}
\raisetag{40pt} 
&\vec{E}^{[1]}(\vec{r},t), \vec{H}^{[2]}(\vec{r},t), \vec{j_e}^{[1]}(\vec{r},t),\\
&\vec{j_g}^{[2]}(\vec{r},t), \rho^{[1]}_e(\vec{r},t), \rho^{[2]}_g(\vec{r},t)
\end{split}
\end{equation} are $P$-uneven, $t$-even,
 \begin{equation} 
\label{eq32cadegh}
\begin{split}
\raisetag{40pt}
&\vec{E}^{[2]}(\vec{r},t), \vec{H}^{[1]}(\vec{r},t), \vec{j_e}^{[2]}(\vec{r},t),\\ &\vec{j_g}^{[1]}(\vec{r},t), \rho^{[2]}_e(\vec{r},t), \rho^{[1]}_g(\vec{r},t)
\end{split}
\end{equation}
 are $P$-uneven, $t$-uneven,
 \begin{equation} 
\label{eq33cadegh}
\begin{split}
\raisetag{40pt}
&\vec{E}^{[3]}(\vec{r},t), \vec{H}^{[4]}(\vec{r},t), \vec{j_e}^{[3]}(\vec{r},t),\\ &\vec{j_g}^{[4]}(\vec{r},t), \rho^{[3]}_e(\vec{r},t), \rho^{[4]}_g(\vec{r},t)
\end{split}
\end{equation}
 are $P$-even, $t$-even,
\begin{equation} 
\label{eq34cadegh}
\begin{split}
\raisetag{40pt}
&\vec{E}^{[4]}(\vec{r},t), \vec{H}^{[3]}(\vec{r},t), \vec{j_e}^{[4]}(\vec{r},t),\\ &\vec{j_g}^{[3]}(\vec{r},t), \rho^{[4]}_e(\vec{r},t), \rho^{[3]}_g(\vec{r},t)
\end{split}
\end{equation}
 are $P$-even, $t$-uneven.  According to definition   of quaternions $\vec{\mathfrak{E}}(\vec{r},t)$, $\vec{\mathfrak{H}}(\vec{r},t)$, $\vec{\mathfrak{j_e}}(\vec{r},t)$, $\vec{\mathfrak{j_g}}(\vec{r},t)$, $\mathfrak{\rho_e}(\vec{r},t)$,
$\mathfrak{\rho_g}(\vec{r},t)$ are quaternions. It means, that EM-field has quaternion structure and dual and hyperbolic dual symmetry of  Maxwell equations will take proper account, if all the  vector and scalar quantities to represent in quaternion form. 
Consequently, we have
\begin{equation}
\label{eq5abcce}
\begin{split}
\raisetag{40pt}
\left[\nabla\times(\vec{\mathfrak{E}}(\vec{r},t)) \right] = 
- \mu_0 \left[\frac{\partial \vec{\mathfrak{H}}(\vec{r},t)}{\partial t} \right] 
- \vec{\mathfrak{j_g}}(\vec{r},t), 
\end{split}
\end{equation}
\begin{equation}
\label{eq6bcdde}
\begin{split}
\raisetag{40pt}
\left[\nabla\times(\vec{\mathfrak{H}}(\vec{r},t)) \right] = 
\epsilon_0 \left[\frac{\partial \vec{\mathfrak{E}}(\vec{r},t)}{\partial t} \right] 
&+ \vec{\mathfrak{j_e}}(\vec{r},t), 
\end{split}
\end{equation}
\begin{equation}
\label{eq7abccde}
(\nabla \cdot (\vec{\mathfrak{E}}(\vec{r},t))) = \mathfrak{\rho_e}(\vec{r},t),
\end{equation}
\begin{equation}
\label{eq8abccde}
(\nabla \cdot (\vec{\mathfrak{H}}(\vec{r},t))) = \mathfrak{\rho_g}(\vec{r},t)
\end{equation}
Therefore, symmetry of Maxwell equations under  dual transformations of both the kinds allows along with  generalization of Maxwell equations themselves   to  extend the field of application of Maxwell equations. It means also, that dual electrodynamics, developed by Tomilchick and co-authors, see for instance \cite{Tomilchick}, obtains additional ground. Basic field equations in dual electrodynamics \cite{Tomilchick}, \cite{Berezin}, being to be written separately for two type of independent photon fields with various parities under space inversion or time reversal, will be isomorphic to Maxwell equations in complex form.  It was in fact shown partly  earlier in  \cite{Tolkachev}, \cite{Berezin}, where  complex charge was taken into consideration. At the same time all aspect of dual symmetry, leading to four-component quaternion form of Maxwell equations seem to be representing for the first time. 

\section{Cavity Dually Symmetric Electrodynamics}

Let us find the conserving quantities, which correspond to dual and hyperbolic dual symmetries of Maxwell equations. It seems to be interesing to realize given task on concrete practically essential example of cavity EM-field. At the same time to built the Lagrangian, which is adequate to given task it seems to be reazonable to solve the following concomitant task - to find dually symmetric  solutions of Maxwell equations. It seems to be  understandable, that the general solutions of differential equations can also possess by the same symmetry, which have starting differential equations, nevetheless dual symmetry of the solutions of Maxwell equations was earlier not found.

\subsection{\textbf{Classical Cavity EM-Field}}

Suppose EM-field in volume rectangular cavity without any matter inside it and made up of perfectly electrically conducting walls. Suppose also, that the field is linearly polarized and without restriction of commonness let us choose the one of two possible polarization of   EM-field electrical component $\vec{E}(\vec{r},t)$ along x-direction. Then the vector-function $E_x(z,t) \vec{e}_x$  can be represented in well known form of Fourier sine series
\begin{equation}
\label{eq1ab}
\vec{E}^{[1]}(\vec{r},t) = 
E_x(z,t) \vec{e}_x = \left[\sum_{\alpha=1}^{\infty}A^{E}_{\alpha}q_{\alpha}(t)\sin(k_{\alpha}z)\right]\vec{e}_x ,
\end{equation}
where $q_{\alpha}(t)$ is amplitude of $\alpha$-th normal mode of the cavity, $\alpha \in N$, $k_{\alpha} = \alpha\pi/L$, $A^{E}_{\alpha}=\sqrt{2 \omega_{\alpha}^2m_{\alpha}/V\epsilon_0}$, $\omega_{\alpha} = \alpha\pi c/L$, $L$ is cavity length along z-axis, $V$ is cavity volume, $m_{\alpha}$ is parameter, which is introduced to obtain the analogy with mechanical harmonic oscillator. Let us remember, that the expansion in Fourier series instead of  Fourier integral expansion is  determined by known diskretness of $\vec{k}$-space, which is the result of finiteness of cavity volume. Particular sine case of Fourier  series is consequence of boundary conditions 
\begin{equation}
\label{eq1abc}
[\vec{n} \times\vec{E}]|_S = 0, (\vec{n} \vec{H})|_S = 0,
\end{equation}
which are held true for the perfect cavity considered. Here $\vec{n}$ is the normal to the surface $S$ of the cavity. It is easily to show, that $E_x(z,t)$ represents itself  a standing wave along z-direction.

Let us analyse the solutions of Maxwell equations for EM-field in a cavity in comparison with known solutions from the literature to pay the attention to some mathematical details, which have however substantial physical conclusions,  allowing to extend our insight to EM-field nature. For given reasons, despite on analysis simplicity, we will produce the consideration in detail.

 Using the equation
\begin{equation}
\label{eq2ab}
\epsilon_0 \frac{\partial\vec{E}(z,t)}{\partial t } = \left[\nabla\times\vec{H}(z,t)\right],
\end{equation}
 we obtain  the expression for magnetic field
 \begin{equation}
\label{eq3ab}
\vec{H}(\vec{r},t) =  \left[\sum_{\alpha=1}^{\infty}A^{E}_{\alpha}\frac{\epsilon_0}{k_{\alpha}}\frac{dq_{\alpha}(t)}{dt}\cos(k_{\alpha}z) + f_{\alpha}(t)\right]\vec{e}_y,
\end{equation}   
where  $\{f_{\alpha}(t)\}$, $ \alpha \in N $, is the set of arbitrary  functions of the time. It is evident, that the expression for $ \vec{H}(\vec{r},t)$ (\ref{eq3ab}) is satisfying to boundary conditions (\ref{eq1abc}). 
The partial solution, in which the functions $\{f_{\alpha}(t)\}$ are identically zero, that is,  $\vec{H}(\vec{r},t)$ is
\begin{equation}
\label{eq3abc}
\vec{H}^{[1]}(\vec{r},t) =  \left[\sum_{\alpha=1}^{\infty}A^{E}_{\alpha} \frac{\epsilon_0}{k_{\alpha}} \frac{dq_{\alpha}(t)}{dt}\cos(k_{\alpha}z)\right] \vec{e}_y,
\end{equation} 
 is always used in all the EM-field literature. 
However even in given case it is evident, that the Maxwellian field is complex field. Really 
 using the equation
\begin{equation}
\label{eq4ab}
\left[ \nabla\times\vec{E}\right] = -\frac{\partial \vec{B}}{\partial t} = -\mu_0 \frac{\partial \vec{H}}{\partial t}
\end{equation}
it is easily to find the class of field functions $\{q_{\alpha}(t)\}$. They will satisfy to  differential equations
\begin{equation}
\label{eq5ab}
\frac{d^2q_{\alpha}(t)}{dt^2} + \frac{k_{\alpha}^2}{\mu_0\epsilon_0} q_{\alpha}(t)=0, \alpha \in N.
\end{equation}
Consequently,  we have
\begin{equation}
\label{eq6ab}
q_{\alpha}(t) = C_{1\alpha} e^{i\omega_{\alpha}t} + C_{2\alpha} e^{-i\omega_{\alpha}t}, \alpha \in N,
\end{equation}
where  $C_{1\alpha}, C_{2\alpha}, \alpha \in N$ are arbitrary constants.
Thus, real-valued free Maxwell field equations result in well known in the theory of differential equations  
situation - the solutions are complex-valued functions. It means, that generally the field functions for free Maxwellian field in the cavity produce complex space. So we obtain additional independent argument, that the  known conception, on the only  real-quantity definiteness of EM-field, has to be corrected. On the other hand, the equation (\ref{eq5ab}) has also the only real-valued general solution, which can be represented in the form
\begin{equation}
\label{eq6abc}
q_{\alpha}(t) = B_{\alpha} \cos(\omega_{\alpha}t + \phi_{\alpha}),
\end{equation}
where  $B_{\alpha}, \phi_{\alpha}, \alpha \in N$ are arbitrary constants. It is substantial, that the functions in real-valued general solution have a definite t-parity. 

Thus, we come independently on the previous consideration in Sec.III and Sec.IV to the conclusion,  that classical Maxwellian EM-field can be both real-quantity definited and complex-quantity definited.
 
It is interesting, that
there is the second physically substantial solution of Maxwell equations. Really, from general expression (\ref{eq3ab}) for the  field $\vec{H}(\vec{r},t)$ 
it  is easily to obtain differential equations  for $\{f_{\alpha}(t)\}$, $ \alpha \in N $, 
\begin{equation}
\label{eq7ab}
\begin{split}
&\frac{d f_{\alpha}(t)}{dt} + A^{E}_{\alpha}\frac{\epsilon_0}{k_{\alpha}}\frac{\partial^2q_{\alpha}(t)}{\partial t^2}\cos(k_{\alpha}z) \\
&- \frac {1}{\mu_0} A^{E}_{\alpha}k_{\alpha}q_{\alpha}(t)\cos(k_{\alpha}z) = 0.
\end{split}
\end{equation}
The  formal solution of given equations 
in general case is
\begin{equation}
\label{eq8ab}
f_{\alpha}(t) =  A^{E}_{\alpha} \cos(k_{\alpha}z)\left[\frac{k_{\alpha}}{\mu_0} \int\limits _{0}^{t} q_{\alpha}(\tau)d\tau -\frac{dq_{\alpha}(t)}{dt}\frac{\epsilon_0}{k_{\alpha}}\right]
\end{equation}
Therefore, we have the second  solution  of Maxwell equations for $\vec{H}(\vec{r},t)$ in the form
\begin{equation}
\label{eq9ab}
\vec{H}^{[2]}(\vec{r},t) = -\left\{\sum_{\alpha=1}^{\infty} A^{H}_{\alpha} q_{\alpha}'(t)\cos(k_{\alpha}z) \right\}\vec{e}_y,
\end{equation}
where $A^{H}_{\alpha}=\sqrt{2 \omega_{\alpha}^2m_{\alpha}/V\mu_0}$.
Similar consideration gives the second  solution for $\vec{E}(\vec{r},t)$ 
\begin{equation}
\label{eq10ab}
\vec{E}^{[2]}(\vec{r},t) = \left\{\sum_{\alpha=1}^{\infty} A^{E}_{\alpha}q_{\alpha}''(t)\sin(k_{\alpha}z)\right\}\vec{e}_x,
\end{equation}
The functions $q_{\alpha}'(t)$ and $q_{\alpha}''(t)$ in relationships (\ref{eq9ab}) and (\ref{eq10ab})  are
\begin{equation}
\label{eq11ab}
\begin{split}
&q_{\alpha}'(t) = {\omega_{\alpha}}\int\limits _{0}^{t} q_{\alpha}(\tau)d\tau\\
&q_{\alpha}''(t) = {\omega_{\alpha}}\int\limits _{0}^{t} q_{\alpha}'(\tau')d\tau'
\end{split}
\end{equation}
correspondingly. Owing to the fact, that the solutions have simple form of harmonic trigonometrical functions, the 
second solution for electric field differs from the first solution the only by sign, that is substantial,  and by inessential integration constants.  Integration constants  can be taken into account by means of redefinition of factor $m_{\alpha}$ in field amplitudes.    It is also evident, that if vector-functions $\vec{E}(\vec{r},t)$ and $\vec{H}(\vec{r},t)$ are the solutions of Maxwell equations, then vector-functions $\hat{T}\vec{E}(\vec{r},t)$ and $\hat{T}\vec{H}(\vec{r},t)$, where $\hat{T}$ is time inversion operator, are also the solutions of Maxwell equations. Moreover, if starting vector-function, to which operator $\hat{T}$ is applied is $t$-even, then there is $t$-uneven  solution, for instance for magnetic component in the form 
\begin{equation}
\label{eq11abc}
\frac{\hat{T}[t \vec{H}(\vec{r},t)]}{t}, 
\end{equation}
where $t$ is time. 
It can be shown in a similar way, that dually symmetric solutions, which are $P$-even and $P$-uneven are also existing.

Therefore, there are the solutions with various combinations of the signs for vector-functions $\vec{E}(\vec{r},t)$ and $\vec{H}(\vec{r},t)$, which are realized simultaneously, that is, their linear combination with coefficients from the field $C$ of complex numbers will represent  
 the solution of Cauchy problem for Maxwell equations in correspondence with  known theorem, that the solution of Cauchy problem  for any systems of homogeneous linear equations in partial derivatives exists and it is unique  in the vicinity of any point of the  initial surface  (in the case, when the point selected is not characteristic point and the function, which determines given hypersurface is continuously differentiable). In other words, we obtain again  the agreement with Maxwell equation symmetry consideration.   Given property of EM-fied  seems to be essential, since it permits passing for the processes, which seemingly are forbidden by CPT-theorem. For example, let us consider the resonance system EM-field plus matter in the cavity, in particular, the so called dressed state of some quasiparticles' system. Suppose, that wave function can be factorized, matter part is $P$- and $t$-even under space  and time inversion transformations, while EM-field part is  $P$-uneven. CPT-invariance will be preserved, since EM-field has simultaneously with $t$-even the  $t$-uneven component, determined by expression  (\ref{eq11abc}).  Therefore $t$-parity of the function $q_{\alpha}'(t)$ can be various, and in the case, if we  choose $t$-parity to be identical to the parity of the function $q_{\alpha}(t)$, the solution will be different in the meaning, that the field vectors will have opposite  $t$-parity in comparison with the  first  solution. It is evident, that boundary conditions are fullfilled for all the cases considered. 

To built the Lagrangian we can choose the following sets of EM-field functions $\{u^{s,\pm}_{\alpha}(x)\}, s = 1, 2, \alpha \in N$, 
\begin{equation}
\label{eq19abc}
\begin{split}
&\{u^{1,\pm}_{\alpha}(x)\} = \{\sqrt{\epsilon_0}A^{E}_\alpha\sin k_\alpha(x_3) [q_\alpha(x_4) \pm i q^{''}_\alpha(x_4)]\}\\
&\{u^{2,\pm}_{\alpha}(x) = \{\sqrt{\mu_0}A^{H}_\alpha\cos k_\alpha(x_3) [-q{'}_\alpha(x_4) \pm \frac {i}{\omega_\alpha}\frac{dq_\alpha(x_4)}{dx_4}]\}
\end{split}
\end{equation} 
The functions $\{u^{s,\pm}_{\alpha}(x)\}, s = 1, 2, \alpha \in N$ are built from the components of   
the expansion in Fourier series of the fields $\vec{E}^{[1]}(\vec{r},t), \vec{E}^{[2]}(\vec{r},t)$ and
$\vec{H}^{[2]}(\vec{r},t), \vec{H}^{[1]}(\vec{r},t)$ correspondingly. At the same time the sets $\{u^{s,\pm}_{\alpha}(x)\}, s = 1, 2, \alpha \in N$ produce at fixed $x$ two  orthogonal countable bases, corresponding to  $s = 1, 2$ in two Hilbert spaces, which are formed by vectors $\mathfrak U^{[s,\pm]}(u^{s,\pm}_{1}(x), u^{s,\pm}_{2}(x), ...)$ for variable $x \in {^1}R_4$.
Really scalar product of two arbitrary vectors $\mathfrak U_i^{[s,\pm]}(u^{s,\pm}_{1}(x_i), u^{s,\pm}_{2}(x_i), ...)$ and $\mathfrak U_j^{[s,\pm]}(u^{s,\pm}_{1}(x_j), u^{s,\pm}_{2}(x_j), ...)$, that is
\begin{equation}
\label{eq19bcd}
\langle \mathfrak U_i^{[s,\pm]}(x_i)\mid \mathfrak U_j^{[s,\pm]}(x_j)\rangle
\end{equation}
is equal to 
\begin{equation}
\label{eq19cde}
\sum_{\alpha = 1}^{\infty}\int\limits _{0}^{L}u^{*s,\pm}_{\alpha}(x_{4,i}, z)u^{s,\pm}_{\alpha}(x_{4,j}, z) dz, s = 1, 2, 
\end{equation}
that means, that it is restricted, since the sum over $s$ represents the energy of the field in restricted volume.  Consequently, the norm of vectors can be defined by the relationship
\begin{equation}
\label{eq19def}
\begin{split}
&\|\mathfrak U^{[s,\pm]}(x)\| = \sqrt{\langle \mathfrak U^{[s,\pm]}(x)\mid \mathfrak U^{[s,\pm]}(x)\rangle} = \\
&\sqrt{\sum_{\alpha = 1}^{\infty}\int\limits _{0}^{L}u^{*s,\pm}_{\alpha}(x_{4,i}, z)u^{s,\pm}_{\alpha}(x_{4,j}, z) dz}, s = 1, 2. 
\end{split}
\end{equation}
Then vector distance is
\begin{equation}
\label{eq19efg}
d(\mathfrak U^{[s,\pm]}(x_i), \mathfrak U^{[s,\pm]}(x_j)) = \|\mathfrak U^{[s,\pm]}(x_i) - \mathfrak U^{[s,\pm]}(x_j)\|.
\end{equation}
So we obtain, that the vectors $\{\mathfrak U^{[s,\pm]}(x)\}$, $x \in {^1}R_4$ produce the space $L_2$ and taking into account the Riss-Fisher theorem it means, that given vector space is complete, that in its turn means, that the spaces of vectors $\{\mathfrak U^{[s,\pm]}(x)\}$, $x \in {^1}R_4, s = 1, 2$, are Hilbert spaces.
Consequently Lagrangian $L(x)$ can be represented in the following form
\begin{equation}
\label{eq20abcd}
\begin{split}
&L(x) = \sum_{s=1}^{2}\sum_{\mu=1}^{4}\sum_{\alpha=1}^{\infty} \frac{\partial u^{s,\pm}_{\alpha}(x)}{\partial x_\mu}\frac{\partial u^{*s,\pm}_{\alpha}(x)}{\partial x_\mu} \\
&- \sum_{s=1}^{2}\sum_{\mu=1}^{4}\sum_{\alpha=1}^{\infty} K(x) u^{s,\pm}_{\alpha}(x) u^{*s,\pm}_{\alpha}(x),
\end{split}
\end{equation}
where $ K(x)$  is factor, depending on the set of variables $ x = \{x_\mu\}, \mu =  \overline {1,4}$. 

Let us find the conserving quantity, corresponding to dual symmetry of Maxwell equations.   Dual transformation, determined by relation (\ref{eq1bc})
is the  transformation the only in the space of field three-dimensional vector-functions $\vec{E}, \vec{H}$, (let us designate it by  $(\vec{E}, \vec{H})$-space) and it does not touch upon the coordinates. It seems to be conveniet to define in given space the reference frame, then the  transformation, given by (\ref{eq1bc}) is the rotation of two component matrix vector-function
\begin{equation}
\label{eq21abcd}
 \|F\| = \left[\begin{array} {*{20}c}  \vec {E} \\ \vec {H} \end{array}\right].
\end{equation}
 Instead of two Hilbert space for two sets of vectors $\{\mathfrak U^{[s,\pm]}(x)\}$, $x \in {^1}R_4, s = 1, 2$ we can also define one Hilbert space for  row matrix vector function set
\begin{equation}
\label{eq22abcd}
\{\|\mathfrak U^{}(x)\|\} = \{\left[\mathfrak U^{[1,\pm]}(x) \mathfrak U^{[2,\pm]}(x)\right]\}
\end{equation}
with the set of  components
\begin{equation}
\label{eq22abcde}
\{\|U_{\alpha}(x)\|\} = \{\left[u^{1,\pm}_{\alpha}(x) u^{2,\pm}_{\alpha}(x)\right]\},
\end{equation}
where $ \alpha \in N$. 
In general case instead parameter $\theta$ we can define rotation angles $\theta_{ik}$, $i, k = \overline {1,3}$ in 2D-planes of $(\vec{E}, \vec{H})$ functional space. It is evident, that $\theta_{ik}$ are antisymmetric under the indices $i, k = \overline {1,3}$.
According to N\"{o}ther theorem, the conserving quantity, corresponding to parameters $\theta_{ik}$ in dual transformations (\ref{eq1bc}), that is at $\theta_{ik}$ = $\theta_{12}$ is determined by 
relations like to (\ref{eq19a}) and (\ref{eq19b}). So,  we obtain
\begin{equation}
\label{eq19bac}
S^{\mu}_{12} = -[\sum_{\alpha=1}^{\infty}\frac{\partial{L}}{\partial(\partial_{\mu}\|U^*_{\alpha}\|)}  \|Y_{\alpha}\|] + c.c.,
\end{equation}
 where $ \mu = \overline {1,4}$  and it was taken into account, that $\|X_{\alpha}\|$ in matrix relation (\ref{eq19bac}), which is like to (\ref{eq19a}) is equal to zero. 
The factor $\frac{\partial{L}}{\partial(\partial_{\mu}\|U^*_{\alpha}\|)}$ in  (\ref{eq19bac})  is row matrix
\begin{equation}
\label{eq21cabde}
\frac{\partial{L}}{\partial(\partial_{\mu}\|U*_{\alpha}\|)} = \left[\frac{\partial{L}}{\partial(\frac{\partial u^{*,1\pm}_\alpha}{\partial x_\mu})} \frac{\partial{L}}{\partial(\frac{\partial u^{*,2\pm}_\alpha}{\partial x_\mu})}\right],
\end{equation}
matrix $\|Y_{\alpha}\|$ is product of matrices $ \|I_{\alpha}\|$  and $ \widetilde{\|U_{\alpha}(x)\|}$, that is 
\begin{equation}
\label{eq23cabde}
\|Y_{\alpha}\| = \|I_{\alpha}\|\left[\begin{array} {*{20}c} u_{\alpha }^{1\pm}  \\ u_{\alpha}^{2\pm}  \end{array}\right],
\end{equation}
where $\|I_{\alpha}\|$ is the matrix, which corresponds to infinitesimal operator of dual  or hyperbolic dual  transformations of $\alpha$-th mode of cavity EM-field. It represents in general case the product of three matrices, corresponding to rotation along three mutually perpendicular axes in 3D functional space above defined. So $\|I_{\alpha}\| =
\|I^1_{\alpha}\| \|I^2_{\alpha}\|\|I^3_{\alpha}\|$. The transformations in the form, which is given by (\ref{eq1bc}) correspond to $\theta_{23} = \theta$, $\theta_{12} = 0, \theta_{31} = 0$, that is  $\|I^2_{\alpha}\| = \|I^3_{\alpha}\| = E$, where $E$ is unit $[2\times 2]$-matrix.  In the absence of  dispersive medium in the cavity $\|I_\alpha\|$ will be independent on $\alpha$. Moreover, it is easily to see, that   
infinitesimal operator with matrix $\|I_\alpha\|$ is the same for dual transformations, determined by   (\ref{eq1bc}) and hyperbolic dual  transformations, determined by   (\ref{eq1bca}). Really $\|I_\alpha\|$ in both the cases is
\begin{equation}
\label{eq62b}
\|I_\alpha\| = \left[\begin{array} {*{20}c}0&1 \\-1&0   \end{array}\right], 
\end{equation}
 for any $\alpha \in N$.

Conserving quantity is
\begin{equation}
\label{eq20cab}
\begin{split}
S^{4}_{12} = 
-\frac{i}{c} \int\{[\sum_{\alpha=1}^{\infty}\frac{\partial{L}}{\partial(\partial_{\mu}\|U^*_{\alpha}\|)}  \|Y_{\alpha}\|] + c.c.\}d^3x
\end{split}
\end{equation}
The structure of (\ref{eq20cab}) unambiguously indicates, that it is the component of  spin tensor \cite{Bogush}, \cite{Shirkov}, to which dual vector component can be set in the correspondence according to relation  
\begin{equation}
\label{eq24cab}
\begin{split}
&S^{4}_i = \varepsilon_{ijk}S^{4}_{jk}  =\\
&-\varepsilon_{ijk}\frac{i}{c} \int\{[\sum_{\alpha=1}^{\infty}\frac{\partial{L}}{\partial(\partial_{\mu}\|U^*_{\alpha}\|)}  \|Y_{\alpha}\|]_{jk} + c.c.\}d^3x, 
\end{split}
\end{equation}
where $\varepsilon_{ijk}$ is completely antisymmetric Levi-Civita 3-tensor.

Therefore we obtain, that  the same physical conserving quantity corresponds to
dual and hyperbolic dual symmetry of Maxwell equations.  Taking into account the expressions for Lagrangian (\ref{eq20abcd}) and for  infinitesimal operator (\ref{eq62b}), in the geometry choosed, when vector $\vec{E}$ is directed along absciss axis, vector $\vec{H}$ is directed along ordinate axis in  $(\vec{E}, \vec{H})$ functional space, we have
\begin{equation}
\label{eq63ac}
S^{\mu}_{12} = \sum_{\alpha=1}^{\infty}[\frac{\partial u^{*,1\pm}_\alpha}{\partial x_\mu}u^{2\pm}_\alpha - \frac{\partial u^{*,2\pm}_\alpha}{\partial x_\mu}u^{1\pm}_\alpha]  + c.c.
\end{equation}
and
\begin{equation}
\label{eq25cab}
S^{4}_3 = \varepsilon_{312}S^{4}_{12}  = -\frac{i}{c} \int\{[\sum_{\alpha=1}^{\infty}\frac{\partial{L}}{\partial(\partial_{\mu}\|U^*_{\alpha}\|)}  \|Y_{\alpha}\|] + c.c.\}d^3x, 
\end{equation}
It is projection of spin on the propagation direction. Therefore we have in given case right away physically significant quantity - spirality. 

The relations (\ref{eq19bac}), (\ref{eq24cab}), (\ref{eq63ac}), (\ref{eq25cab}) show, that  spin of classical relativistic EM-field in the cavity and, correspondingly, spirality are additive
quantities and they represent the sum of cavity spin and spirality  modes. On the connection  of the conserving quantity, which is  invariant of dual symmetry with spin was indicated in \cite{Tomilchick}, where free EM-field was considered with traditional Lagrangian, which uses vector potentials to be field functions.
The result obtained together with aforecited result in \cite{Tomilchick} lift 
dilemma on the necessity of using of given quantity by consideration of classical 
EM-field. Really, the situation was to some extent paradoxical, and it can be displayed by the following conversation between two disputant
physicists. "Spin  exists" - has insisted the first, referring on the appearance of additional tensor component in total tensor of moment - intrinsic moment - to be consequence of Minkowsky space symmetry under Lorentz transformations, "Spin  does not exists" - has insisted the second, referring on the metrized tensor of the moment, in which spin part is equal to zero \cite{Bogush} in distinction from canonical tensor. In other words, both disputants were in one's own way right. Dual symmetry leads to unambiguous conclusion "Spin  exists" and has to be taken into consideration by the solution of tasks, concerning both classical and quantum electrodynamics. Moreover spin takes on special leading significance among the physical characteristics of EM-field, since the only spin (spirality in the simplest case above considered) combine two subsystems of photon fields, that is the subsystem of two fields, which have definite $P$-parity  (even and uneven) with the subsystem of two fields, which have definite $t$-parity (also even and uneven) into one system. In fact we obtain the proof for four component structure of EM-field to be a single whole, that is confirmation along with the possibility of the representation of EM-field in four component quaternion form, given by (\ref{eq5abcce}), (\ref{eq6bcdde}), (\ref{eq7abccde}), (\ref{eq8abccde}),
the necessity of given representation. It
extends the overview on the nature of EM-field itself. It seems to be remarkable, that given result on the special leading significance of spin is in agreement with result in \cite{D_Yearchuck_A_Dovlatova}, where was shown,  that spin is quaternion vector of the state in Hilbert space, defined under ring of quaternions, of any quantum system (in the
frame of the chain model considered) interacting with EM-field.

It is  interesting, that the charge und current, being to be the components of 4-vector, which are transformed by corresponding representation of Lorentz group, are invariants of hyperbolic dual transformations, that is, they are also Lorentz invariants in the case, when both charge und current  are taken separately by $|\vec {E}| = |\vec {H}|$. The proof is evident, if to take into account, that  Lorentz transformations are particular case of hyperbolic dual transformations. It is seen immediately from the expressions for 4-current  and it means, that observers in various inertial frames will register the same value of the charge in correspondence with conclusion in \cite{Tomilchick}. It is connected with invariance of Lagrange equations and   expressions for 4-current by multiplication of field functions on arbitrary complex number, established in Sec.1, since  by hyperbolic dual transformations the multiplication of field functions on some  complex number takes place.

\subsection{Connection between Gauge Invariance of  EM-field and Analicity of its  Vector-Functions}

The methods of theory of function of complex variable seem to be also useful along with algebraic methods for  the study of complex fields. The first example is the  conclusion, that Maxwell equations for free EM-field, for which electric and magnetic vector-functions are suggested to be real vector-functions, represent themselves the analicity condition for the complex-valued vector-function 
\begin{equation}
\label{eq53q}
\vec{F}(\vec{r},t) \equiv \vec{H}_r(\vec{r},t) - i \vec{E}_r(\vec{r},t)
\end{equation}
 of two variables $\vec{r}$ and $t$, where $\vec{r}$ is variable, which belong to any spacelike hypersurface $V$ $\subset {^{1}R_4}$, that is $\vec{r} \in R^3$, $t\in (0,\infty)$ is time. The proof is evident, it is sufficient to write down Cauchy-Riemann conditions for complex-valued vector-function (\ref{eq23a}).

It seems to be interesting to ascertain, whether is there the connection between symmetry of dynamical systems, in particular, between gauge symmetry, and analytical properties of quantities, which are invariant under corresponding symmetry qroups. Let us consider  
 the complex-valued function $Q(\vec{r},t)$ = $Q_1(\vec{r},t) + iQ_2(\vec{r},t)$, which are defined by (\ref{eq22}, \ref{eq23}), if integration limits  in (\ref{eq22}, \ref{eq23}) are  variable, that is we have then  the function of the same two variables $\vec{r}$ and $t$,  $\vec{r} \in R^3$, $t\in (0,\infty)$. At the same time it is also function of field functions $u(x)$, which satisfy initially to Lagrange equations, that is differential equations of the second order in partial derivatives. It follows from definition of differential equation solutions, that the field functions $u(x)$ are continuously differentiable functions, their first partial derivatives are also continuously differentiable functions and the second partial derivatives are continuous functions. So integrands in (\ref{eq22}, \ref{eq23}) are continuous functions of variables $\vec{r}$ and $t$. It is sufficient for variable integration limit  differentiation of integrals in relationships (\ref{eq22}, \ref{eq23}). Let us introduce complex vector-scalar variable $z = \vec{r} + ict$. 
Then the following statement takes place

8.\textit{Gauge-invariant complex-valued quantity, that is complex charge, represents itself analytical function in complex "plane" of variable} $z = \vec{r} + ict$ \textit{for any complex relativistic classical field}. 

To prove the statement, it is sufficient to show, that $ReQ(\vec{r},t)$ and $Im Q(\vec{r},t)$ of the function $Q(\vec{r},t) = Q_1(\vec{r},t) + iQ_2(\vec{r},t)$ are satysfying to Cauchy-Riemann conditions, that is, the following relationships take place
\begin{equation}
\label{eq6q}
\frac{\partial Q_1(\vec{r},t)}{\partial \vec{r}} = \frac{\partial Q_2(\vec{r},t)}{\partial t}, 
\end{equation}
\begin{equation}
\label{eq7q}
\frac{\partial Q_1(\vec{r},t)}{\partial t} = - \frac{\partial Q_2(\vec{r},t)}{\partial \vec{r}}.
\end{equation}
Let us solve (\ref{eq6q}) and (\ref{eq7q}) regarding to $Q_2(\vec{r},t)$ (the quantity $Q_1(\vec{r},t)$ is considered to be given). It is apparent, that

\begin{equation}
\label{eq54a}
\begin{split}
\raisetag{40pt}
&\frac{\partial Q_1(\vec{r},t)}{\partial \vec{r}} = \left.-\left[\frac{\partial{L(t,\vec{r}')}}{\partial \left(\frac{\partial u_i}{\partial x_4}\right)} u_{i} - \frac{\partial{L(t,\vec{r}')}}{\partial \left(\frac{\partial u_i}{\partial x_4}\right)^{*}} u_{i}^*\right]\right|_{\vec{r}} = \\ 
&\frac{\partial Q_2(\vec{r},t)}{\partial t}. 
\end{split}
\end{equation}
Consequently $Q_2(\vec{r},t)$ is
\begin{equation}
\label{eq55a}
\begin{split}
\raisetag{40pt}
&Q_2(\vec{r},t) = -\int\limits_{(t)}\frac{\partial{L(\vec{r},t')}}{\partial \left(\frac{\partial u_i}{\partial x_4}\right)} u_{i}dt' - \\
&\int\limits_{(t)}\frac{\partial{L(\vec{r},t')}}{\partial \left(\frac{\partial u_{i}^*}{\partial x_4}\right)} u_{i}^*dt' + f(\vec{r}).
\end{split}
\end{equation}
The equation for determination of $f(\vec{r})$ is
\begin{equation}
\label{eq56a}
\begin{split}
\raisetag{40pt}
&\frac{d f(\vec{r})}{d \vec{r}} = \frac{\partial}{\partial \vec{r}}\int\limits_{(t)}\left[\frac{\partial{L(\vec{r},t')}}{\partial \left(\frac{\partial u_i}{\partial x_4}\right)} u_{i} - \frac{\partial{L(\vec{r},t')}}{\partial \left(\frac{\partial u_{i}^*}{\partial x_4}\right)} u_{i}^*\right]dt' \\
&+ \frac{\partial}{\partial t} \int\limits_{(\vec{r})}\left[\frac{\partial{L(t,\vec{r}')}}{\partial \left(\frac{\partial u_i}{\partial x_4}\right)} u_{i} - \frac{\partial{L(t,\vec{r}')}}{\partial \left(\frac{\partial u_{i}^*}{\partial x_4}\right)} u_{i}^*\right]d\vec{r}'. 
\end{split}
\end{equation} 
So we have 
\begin{equation}
\label{eq57a}
\begin{split}
\raisetag{40pt}
&Q_2(\vec{r},t) =\\
&\int\limits_{(\vec{r})}\left\{\frac{\partial}{\partial t} \int\limits_{(\vec{r}'')}\left[\frac{\partial{L(\vec{r}',t)}}{\partial \left(\frac{\partial u_i}{\partial x_4}\right)}u_{i} - \frac{\partial{L(\vec{r}',t)}}{\partial \left(\frac{\partial u_{i}^*}{\partial x_4}\right)}u_{i}^*\right]d\vec{r}'\right\}d\vec{r}''.
\end{split}
\end{equation} 
Then, in suggestion, that dynamic system studied is autonomous, that is
$L(\vec{r},t) = L(\vec{r})$,  occupies volume $ v \subset R_3$ and taking into account the coincidence of integration ranges $(\vec{r}) = (\vec{r}'')$, we will have
\begin{equation}
\label{eq58a}
\begin{split}
\raisetag{40pt}
&Q_2(\vec{r},t) = \\
&v \int\limits_{(\vec{r})}\left[\frac{\partial{L(\vec{r}')}}{\partial \left(\frac{\partial u_i}{\partial x_4}\right)} \frac{\partial u_{i}}{\partial t} - \frac{\partial{L(\vec{r}')}}{\partial \left(\frac{\partial u_{i}^*}{\partial x_4}\right)}\frac{\partial u_{i}^*}{\partial t}\right]d\vec{r}'.
\end{split}
\end{equation} 
Further, taking into consideration, that general solution of general relativistic equation is superposition of monochromatic plane waves, which have the view 
$ u_i(t) \sim e^{-i \frac{\mathcal E}{\hbar} t}$, and making a transformation of variable $t \rightarrow ict = x_4$ in the simplest case of one plane wave we  obtain 
\begin{equation}
\label{eq59q}
\begin{split}
\raisetag{40pt}
&Q_2(\vec{r},x_4) = \\
&\frac{v \mathcal E }{\hbar c} \int\limits_{(\vec{r})}\left[\frac{\partial {L(\vec{r}')}}{\partial \left(\frac{\partial u_i}{\partial x_4}\right)} u_{i}(\vec{r}',x_4)\right. 
+ \left.\frac{\partial {L(\vec{r}')}}{\partial \left(\frac{\partial u_{i}^*}{\partial x_4}\right)} u_{i}^*(\vec{r}',x_4)\right] d\vec{r}'. 
\end{split}
\end{equation} 
We see, that relationships (\ref{eq22}) and (\ref{eq59q}) are coinciding to scaling factor. They will coincide fully, if to make a transformation of parameter $\beta \rightarrow \beta ^{'} = \beta \frac{v \mathcal E}{\hbar c}$. The statement is proved. 

The converse can also be proven and can be employed for independent establishing of existence of some physical quantities in complex fields. For the example studied the suggestion on analicity of charge function of EM-field leads to existence of two quantities - real (electric) and imaginary (magnetic) component of the charge, which are invariant under gauge transformations.

\subsection{\textbf{Quantized Cavity EM-Field}}
The quantization of EM-field  was proposed for the first time still at the earliest stage of quantum physics in the works \cite{Born},\cite{Born_Heisenberg}, where quantum theory of dipole radiation was considered and the energy fluctuations in radiation field of blackbody have been calculated. The idea of Born-Jordan EM-field quantization is regarding of EM-field components to be matrices. At the same time quite another idea - to set up in the 
correspondence to each mode of radiation field the quantized harmonic oscillator, was proposed for the first time by Dirac  \cite{P.Dirac} and it is widely used in quantum electrodynamics (QED) including quantum optics \cite{Scully}, it is canonical quantization. Nevetheless at present in EM-field theory the first idea of quantization is also used. For instance, matrix representation of Maxwell equations in quantum optics \cite{Scully} corresponds to given idea.

EM-field potentials are used to be field functions by canonical quantization. At the same time to describe free  EM-field   it is sufficient to choose immediately the observable quantities - vector-functions $\vec{E}(\vec{r},t)$ and $\vec{H}(\vec{r},t)$ - to be field functions.
We use further given idea by EM-field quantization.

\subsubsection{\textbf{Time-Local Quantization of Cavity EM-Field}}

 We can start like to canonical quantization, from classical Hamiltonian, which for the first partial classical solution of Maxwell equations is

\begin{equation}
\label{eq12ab}
\begin{split}
&\mathcal{H}^{[1]}(t) = \frac{1}{2}\iiint\limits_{(V)}\left[\epsilon_0E_x^2(z,t)+\mu_0H_y^2(z,t)\right]dxdydz\\
&= \frac{1}{2}\sum_{\alpha=1}^{\infty}\left[m_{\alpha}\nu_{\alpha}^2q_{\alpha}^2(t) + \frac{p_{\alpha}^2(t)}{m_{\alpha}} \right],
\end{split}
\end{equation}
where
\begin{equation}
\label{eq42abc}
p_{\alpha} = m_{\alpha} \frac{dq_{\alpha}(t)}{dt}.
\end{equation}

 So, taking into consideration the relationship for Hamiltonian $\mathcal{H}^{[1]}(t)$ we set in correspondence to canonical variables ${q}_{\alpha}(t), {p}_{\alpha}(t)$, determined by the first partial solution of Maxwell equations,  the operators by usual way
\begin{equation}
\label{eq25ab}
\begin{split}
&\left[\hat {p}_{\alpha}(t) , \hat {q}_{\beta}(t)\right] = i\hbar\delta_{{\alpha}\beta}\\
&\left[\hat {q}_{\alpha}(t) , \hat {q}_{\beta}(t)\right] = \left[\hat {p}_{\alpha}(t) , \hat {p}_{\beta}(t)\right] = 0,
\end{split}
\end{equation}
where
$\alpha, \beta \in N$.
Introducing the operator functions  of time $\hat{a}_{\alpha}(t)$ and $ \hat{a}^{+}_{\alpha}(t)$
\begin{equation}
\label{eq26ab}
\begin{split}
&\hat{a}_{\alpha}(t) = \frac{1}{ \sqrt{ 2 \hbar  m_{\alpha} \omega_{\alpha}}} \left[ m_{\alpha} \omega_{\alpha}\hat {q}_{\alpha}(t) + i \hat {p}_{\alpha}(t)\right]\\
&\hat{a}^{+}_{\alpha}(t) = \frac{1}{ \sqrt{ 2 \hbar  m_{\alpha} \omega_{\alpha}}} \left[ m_{\alpha} \omega_{\alpha}\hat {q}_{\alpha}(t) - i \hat {p}_{\alpha}(t)\right],
\end{split}
\end{equation}
we obtain the operator functions of canonical variables in the form
\begin{equation}
\label{eq27ab}
\begin{split}
&\hat {q}_{\alpha}(t) = \sqrt{\frac{\hbar}{2 m_{\alpha} \omega_{\alpha}}} \left[\hat{a}^{+}_{\alpha}(t) + \hat{a}_{\alpha}(t)\right]\\
&\hat {p}_{\alpha}(t) = i \sqrt{\frac{\hbar m_{\alpha} \omega_{\alpha}}{2}} \left[\hat{a}^{+}_{\alpha}(t) - \hat{a}_{\alpha}(t)\right]. 
\end{split}
\end{equation}
Then EM-field  operator functions are obtained right away and they are
\begin{equation}
\label{eq28ab}
\hat{\vec{E}}(\vec{r},t) = \{\sum_{\alpha=1}^{\infty} \sqrt{\frac{\hbar \omega_{\alpha}}{V\epsilon_0}} \left[\hat{a}^{+}_{\alpha}(t) + \hat{a}_{\alpha}(t)\right] \sin(k_{\alpha} z)\} \vec{e}_x,
\end{equation}

\begin{equation}
\label{eq29ab}
\hat{\vec{H}}(\vec{r},t) = i \{\sum_{\alpha=1}^{\infty} \sqrt{\frac{\hbar \omega_{\alpha}}{V\mu_0}} \left[\hat{a}^{+}_{\alpha}(t) - \hat{a}_{\alpha}(t)\right] \cos(k_{\alpha} z)\} \vec{e}_y,
\end{equation} 
Taking into account the relationships (\ref{eq28ab}), (\ref{eq29ab}) and Maxwell equations, it is easily to find  an explicit form for the dependencies of operator functions   $\hat{a}_{\alpha}(t)$ and $ \hat{a}^{+}_{\alpha}(t)$ on the time. They are
\begin{equation}
\label{eq30ab}
\begin{split}
&\hat{a}^{+}_{\alpha}(t) = \hat{a}^{+}_{\alpha}(t = 0) e^{i\omega_{\alpha}t},\\
&\hat{a}_{\alpha}(t) = \hat{a}_{\alpha}(t = 0) e^{-i\omega_{\alpha}t},
\end{split}
\end{equation}
where $\hat{a}^{+}_{\alpha}(t = 0), \hat{a}_{\alpha}(t = 0)$ are constant, complex-valued in general case, operators. 

Physical sense of operator time dependent functions $\hat{a}^{+}_{\alpha}(t)$ and $\hat{a}_{\alpha}(t)$ is well known. They are creation  and annihilation operator of the $\alpha$-mode photon.
They are continuously differentiable operator functions of time. It means, that the time of  photon creation (annihilation) can be determined strictly, at the same time operator  functions $\hat{a}^{+}_{\alpha}(t)$ and $\hat{a}_{\alpha}(t)$  do not curry any information on the place, that is on space coordinates of given event.

It seems to be essential, that complex exponential dependencies in (\ref{eq30ab}) cannot be replaced by the real-valued harmonic trigonometrical functions. Really, if to suggest, that 
\begin{equation}
\label{eq31ab}
\hat{a}^{+}_{\alpha}(t) = \hat{a}^{+}_{\alpha}(t = 0)\cos\omega_{\alpha}t,
\end{equation}
then we obtain, that the following relation has to be taking place
\begin{equation}
\label{eq32ab}
\begin{split}
&[\hat{a}^{+}_{\alpha}(t = 0) - \hat{a}_{\alpha}(t = 0)]^{-1}[\hat{a}^{+}_{\alpha}(t = 0)\\
& + \hat{a}_{\alpha}(t = 0)] = \tan\omega_{\alpha}t.
\end{split}
\end{equation}
We see, that left-hand side in relation (\ref{eq32ab})  does not depend on time, right-hand side is depending. The contradiction obtained establishes an assertion.
Therefore, the quantized Maxwellian EM-field is  complex-valued  field in full correspondence with pure algebraic conclusion in Sec.III. 

Consequently, there is difference between classical and  quantized EM-fields, since classical EM-field can be determined by both complex-valued and real-valued functions. 
The fields $\vec{E}^{[2]}(\vec{r},t), \vec{H}^{[2]}(\vec{r},t)$ can be quantized in much the same way.
The operators $\hat{a}{''}_{\alpha}(t)$, $\hat{a}{''}^{+}_{\alpha}(t)$ are introduced analogously to (\ref{eq26ab}).
\begin{equation}
\label{eq33ab}
\begin{split}
&\hat{a}{''}_{\alpha}(t) = \frac{1}{ \sqrt{ 2 \hbar  m_{\alpha} \omega_{\alpha}}} \left[ m_{\alpha} \omega_{\alpha}\hat {q}{''}_{\alpha}(t) + i \hat {p}{''}_{\alpha}(t)\right]\\
&\hat{a}{''}^{+}_{\alpha}(t) = \frac{1}{ \sqrt{ 2 \hbar  m_{\alpha} \omega_{\alpha}}} \left[ m_{\alpha} \omega_{\alpha}\hat {q}{''}_{\alpha}(t) - i \hat {p}{''}_{\alpha}(t)\right]
\end{split}
\end{equation}
For the operators of field function we obtain
\begin{equation}
\label{eq34ab}
\begin{split}
&\hat{\vec{E}}^{[2]}(\vec{r},t) = \\
&\{\sum_{\alpha=1}^{\infty} \sqrt{\frac{\hbar \omega_{\alpha}}{V\epsilon_0}} \left[\hat{a}{''}^{+}_{\alpha}(t) + \hat{a}{''}_{\alpha}(t)\right] \sin(k_{\alpha} z)\} \vec{e}_1,
\end{split}
\end{equation}
\begin{equation}
\label{eq35ab}
\begin{split}
&\hat{\vec{H}}^{[2]}(\vec{r},t) = \\
&\{\sum_{\alpha=1}^{\infty} \sqrt{\frac{\hbar \omega_{\alpha}}{V\mu_0}} (-i) \left[\hat{a}{''}^{+}_{\alpha}(t) - \hat{a}{''}_{\alpha}(t)\right] \cos(k_{\alpha} z)\} \vec{e}_2.
\end{split}
\end{equation}
 In accordance with definition of complex quantities we  can built the following combination of solutions, satisfying Maxwell equtions
\begin{equation}\label{eq36ab}
(\vec{E}^{[1]}(\vec{r},t), \vec{E}^{[2]}(\vec{r},t)) \rightarrow \vec{E}^{[1]}(\vec{r},t) +  i \vec{E}^{[2]}(\vec{r},t) = \vec{E}(\vec{r},t),
\end{equation}
\begin{equation}\label{eq37ab}
(\vec{H}^{[2]}(\vec{r},t), \vec{H}^{[1]}(\vec{r},t)) \rightarrow \vec{H}^{[2]}(\vec{r},t) +  i \vec{H}^{[1]}(\vec{r},t) = \vec{H}(\vec{r},t).
\end{equation}
Consequently, the electric and magnetic field operators for quantized EM-field, corresponding to general solution of Maxwell equations,  are
\begin{equation}\label{eq33}
\begin{split}
&\hat{\vec{E}}(\vec{r},t) =  \{\sum_{\alpha=1}^{\infty} \sqrt{\frac{\hbar \omega_{\alpha}}{V\epsilon_0}} \{\left[\hat{a}^{+}_{\alpha}(t) + \hat{a}_{\alpha}(t)\right]\\
& + i \left[\hat{a}{''}_{\alpha}(t) + \hat{a}{''}^{+}_{\alpha}(t)\right]\} \sin(k_{\alpha} z)\} \vec{e}_x,
\end{split}
\end{equation}
and
\begin{equation}\label{eq34}
\begin{split}
&\hat{\vec{H}}(\vec{r},t) =  \{\sum_{\alpha=1}^{\infty} \sqrt{\frac{\hbar \omega_{\alpha}}{V\mu_0}}\{ \left[\hat{a}^{}_{\alpha}(t) - \hat{a}^{+}_{\alpha}(t)\right] \\
& + i \left[\hat{a}{''}_{\alpha}(t) - \hat{a}{''}^{+}_{\alpha}(t)\right] \} \cos(k_{\alpha} z) \} \vec{e}_y,
\end{split}
\end{equation}
It is substantial, that both field operators $\hat{\vec{E}}(\vec{r},t)$ and $\hat{\vec{H}}(\vec{r},t)$ are Hermitian operators.

The method of EM-field quantization above considered is in fact development of canonical quantization, proposed by Dirac. Further development can be made, if to 
take into account the independence and equal rights of all the coordinates $x_\mu,
\mu = \overline {1,4}$ in Minkowsky space $^{1}R_4$. Really all physical events are taking place on finite segment of time. It leads in application to electrodynamics to diskretness of $\omega$- space of possible light frequences like to diskretness of $\vec{k}$-space, which is the result of finiteness of cavity volume. It is interesting that, Dirac himself has in \cite{P.Dirac} written, that the theory proposed is not strictly relativistic, since the time everywhere is considered to be $c$-number instead of to consider it symmetrically with the space coordinates. From here follows unambiguously, that quantum electrodynamics, based on Dirac canonical EM-field quantization method is not fully relativistic and correspondingly it is not fully quantum theory. We will show in the next subsubsections the way to obtain fully quantum theory of electrodynamics.

\subsubsection{\textbf{Space-Local  Quantization of Cavity EM-Field}} 
We will consider for the simplicity the dependence of EM-field vector-functions the only on z-space coordinate, which is choosed in propagation direction in $R_3$ $\in$ $^{1}R_4$. The generalization on 3D-case is simple and will be not considered. 
Taking into account the independence and equal rights of all the coordinates $x_\mu,
\mu = \overline {1,4}$ in Minkowsky space $^{1}R_4$ we can also make Fourier transform on the segment [0, T], where $T$ is fixed time value, that is to represent the EM-field vector-functions in the form
\begin{equation}\label{eq35}
E^{[1]}_x(z,t)\vec{e}_x = \left[ \sum_{\alpha=1}^{\infty}A'_{\alpha}q_{\alpha}(z)\sin(\omega_{\alpha}t)\right] \vec{e}_x,
\end{equation}
\begin{equation}\label{eq36}
\begin{split}
& {H}^{[1]}_y(z,t)\vec{e}_y = \\
& \left\{ - \epsilon_0 \sum_{\alpha=1}^{\infty}\left[ A'_{\alpha}\omega_{\alpha}\cos(\omega_{\alpha}t)\int\limits_{0}^{z} q_{\alpha}(z')dz' + H_{\alpha 0}(t)\right]\right\}\vec{e}_y,
\end{split}
\end{equation}
where $q_{\alpha}(z)$,$\alpha \in N$,  is  $\alpha$-th normal mode of the 4-dimensional cavity, which include time coordinate along with space coordinates, 
\begin{equation}\label{eq37}
k_{\alpha} = \frac{\alpha\pi}{cT},  A'_{\alpha} = \sqrt{\frac{2 \omega_{\alpha}^2 m_{\alpha}}{T \epsilon_0}}, \omega_{\alpha} = \frac{\alpha \pi}{T},
\end{equation}
  $\{H_{\alpha 0}(t)\}$, $\alpha \in N$, is the set of arbitrary  functions of the time. Then the Hamiltonian can be obtained taking into account the expressions for $ E^{[1]}_x(z,t)$, ${H}^{[1]}_y(z,t)$ and
integrating.  So we will have
\begin{equation}\label{eq38}
G^{[1]}(z) = \frac{1}{2}\sum_{\alpha=1}^{\infty}\{m_{\alpha}\omega^{2}_{\alpha}\left[\frac{dq_{\alpha}'(z)}{dz}\right]^{2} + \frac{1}{c^2} \omega^{4}_{\alpha}m_{\alpha}\left[q_{\alpha}'(z)\right]^{2}\},
\end{equation}
where the case with  $\{H_{\alpha 0}(t)\} \equiv 0$, $t \in [0, T]$ for all $\alpha \in N$  is choosed and
\begin{equation}\label{eq39}
 q_{\alpha}'(z) = \int\limits_{0}^{z} q_{\alpha}(z')dz'.
\end{equation}
By redefition of  the variables in accordance with relations
\begin{equation}\label{eq40a}
\begin{split}
&q_{\alpha}''(z) = \frac{1}{c}\omega_{\alpha} q_{\alpha}'(z),\\
&p_{\alpha}''(z) = m_{\alpha}\omega_{\alpha} \frac{dq_{\alpha}'(z)}{dz},
\end{split}
\end{equation}
the  Hamiltonian $G^{[1]}(z)$ will have the canonical form
\begin{equation}\label{eq41}
G^{[1]}(z) = \frac{1}{2}\sum_{\alpha=1}^{\infty}\{\frac{[p_{\alpha}''(z)]^2}{m_{\alpha}} + m_{\alpha}\omega^{2}_{\alpha}[q_{\alpha}''(z)]^{2}\}.
\end{equation}
It means, that
space coordinates' dependent quantization of cavity EM-field can be  realized in a similar manner with above described time dependent quantization. 
 So, we can define quite analogously the quantization rules by the relationships
\begin{equation}\label{eq42}
\begin{split}
&\left[\hat{p}{''}_{\alpha}(z) , \hat {q}{''}_{\beta}(z)\right] = i\lambda_{0}\delta_{{\alpha}\beta}\\
&\left[\hat {q}{''}_{\alpha}(z) , \hat {q}{''}_{\beta}(z)\right] = \left[\hat {p}{''}_{\alpha}(z) , \hat {p}{''}_{\beta}(z)\right] = 0,
\end{split}
\end{equation}
where $\alpha,\beta \in N$, $\lambda_{0}$ is analogue of Planck constant.  It is evident from $\lambda_{0}$-definition by (\ref{eq42}), that $\lambda_{0}$ and Planck constant have the same dimension, however their numerical coincidence seems to be unobvious, since Planck constant characterizes the "seizure" of the time by propagating of EM-field, while $\lambda_{0}$ characterises the "seizure" of the space.

The operators $\hat{a}{''}_{\alpha}(z)$, $\hat{a}{''}^{+}_{\alpha}(z)$ are defined also analogously to operators $\hat{a}{}_{\alpha}(t)$, $\hat{a}{}^{+}_{\alpha}(t)$ and they are
\begin{equation}\label{eq43}
\begin{split}
&\hat{a}{''}_{\alpha}(z) = \frac{1}{ \sqrt{ 2 m_{\alpha} \lambda_{0}  \omega_{\alpha}}} \left[m_{\alpha} \omega_{\alpha}\hat {q}{''}_{\alpha}(z) + i \hat {p}{''}_{\alpha}(z)\right]\\
&\hat{a}{''}^{+}_{\alpha}(z) = \frac{1}{ \sqrt{ 2 m_{\alpha} \lambda_{0} \omega_{\alpha}}} \left[m_{\alpha} \omega_{\alpha}\hat {q}{''}_{\alpha}(z) - i \hat {p}{''}_{\alpha}(z)\right].
\end{split}
\end{equation}
The dependencies of given scalar operator functions on coordinate $z$ in an explicit form for Maxwellian EM-field can be easily obtained by means of solutions of Maxwell equations and they are
\begin{equation}\label{eq44}
\begin{split}
&\hat{a}^{+}_\alpha(z) = \hat{a}^{+}_\alpha (0) e^{ik_\alpha z}\\
&\hat{a}_\alpha(z) = \hat{a}_\alpha(0) e^{-ik_\alpha z},
\end{split}
\end{equation}
where $\hat{a}^{+}_{\alpha}(0), \hat{a}_{\alpha}(0)$ are constant, complex-valued in general case, operators.
Let us remark in passing, that the dependencies (\ref{eq44}) on $z$-coordinate are similar to  dependencies $\hat{a}^{+}_{\alpha}(t), \hat{a}_{\alpha}(t)$ on time, which are given by  (\ref{eq30ab}).

From relationships  (\ref{eq43}) we obtain the expressions for operators of canonical variables $\hat {q}{''}_{\alpha}(z)$ and $\hat {p}{''}_{\alpha}(z)$ in the form
\begin{equation} \label{eq45}
\begin{split}
&\hat {q}{''}_{\alpha}(z) = \sqrt{\frac{\lambda_{0}}{2 m_{\alpha}\omega_{\alpha}}} \left[\hat{a}{''}^{+}_{\alpha}(z) + \hat{a}{''}_{\alpha}(z)\right]\\
&\hat {p}{''}_{\alpha}(z) = i \sqrt{\frac{m_{\alpha}\lambda_{0} \omega_{\alpha}}{2}} \left[\hat{a}{''}^{+}_{\alpha}(z) - \hat{a}{''}_{\alpha}(z)\right]. 
\end{split}
\end{equation}
Then it is easily to show, that Hamilton operator $\hat{G}^{[1]}(z)$ can be represented in the simple form
\begin{equation}\label{eq46}
\hat{G}^{[1]}(z) = \sum_{\alpha=1}^{\infty}\lambda_{0} \omega_{\alpha}\left[\hat{a}{''}^{+}_{\alpha}(z)\hat{a}{''}_{\alpha}(z) + \frac{1}{2}\right],
\end{equation}
which determines physical meaning of the operators
$\hat{a}{''}^{+}_{\alpha}(z)$ and $\hat{a}{''}_{\alpha}(z)$. It is evident, that they are operators of creation and annihilation of the photon at space coordinate $z$. So, we see, that it is possible by space coordinates' dependent quantization   to determine the place of photon creation (annihilation), however it is impossible to determine the time of photon creation (annihilation). Therefore we have reverse picture to the case of  the time dependent quantization, where (see previous Subsection)  it is possible to determine the time of photon creation (annihilation) and it is impossible to determine the place of photon creation (annihilation). The view of (\ref{eq46}), which is coinciding with view of known expressions for canonical  quantization, if $\lambda_{0}$ to replace by $\hbar$, confirms the conclusion, that dimension of constant of space coordinates' dependent quantization and dimension of Planck  constant are identical, that is $[\lambda_{0}]$ = $[\hbar]$.

From relationships  (\ref{eq43}) and (\ref{eq42}) we can obtain the expressions for commutation relations of the creation and annihilation operators $\hat{a}{''}^{+}_{\alpha}(z)$ and $\hat{a}{''}_{\alpha}(z)$. They are
\begin{equation}\label{eq47}
[\hat{a}{''}_{\alpha}(z),\hat{a}{''}^{+}_{\beta}(z)] = \hat{e}\delta_{\alpha\beta},
\end{equation}
where $\hat{e}$ is unit operator, $\alpha,\beta \in N$.

It seems to be evident, that the second case of EM-field quantization, that is   space coordinates' dependent quantization is acceptable for the quantization of any Coulomb field, which has nonzeroth curl, that takes place in  1D and in 2D systems. It was passed earlier for impossible  to quantize any Coulomb field, see for example \cite{Dutra}. The quantization of Coulomb field in lowdimensional aforesaid systems corresponds to the presence of own life of radiation Coulomb field in given systems, that is Coulomb field in lowdimensional  systems has the character of radiation field and it can exist without the sources, which have created given field.  Given conclusion seems to be substantial to gain a better understanding, for instance, of the properties of organic conductors, perfect nanowires and nanotubes, graphene and the systems like them, including 1D and 2D biological subsystems. 

The expressions for 
the operators of vector-functions of EM-field are similar in their structure to expressions, given by (\ref{eq33}), (\ref{eq34}) and they are
\begin{equation}\label{eq48}
\hat{\vec{E}}^{[1]}(\vec{r},t) = \{i\sum_{\alpha=1}^{\infty}  \sqrt{\frac{\lambda_{0}\omega_{\alpha}}{T\epsilon_0}} \sin\omega_{\alpha}t \left[\hat{a}{''}^{+}_{\alpha}(z) - \hat{a}{''}_{\alpha}(z)\right]\}\vec{e}_x
\end{equation}
 and
\begin{equation}\label{eq49}\begin{split}
&\hat{\vec{H}}^{[1]}(\vec{r},t) =\\ 
&\{-\sum_{\alpha=1}^{\infty}\sqrt{\frac{\lambda_{0}\omega_{\alpha}}{T\mu_0}} \cos\omega_{\alpha}t\left[\hat{a}{''}^{+}_{\alpha}(z) + \hat{a}{''}_{\alpha}(z)\right]\}\vec{e}_y.\end{split}
\end{equation}  
We see, that the field operators $\hat{\vec{E}}^{[1]}(\vec{r},t)$ $\hat{\vec{H}}^{[1]}(\vec{r},t)$ are local operators in the space $R_3$, that allows to enter the photon wave function in coordinate representation, that is, to solve the problem, which was accepted to be unsolvable in the principle \cite{Scully}, \cite{Akhiezer}, \cite{Dutra}. 

\subsubsection{\textbf{Space-Time Local Quantization of Cavity EM-Field}} 

Let us consider general case, corresponding to discrete both  $\omega$-space of possible light frequencies  and  $\vec{k}$-space of light wave vectors, which are  result of finiteness of 4-cavity space volume and time segment. Let us find the relations for EM-field vector-functions.  In the case of cavity electrodynamics considered  we have two $1D$ ranges of variables $t$  and $z$, which belong to segments $t \in [0, T]$, $z \in [0, L]$, that is, there is in fact  to be given 2D-range D(t,z), which can be considered to be definitional domain of vector-functions $\vec{E}(\vec{r},t)$ and $\vec{H}(\vec{r},t)$ of two  variables $t$  and $z$. In the case, when given functions are absolutely integrable over both the variables $t$  and $z$, they can be represented in the form of  multiple series, given by the relations
\begin{equation}\label{eq50}
\vec{E}(\vec{r},t) = \{\sum_{\alpha=1}^{\infty}\sum_{\beta=1}^{\infty} A^{'E}_{\alpha \beta} q_{\alpha}(t) q_{\beta}(z)\}\vec{e}_x,
\end{equation}
and 
\begin{equation}\label{eq51}
\vec{H}(\vec{r},t) = \{-\sum_{\alpha=1}^{\infty}\sum_{\beta=1}^{\infty} A^{'H}_{\alpha \beta} \frac{dq_{\alpha}(t)}{dt}\int\limits _{0}^{z} q_{\beta}(z')dz'\}\vec{e}_y, 
\end{equation}
where $\{q_{\alpha}(t)\}$, $\{q_{\beta}(z)\}$, $\alpha$, $\beta$ $\in$ $N$ are two systems of orthogonal functions, $A^{'E}_{\alpha \beta}$, $A^{'H}_{\alpha \beta}$ are coefficients in given expansions, which depend on both the indices $\alpha$ and $\beta$. Both two-fold series will be two-fold Fourier series, if the sets  $\{q_{\alpha}(t)\}$, $\{q_{\beta}(z)\}$ are two orthogonal systems of harmonical trigonometric functions. It is evident, that the sets $\{q_{\alpha}(t)\}$, $\{q_{\beta}(z)\}$ are independent each other and produce bases with  $\aleph_0$ dimension
in the metrizable complete spaces $L_2$, which are therefore Hilbert spaces.  It follows from physical meaning in the case of definite direction of propagation, that between the bases $\{q_{\alpha}(t)\}$, $\{q_{\beta}(z)\}$ and correspondingly between both Hilbert spaces the mapping
\begin{equation}\label{eq52}
\Gamma: \{q_{\alpha}(t)\} \rightarrow \{q_{\beta}(z)\}
\end{equation}
is isomorphism, at that if there is preferential (propagation) direction in $^1R_4$-space, both the sets have to be ordered in correspondence with running numbers. 
It means, that
\begin{equation}\label{eq53}
\begin{split}
&\vec{E}(\vec{r},t) = \{\sum_{\alpha=1}^{\infty}\sum_{\beta=1}^{\infty} A^{'E}_{\alpha \beta} q_{\alpha}(t) q_{\beta}(z)\}\vec{e}_x =\\ &\{\sum_{\alpha=1}^{\infty} A^{''E}_{\alpha} q_{\alpha}(t) q_{\alpha}(z)\}\vec{e}_x
\end{split}
\end{equation}
and
\begin{equation}
\label{eq54}
\begin{split}
&\vec{H}(\vec{r},t) = \{\sum_{\alpha=1}^{\infty}\sum_{\beta=1}^{\infty} A^{'H}_{\alpha \beta}\frac{dq_{\alpha}(t)}{dt}\int\limits_{0}^{z}  q_{\beta}(z')dz' \}\vec{e}_y = \\
&\{\sum_{\alpha=1}^{\infty} A^{'H}_{\alpha}\frac{dq_{\alpha}(t)}{dt} \int\limits_{0}^{z}q_{\alpha}(z')dz'\}\vec{e}_y,
\end{split}
\end{equation}
where $A^{''E}_{\alpha}$, $A^{'H}_{\alpha}$ are coefficients in given expansions, which depend now the only on index $\alpha$. We have considered the mathematical aspect. Physically the insert of Kronecker symbol $\delta_{\alpha\beta}$ in double sum in (\ref{eq50}), (\ref{eq51}) corresponds to renumbering of the massive $\{\beta\}$ in that way, in order to $\alpha$ and $\beta$ were running  the sets $\{\alpha\}$ and $\{\beta\}$ synchronously one after another with number growth. It is additional requirement, since, although both the sets $\{\alpha\}$ and $\{\beta\}$ have the same cardinal number $\aleph_0$ and although the mapping (\ref{eq52}) in view of its biectivity gives one-to-one relation
between both the sets, it can be realized along with synchronous running above indicated by infinite number of asynchronous running. The choice of synchronous running is determined by causality principle - the photons by  their propagation synchronously "lock on" the space and the time.
It means in its turn, that complete local quantization of EM-field becomes to be possible. 

It can be shown, that along with expression for $\vec{H}(\vec{r},t)$, given by (\ref{eq54}),  $z$-coordinate part can be represented in more symmetrical  form like to $t$-coordinate part in (\ref{eq3abc}), that is
\begin{equation}
\label{eq55}
\begin{split}
\vec{H}(\vec{r},t) = 
\{\sum_{\alpha=1}^{\infty} A^{''H}_{\alpha}\frac{1}{\omega_{\alpha}}\frac{dq_{\alpha}(t)}{dt} \left(\frac{1}{k_{\alpha}}\frac{dq_{\alpha}(z)}{dz}\right)\}\vec{e}_y,
\end{split}
\end{equation}
where  $t \in [0, T]$, $z \in [0, L]$ and $A^{''H}_{\alpha}$ is
\begin{equation}
\label{eq56}
A^{''H} = \sqrt{\frac{2\omega^2_{\alpha} m_{\alpha}}{\mu_{0} V T}}.
\end{equation}
 For 
$\vec{E}(\vec{r},t)$ we retain the relation, given by (\ref{eq53}), in which  $t \in [0, T]$, $z \in [0, L]$ and $A^{''E}_{\alpha}$ is
 \begin{equation}
\label{eq57}
A^{''E}_{\alpha} = \sqrt{\frac{2\omega^2_{\alpha} m_{\alpha}}{\epsilon_{0} V T}}.
\end{equation}

It seems to be essential, that the segment $[0, T]$ is not arbitrary, $T$ has to be equal to $\frac{L}{c}$, that ensures the synchronization above discussed of the EM-field propagation in the space and in the time. Then discretness of $\omega$-space will correspond to discretness of $\vec{k}$-space.

Let us designate
\begin{equation}
\label{eq58}
\begin{split}
&q_{\alpha}(z)q_{\alpha}(t) = q_{\alpha}(z,t),\\ 
&m_{\alpha}\frac{dq_{\alpha}(t)}{dt} = p_{\alpha}(t),\\
&\frac{1}{k_{\alpha}}\frac{dq_{\alpha}(z)}{dz} = p_{\alpha}(z),\\
&p_{\alpha}(z)p_{\alpha}(t) = p_{\alpha}(z,t).
\end{split}
\end{equation}
Then classical Hamiltonian density is 
\begin{equation}
\label{eq59}
\begin{split}
&\mathfrak{W}(z,t) = \frac{1}{2} \{ \epsilon_0 \left[\sum_{\alpha=1}^{\infty} \sqrt{\frac{2 \omega^2_{\alpha} m_{\alpha}}{\epsilon_{0} V T}} q_{\alpha}(z,t)\right]^2 + \\
&\mu_0 \left[\sum_{\alpha=1}^{\infty} \sqrt{\frac{2 \omega^2_{\alpha} m_{\alpha}}{\mu_{0} V T}} \frac{1}{\omega_{\alpha} m_{\alpha}} p_{\alpha}(z,t) \right]^2 \} = \\
&\frac{1}{2} \{\sum_{\alpha=1}^{\infty} \frac
{2 \omega^2_{\alpha} m_{\alpha}}{V T} q^2_{\alpha}(z,t)  +  \\
&\sum_{\alpha\neq\beta}\sum_{\beta=1}^{\infty}\frac{2 \omega_{\alpha} \omega_{\beta}}{V T} \sqrt{m_{\alpha} m_{\beta}}  q_{\alpha}(z,t)
q_{\beta}(z,t)  +  \\
&\sum_{\alpha=1}^{\infty} \frac
{2}{V T} \frac{1}{m_{\alpha}} p^2_{\alpha}(z,t)  +  \\
&\sum_{\alpha\neq\beta}^{}\sum_{\beta=1}^{\infty}\frac{2}{V T \sqrt{m_{\alpha} m_{\beta}}} p_{\alpha}(z,t) p_{\beta}(z,t)\}
\end{split}
\end{equation}  

It seems to be evident, that by integration over 4-volume both  the items with double sum will give contribution, which is equal to zero.  It is consequence of 
orthogonality of the functions $\{q_{\alpha}(t)\}$, $\{q_{\beta}(z)\}$. 
It means, that the Hamiltonian density can be choosed in the canonical form
\begin{equation}
\label{eq60q}
\begin{split}
\mathfrak{W}^{[1]}(z,t) = \frac{1}{V T}\sum_{\alpha=1}^{\infty} \left[ 
m_{\alpha} \omega^2_{\alpha} q^2_{\alpha}(z,t)  +    
 \frac{p^2_{\alpha}(z,t)}{m_{\alpha}} \right]    
\end{split}
\end{equation}
Then following Dirac canonical quantization method, we have 
\begin{equation}
\label{eq61a}
\begin{split}
&\left[\hat {p}_{\alpha}(z, t) , \hat {q}_{\beta}(z, t)\right] = i\hat{g}^{(1)}(z, t) \delta_{{\alpha}\beta}\\
&\left[\hat {q}_{\alpha}(z, t) , \hat {q}_{\beta}(z, t)\right] = \left[\hat {p}_{\alpha}(z, t) , \hat {p}_{\beta}(z, t)\right] = 0,
\end{split}
\end{equation}
It is substantial, that instead scalar value we have $\hat{g}^{(1)}(z, t)$, that is operator function of the variables $z$ and $t$.  Really, taking into account
(\ref{eq58}), we obtain
\begin{equation}
\label{eq64a}
\begin{split}
&[\hat{p}_{\alpha}(z,t),\hat{q}_{\beta}(z,t)] = [\hat{p}_{\alpha}(z) \hat{p}_{\alpha}(t),
\hat{q}_{\beta}(z)\hat{q}_{\beta}(t)] = \\ &i\hbar\delta_{\alpha\beta}\hat{p}_{\alpha}(z)\hat{q}_{\beta}(z) + i \lambda_0 \delta_{\alpha\beta} \hat{p}_{\alpha}(t) \hat{q}_{\beta}(t)
\end{split}
\end{equation}
Therefore, $\hat{g}^{(1)}(z, t)$ is
\begin{equation}
\label{eq65a}
\hat{g}^{(1)}(z, t) = i \delta_{\alpha\beta} [\hbar \hat{p}_{\alpha}(z) \hat{q}_{\beta}(z) +  \lambda_0  \hat{p}_{\alpha}(t) \hat{q}_{\beta}(t)]
\end{equation}
It is seen, that $\hat{g}^{(1)}(z, t)$ is dependent on both the sequence of indices $\alpha$, $\beta$ (in distinction from usual case) and on the sequence of operator functions in (\ref{eq65a}). In other words there are else three operator functions of analogous structure. They are
\begin{equation}
\label{eq66a}
\hat{g}^{(2)}(z, t) = -i \delta_{\alpha\beta} [\hbar \hat{p}_{\alpha}(z) \hat{q}_{\beta}(z) +  \lambda_0  \hat{p}_{\alpha}(t) \hat{q}_{\beta}(t)]\end{equation}
\begin{equation}
\label{eq62a}
\hat{g}^{(3)}(z, t) = i \delta_{\alpha\beta} [\hbar \hat{p}_{\beta}(z) \hat{q}_{\alpha}(z) +  \lambda_0  \hat{p}_{\beta}(t) \hat{q}_{\alpha}(t)]\end{equation}
\begin{equation}
\label{eq73}
\hat{g}^{(4)}(z, t) = -i \delta_{\alpha\beta} [\hbar \hat{p}_{\beta}(z) \hat{q}_{\alpha}(z) +  \lambda_0  \hat{p}_{\beta}(t) \hat{q}_{\alpha}(t)]
\end{equation}
It seems to be convenient to define symmetrized operator $\hat{g}$ by all the four $\hat{g}^{(j)}(z, t)$ , $j = \overline {1,4}$,  functions,
that is 
\begin{equation}\label{eq67a}
\hat{g} = \frac{1}{4}\sum_{j=1}^{4}\hat{g}^{(j)}(z, t) = -\hbar \lambda_0 \hat{e}, 
\end{equation}
which is scalar, multiplied on unit operator. So  
\begin{equation}\label{eq68a}
g = -\hbar \lambda_0
\end{equation}
The operator functions   $\hat{a}_{\alpha}(z,t)$ and $ \hat{a}^{+}_{\alpha}(z,t)$
\begin{equation}
\label{eq63a}
\begin{split}
&\hat{a}_{\alpha}(z, t) = \frac{1}{ \sqrt{ 2 \hbar \lambda_0  m_{\alpha} \omega_{\alpha}}} \left[ m_{\alpha} \omega_{\alpha}\hat {q}_{\alpha}(z, t) + i \hat {p}_{\alpha}(z, t)\right]\end{split}
\end{equation}
\begin{equation}
\label{eq69a}
\begin{split}
\hat{a}^{+}_{\alpha}(z, t) = \frac{1}{ \sqrt{ 2 \hbar \lambda_0  m_{\alpha} \omega_{\alpha}}} \left[ m_{\alpha} \omega_{\alpha}\hat {q}_{\alpha}(z, t) - i \hat {p}_{\alpha}(z, t)\right].
\end{split}
\end{equation}
Then the operator functions of canonical variables have the form
\begin{equation}
\label{eq70a}
\begin{split}
&\hat {q}_{\alpha}(z, t) = \sqrt{\frac{\hbar \lambda_0 }{2 m_{\alpha} \omega_{\alpha}}} \left[\hat{a}^{+}_{\alpha}(z, t) + \hat{a}_{\alpha}(z, t)\right]
\end{split}
\end{equation}
\begin{equation}
\label{eq71}
\begin{split}
&\hat {p}_{\alpha}(z, t) = i \sqrt{\frac{\hbar \lambda_0  m_{\alpha} \omega_{\alpha}}{2}} \left[\hat{a}^{+}_{\alpha}(z, t) - \hat{a}_{\alpha}(z, t)\right]. 
\end{split}
\end{equation}
It is easily to show, that operator functions $\hat{a}_{\alpha}(z,t)$ and $ \hat{a}^{+}_{\beta}(z,t)$ satisfy the following relation
\begin{equation}
\label{eq72}
[\hat{a}_{\alpha}(z, t), \hat{a}^{+}_{\beta}(z, t)] = -i \delta_{\alpha\beta} \hat{e}
\end{equation}
Taking into account the expressions for $\vec{E}(\vec{r},t)$ and $\vec{H}(\vec{r},t)$, given by (\ref{eq53}), (\ref{eq55}), we have for operators of EM-field vector functions
\begin{equation}\label{eq74}
\begin{split}
&\hat{\vec{E}}(\vec{r},t) = \{\sum_{\alpha=1}^{\infty} A^{''E}_{\alpha} \hat{q}_{\alpha}(t) \hat{q}_{\alpha}(z)\}\vec{e}_x = \\ &\{\sum_{\alpha=1}^{\infty} A^{''E}_{\alpha} \hat{q}_{\alpha}(z, t) \}\vec{e}_x = \\
&\{\sum_{\alpha=1}^{\infty} A^{''E}_{\alpha}\sqrt{\frac{\hbar \lambda_0 }{2 m_{\alpha} \omega_{\alpha}}} [\hat{a}^{+}_{\alpha}(z, t) + \hat{a}_{\alpha}(z, t)]\}\vec{e}_x
\end{split}
\end{equation}
and
\begin{equation}
\label{eq75}
\begin{split}
&\hat{\vec{H}}(\vec{r},t) = 
\{\sum_{\alpha=1}^{\infty} A^{''H}_{\alpha}\frac{1}{\omega_{\alpha}}\frac{d\hat{q}_{\alpha}(t)}{dt} (\frac{1}{k_{\alpha}}\frac{d\hat{q}_{\alpha}(z)}{dz})\}\vec{e}_y = \\
&\{\sum_{\alpha=1}^{\infty} A^{''H}_{\alpha}
\frac{1}{m_{\alpha}\omega_{\alpha}}\hat{p}_{\alpha}(t)
\hat{p}_{\alpha}(z) )\}\vec{e}_y,
\end{split}
\end{equation}
which using the relations (\ref{eq58}) (\ref{eq71}) can be rewritten
\begin{equation}
\label{eq76}
\begin{split}
&\hat{\vec{H}}(\vec{r},t) = \{\sum_{\alpha=1}^{\infty} A^{''H}_{\alpha}
\frac{1}{m_{\alpha}\omega_{\alpha}}
\hat{p}_{\alpha}(z, t)\}\vec{e}_y = \\
&i \{\sum_{\alpha=1}^{\infty} A^{''H}_{\alpha}
 \sqrt{\frac{\hbar \lambda_0}{2 m_{\alpha} \omega_{\alpha}}} [\hat{a}^{+}_{\alpha}(z, t) -\\
& \hat{a}_{\alpha}(z, t)]\}\vec{e}_y
\end{split}
\end{equation}
Therefore by means of operator functions $\hat{a}^{+}_{\alpha}(z, t)$, $\hat{a}_{\alpha}(z, t)$ the local quantization of EM-field is realized, which allows to determine simultaneously along with time of creation (annihilation) of photons the space coordinate of given process.

\subsection{\textbf{Cavity 4-Currents}}
It represents the interest to calculate the 4-currents for given task. 
Let us  place all the vector-functions in pairs in accordance with their parity. Then we have the following pairs
\begin{equation}
\label{eq15abc}
(\vec{E}^{[1]}(\vec{r},t), \vec{E}^{[2]}(\vec{r},t)),
(\vec{H}^{[2]}(\vec{r},t), \vec{H}^{[1]}(\vec{r},t))
\end{equation}
in which both the $\vec{E}$-vectors and $\vec{H}$-vectors have the same space parity (polar and axial correspondingly) and differ each other by t-parity, t-even and t-uneven  in accordance with their numbers in pairs. It means, that they trasform like to $x_4$ and $x_1$ coordinates in $^1R_4$. 
In a similar manner can be set the vector-functions with opposite to the vector-functions in (\ref{eq15abc}) space parity
\begin{equation}
\label{eq15bcd}
(\vec{E}^{[3]}(\vec{r},t), \vec{E}^{[4]}(\vec{r},t)),
(\vec{H}^{[4]}(\vec{r},t), \vec{H}^{[3]}(\vec{r},t)).
\end{equation}
Then taking into account the definition of complex quantities to be pair of real defined quantities, taken in fixed order, we come in a natural way once again  to concept of  complex vector-functions, which describe Maxwellian EM-field equations.  In other words, we have in fact the quantities
\begin{equation}
\label{eq16ab}
\begin{split}
&\vec{E}^{[1]}(\vec{r},t) +  i \vec{E}^{[2]}(\vec{r},t) = \vec{E}_{c,p}(\vec{r},t),\\
&\vec{H}^{[2]}(\vec{r},t) +  i \vec{H}^{[1]}(\vec{r},t) = \vec{H}_{c,a}(\vec{r},t),
\end{split}
\end{equation}
and
\begin{equation}
\label{eq17ab}
\begin{split}
&\vec{E}^{[3]}(\vec{r},t) +  i \vec{E}^{[4]}(\vec{r},t) = \vec{E}_{c,a}(\vec{r},t),\\
&\vec{H}^{[4]}(\vec{r},t) +  i \vec{H}^{[3]}(\vec{r},t) = \vec{H}_{c,p}(\vec{r},t),
\end{split}
\end{equation}
where complex plane put in correspondence to (y, z) real plane, subscripts a and p mean axial and polar respectively.
 It seems to be convenient to determine  the space of EM-field vector-functions  under the ring of quaternions with another basis in comparison with basis, given by (\ref{eq15cacdegh}). We will use now the quaternion basis $\{e_i\}, i = \overline{0,3}$ with algebraic operations between elements, satisfying to relationships 
\begin{equation}
\label{eq40}
e_i e_j = \varepsilon_{ijk}e_k + \delta_{ij} e_{_0}, e_{_0} e_i = e_i, {e_{_0}}^2 = e_{_0}, i,j,k =\overline{1,3},
\end{equation}
where $\varepsilon_{ijk}$ is completely antisymmetric Levi-Civita 3-tensor. 

Let us define the vector biquaternion 
\begin{equation}
\label{eq17abc}
\vec{\Phi} = (\vec{E}^{[1]} + \vec{H}^{[2]}) + i (\vec{H}^{[1]} + \vec{E}^{[2]}),
\end{equation}
 which can be represented to be the sum of the biquaternions
\begin{equation}
\label{eq17bcd} 
\vec{\Phi} = \vec{F} + \vec{\tilde{F}},
\end{equation}
where $\vec{F} =\vec{E}^{[1]} + i (\vec{H}^{[1]}, \vec{\tilde{F}} = \vec{H}^{[2]} + i \vec{E}^{[2]}$.
 Then Maxwell equations for instance for two free photon fields with different t-parity are
\begin{equation}
\label{eq17cde} 
\nabla\vec{\Phi} = 0.
\end{equation}

  The generalized Maxwell equations in quaternion form with quaternion basis, given by (\ref{eq15cacdegh}), can also be rewritten in fully quaternion form, if to use both the bases. It seems to be consequence of independence of basis definition for both the quaternion forms.

\subsubsection{\textbf{Classical Cavity 4-Currents}}

It is evident, that  
\begin{equation}
\label{eq18ab}
{j_{\mu,\pm}}(x) = j^{(1)}_{\mu,\pm}(x) + i j^{(2)}_{\mu,\pm}(x),
\end{equation}
where subscript $\pm$ corresponds to two possibilities for definition of complex vector-functions. Along with relationships (\ref{eq16ab}), (\ref{eq17ab}) they can be defined by the change of addition sign   in (\ref{eq16ab}), (\ref{eq17ab}) into opposite.
The quantity $j^{(1)}_{\mu,\pm}(x)$ is well known quantity, and it is determined by
\begin{equation}
\label{eq19ab}
\begin{split}
&j^{(1)}_{\mu,\pm}(x) = -\frac{i e}{\hbar c}\sum_{\alpha = 1}^{\infty}\sum_{s=1}^{2}\left[ \frac{\partial{L(x)}}{\partial(\partial_{\mu}u^{s,\pm}_\alpha(x))} u^{s,\pm}_{\alpha}(x)\right] \\
&+ \frac{i e}{\hbar c}\sum_{\alpha = 1}^{\infty}\sum_{s=1}^{2}\left[ \frac{\partial{L(x)}}{\partial(\partial_{\mu}u^{*s,\pm}_\alpha(x))} u^{*s,\pm}_{\alpha}(x)\right],
\end{split}
\end{equation}
where $L(x)$ is Lagrange function and  $u^{s,\pm}_{\alpha}(x), s = 1,2$ are
\begin{equation}
\label{eq19abcd}
\begin{split}
&u^{1,\pm}_{\alpha}(x) = \sqrt{\epsilon_0}A^{E}_\alpha \sin k_\alpha(x_3) [q_\alpha(x_4) \pm i q^{''}_\alpha(x_4)]\\
&u^{2,\pm}_{\alpha}(x) = \sqrt{\mu_0}A^{H}_\alpha \cos k_\alpha(x_3) [-q{'}_\alpha(x_4) \pm i \frac {1}{\omega_\alpha}\frac{dq_\alpha(x_4)}{dx_4}]
\end{split}
\end{equation} 
The functions $u^{s,\pm}_{\alpha}(x), s = 1,2, \alpha \in N$ are built from the components of   
the expansion in Fourier series of the fields $\vec{E}^{[1]}(\vec{r},t), \vec{E}^{[2]}(\vec{r},t)$ and
$\vec{H}^{[2]}(\vec{r},t), \vec{H}^{[1]}(\vec{r},t)$ correspondingly.   

To determine the current density $j^{(2)}_{\mu,\pm}(x)$ we have to take into consideration, that gauge symmetry group of EM-field is two-parametric group $\Gamma(\alpha,\beta) = U_{1}(\alpha) \otimes \mathfrak R(\beta)$, where $\mathfrak R(\beta)$ is abelian multiplicative group of real numbers (excluding zero). It leads also to existence for EM-field of complex 4-current densities including complex charge density component. 
 
The current density $j^{(2)}_{\mu,\pm}(x)$ is given by the expression
\begin{equation}
\label{eq21ab}
\begin{split}
&j^{(2)}_{\mu,\pm}(x) = -\frac{ie}{\hbar c}\sum_{\alpha = 1}^{\infty}\sum_{s=1}^{2}\left[ \frac{\partial{L(x)}}{\partial(\partial_{\mu}u^{s,\pm}_\alpha(x))} u^{s,\pm}_{\alpha}(x)\right] \\
&- \frac{ie}{\hbar c}\sum_{\alpha = 1}^{\infty}\sum_{s=1}^{2}\left[ \frac{\partial{L(x)}}{\partial\partial_{\mu}u^{*s,\pm}_\alpha(x)} u^{*s,\pm}_{\alpha}(x)\right].
\end{split}
\end{equation}

It can be easily shown, that $j_{3}^{1,\pm}(\vec{r},t)$ is always equal to zero for any set of twice continuously differentiable functions  $\{q_{\alpha}(t)\}, \alpha \in N$. The expression for arbitrary set of twice continuously differentiable functions  $\{q_{\alpha}(t)\}, \alpha \in N$, for $j_{3}^{2,\pm}(\vec{r},t)$ is
\begin{equation}
\label{eq21abc}
\begin{split}
&j_{3}^{2,\pm}(\vec{r},t) = -\frac{2ie}{\hbar c^2V}\sum_{\alpha = 1}^{\infty}m_{\alpha}\omega_{\alpha}^3\sin 2k_{\alpha}z \times\\
&\{\left[i|q_{\alpha}(t)\pm
i\omega_{\alpha}^2 \int\limits_{0}^{t} \int\limits_{0}^{t{''}}q_{\alpha}(t')dt'dt{''}|^2\right]\\ - &\left[|\omega_{\alpha}\int\limits _{0}^{t}q_{\alpha}(t')dt' \mp \frac{i}{\omega_{\alpha}}\frac{dq_{\alpha}(t)}{dt}|^2\right]\}.
\end{split}
\end{equation}
The relationship (\ref{eq21abc}) is true  for both the variants in superposition
\begin{equation}
\label{eq21bcd}
\vec{H}^{[i]}(\vec{r},t) +  i \vec{H}^{[j]}(\vec{r},t) = \vec{H}^{[ij]}(\vec{r},t), i \neq j, i,j = 1,2.
\end{equation}
Taking into account relationship (\ref{eq6ab}), that is the set $\{q_{\alpha}(t)\}, \alpha \in N$, which satisfy the Maxwell equations   we  will have
\begin{equation}
\label{eq22ab}
\begin{split}
&j_{3}^{2,\pm}(\vec{r},t)= -\frac{8ie}{\hbar c^2V}\sum_{\alpha = 1}^{\infty}m_{\alpha}\omega_{\alpha}^3\sin 2k_{\alpha}z \times\\
&[C_{1\alpha}C^*_{2\alpha}e^{2i\omega_{\alpha}t} + C^*_{1\alpha}C_{2\alpha}e^{-2i\omega_{\alpha}t}],
\end{split}
\end{equation}
the expression for arbitrary set of twice continuously differentiable functions  $\{q_{\alpha}(t)\}, \alpha \in N$, for $j_{4}^{1,\pm}(\vec{r},t)$ is
\begin{equation}
\label{eq23ab}
\begin{split}
&j_{4}^{1,\pm}(\vec{r},t) = -\frac{2e}{\hbar c^2V}\sum_{\alpha = 1}^{\infty}m_{\alpha}\omega_{\alpha}^2\{\sin^2k_{\alpha}z \times\\
&[\frac{dq^*_{\alpha}(t)}{dt}\mp i\frac{dq^{*''}_{\alpha}(t)}{dt}][q_{\alpha}(t) \pm q^{''}_{\alpha}(t)]\\ +
&[\frac{dq_{\alpha}(t)}{dt} \pm i\frac{dq^{''}_{\alpha}(t)}{dt}][- q^*_{\alpha}(t) \pm q^{*''}_{\alpha}(t)]\\
&+ \cos^2k_{\alpha}z [\frac{1}{\omega_{\alpha}}\frac{d^2q^*_{\alpha}(t)}{dt^2} \pm i\omega_{\alpha} q^*_{\alpha}(t)] \times\\
&[\frac{1}{\omega_{\alpha}}\frac{dq_{\alpha}(t)}{dt}
\mp i\omega_{\alpha}\int\limits_{0}^{t}q_{\alpha}(t{'}dt{'}] \\ + &[\frac{-1}{\omega_{\alpha}}\frac{d^2q_{\alpha}(t)}{dt^2} \pm i\omega_{\alpha}q_{\alpha}(t)] \times\\
&[\frac{1}{\omega_{\alpha}}\frac{dq^*_{\alpha}(t)}{dt}
\mp i\omega_{\alpha}\int\limits_{0}^{t}q^*_{\alpha}(t{'})dt{'}]\},
\end{split}
\end{equation}
where $q^{''}_{\alpha}(t) = \omega_{\alpha}^2 \int\limits_{0}^{t} \int\limits_{0}^{t{''}}q_{\alpha}(t')dt'dt{''}$.
It is evident from relationship (\ref{eq23ab}), that in the case  of real-valued sets of twice continuously differentiable functions $\{q_{\alpha}(t)\}, \alpha \in N$, $j_{4}^{1,\pm}(\vec{r},t)$ is equal to zero. For complex-valued functions, determined by (\ref{eq6ab}), we will have
\begin{equation}
\label{eq23abc}
j_{4}^{1,\pm}(\vec{r},t) = \frac{8ie}{\hbar c^2V}\sum_{\alpha = 1}^{\infty}m_{\alpha}\omega_{\alpha}^3 (|C_{1\alpha}|^2 - |C_{2\alpha}|^2).
\end{equation}
It is seen from (\ref{eq23abc}), that $j_{4}^{1,\pm}(\vec{r},t)$ in the  case of Maxwellian EM-field is constant, which is equal to zero at $|C_{1\alpha}| = |C_{2\alpha}|$, that is for all real-valued functions and for complex-valued functions $\{q_{\alpha}(t)\}, \alpha \in N$, which differ each other by arguments of constants $C_{1\alpha}$ and $C_{2\alpha}$. Further, for  the current density $j^{(4)}_{2,\pm}(x)$ we have
\begin{equation}
\label{eq23bcd}
\begin{split}
&j_{4}^{2,\pm}(\vec{r},t) = -\frac{2e}{\hbar c^2V}\sum_{\alpha = 1}^{\infty}\{m_{\alpha}\omega_{\alpha}^2\sin^2k_{\alpha}z \frac{d}{dt}(|q_{\alpha}(t)|^2)  + \\
&\omega_{\alpha}^4\frac{d}{dt}(|\int\limits_{0}^{t} \int\limits_{0}^{t{''}}q_{\alpha}(t')dt'dt{''}|^2) \mp  \frac{d}{dt}[q_{\alpha}(t)\int\limits_{0}^{t} \int\limits_{0}^{t{''}}q^*_{\alpha}(t')dt'dt{''}]\\
& \times i\omega_{\alpha}^2\pm
\frac{d}{dt}[q^*_{\alpha}(t)\int\limits_{0}^{t} \int\limits_{0}^{t{''}}q_{\alpha}(t')dt'dt{''}]i\omega_{\alpha}^2
+ m_{\alpha}\omega_{\alpha}^2\cos^2k_{\alpha}z \\
&\times [\frac{1}{\omega_{\alpha}^2}\frac{d}{dt}(|\frac{dq_{\alpha}(t)}{dt}|^2)                                                                                                \pm i \frac{d}{dt}(\frac{dq^*_{\alpha}(t)}{dt}\int\limits_{0}^{t}q_{\alpha}(t')dt')\\
&+ \omega_{\alpha}^2 \frac{d}{dt}(|\int\limits_{0}^{t}q_{\alpha}(t')dt'|^2) \mp i \frac{d}{dt}(\frac{dq_{\alpha}(t)}{dt}\int\limits_{0}^{t}q^*_{\alpha}(t')dt')]\}.
\end{split}
\end{equation}
For complex-valued functions, determined by (\ref{eq6ab}), we obtain
\begin{equation}
\begin{split}
\label{eq23ccd}
&j_{4}^{2,\pm}(\vec{r},t) = \frac{8ie}{\hbar c^2V}\sum_{\alpha = 1}^{\infty}m_{\alpha}\omega_{\alpha}^3 \cos 2k_{\alpha}z \times\\
&[C_{1\alpha}C^*_{2\alpha}e^{2i\omega_{\alpha}t} - C^*_{1\alpha}C_{2\alpha}e^{-2i\omega_{\alpha}t}].
\end{split}
\end{equation}

It can be shown, that continuity equation
\begin{equation}
\label{eq24ab}
\frac{\partial j^{\pm}_{\mu} (x)}{\partial x_\mu} = 0 
 \end{equation}
is fulfilled  for both general  case and for Maxwellian EM-field functions considered. 
\subsection{\textbf{Quantized Cavity 4-Currents}}

Let us calculate the 4-current densities, which correspond to quantized dually symmetric EM-field, that is to the field, which consist of two components with even and  uneven parities under time reversal or space inversion of both the EM-field vector functions $\vec{E}(\vec{r},t)$ and $\vec{H}(\vec{r},t)$. Let us consider for distinctness the case of two-component EM-field, in which $\vec{E}(\vec{r},t)$- components and $\vec{H}(\vec{r},t)$- components have the same $P$-parity (uneven and even corresondingly) and differ each other by $t$-parity. Given choose corresponds to classical consideration in previous subsection, that allows to compare the results for classical and quantized dually symmetric EM-field. Consequently, we can use the set of  EM-field
vector functions, analogous to (\ref{eq19abcd}), in which the operator functions are set up in conformity to canonical variables. 
 
$\hat{\vec{u}}^{s,\pm}_{\alpha}(x), s = 1, 2$ are
\begin{equation}
\label{eq29abcd}
\begin{split}
&\hat{\vec{u}}^{1,\pm}_{\alpha}(x) = \sqrt{\epsilon_0}A^{E}_\alpha \sin k_\alpha(x_3) [\hat{q}_\alpha(x_4) \pm i \hat{q}^{''}_\alpha(x_4)]\vec{e}_1\\
&\hat{\vec{u}}^{2,\pm}_{\alpha}(x) = \sqrt{\mu_0}A^{H}_\alpha \cos k_\alpha(x_3)\times \\
&[-\hat{q}{'}_\alpha(x_4) \pm i \frac {1}{\omega_\alpha}\frac{d\hat{q}_\alpha(x_4)}{dx_4}]\vec{e}_2,
\end{split}
\end{equation}
where $\vec{e}_1 \equiv \vec{e}_x$, $\vec{e}_2 \equiv \vec{e}_y$,    $\hat{q}{'}_\alpha(x_4)$,  $\hat{q}^{''}_\alpha(x_4)$ are operator functions, which are setting up in the conformity  to classical variables ${q}{'}_\alpha(x_4)$,  ${q}^{''}_\alpha(x_4)$, defined by (\ref{eq11ab}). They are
\begin{equation}
\label{eq41ab}
\begin{split}
&\hat{q}_{\alpha}'(t) = {\omega_{\alpha}}\int\limits _{0}^{t} \hat{q}_{\alpha}(\tau)d\tau\\
&\hat{q}_{\alpha}''(t) = {\omega_{\alpha}}\int\limits _{0}^{t} \hat{q}_{\alpha}'(\tau')d\tau'
\end{split}
\end{equation}
correspondingly.
The functions $\hat{u}^{s,\pm}_{\alpha}(x), s = 1,2, \alpha \in N$ can be built from the components of   
the expansion in Fourier series of quantized dually symmetric EM-field, which consist of two components with even and  uneven parities under time reversal, that is, of    $\hat{\vec{E}}^{[1]}(\vec{r},t), \hat{\vec{E}}^{[2]}(\vec{r},t)$ and
$\hat{\vec{H}}^{[2]}(\vec{r},t), \hat{\vec{H}}^{[1]}(\vec{r},t)$, given by (\ref{eq34}).   Therefore, we have
\begin{equation}
\label{eq42ab}
\begin{split}
&\hat{\vec{u}}^{1,\pm}_{\alpha}(\vec{r},t) = \sqrt{\frac{\hbar \omega_{\alpha}}{V}} \{\left[\hat{a}_{\alpha}(t) + \hat{a}^{+}_{\alpha}(t)\right]\\
& + i \left[\hat{a}{''}_{\alpha}(t) + \hat{a}{''}^{+}_{\alpha}(t)\right]\} \sin(k_{\alpha} z)\} \vec{e}_x,
\end{split}
\end{equation}
\begin{equation}
\label{eq43ab}
\begin{split}
&\hat{\vec{u}}^{2,\pm}_{\alpha}(\vec{r},t) =\sqrt{\frac{\hbar \omega_{\alpha}}{V}}\{ \left[\hat{a}^{}_{\alpha}(t) - \hat{a}^{+}_{\alpha}(t)\right] \\
& + i \left[\hat{a}{''}_{\alpha}(t) - \hat{a}{''}^{+}_{\alpha}(t)\right] \} \cos(k_{\alpha} z) \} \vec{e}_y,
\end{split}
\end{equation}
where superscript $\pm$ means, that in (\ref{eq36ab}) and (\ref{eq37ab}) by definition of complex   EM-field vector function operators $\hat{\vec{E}}(\vec{r},t)$ and $\hat{\vec{H}}(\vec{r},t)$ along with the sign plus, the sign minus   can  be used. We also consider the case of $\hat{\vec{H}}(\vec{r},t)$ formation along with given by (\ref{eq37ab}) (with both the signs in the sum) the following case
\begin{equation}\label{eq44ab}
(\hat{\vec{H}}^{[1]}(\vec{r},t), \hat{\vec{H}}^{[2]}(\vec{r},t)) \rightarrow \hat{\vec{H}}^{[1]}(\vec{r},t) \pm i \hat{\vec{H}}^{[2]}(\vec{r},t) = \hat{\vec{H}}^{\pm}(\vec{r},t).
\end{equation}
It seems to be evident, that 4-current density operator can be determined by the expressions, 
coinciding with classical relations (\ref{eq18ab}), (\ref{eq19ab}), (\ref{eq21ab}), in which  all the physical quantities  are operators.  
For the operator $\hat{j}_{3}^{1,\pm}(\vec{r},t)$ we have
\begin{equation}
\label{eq45ab}
\begin{split}
&\hat{j}_{3}^{1,\pm}(\vec{r},t) = Re\hat{j}_{3}^{\pm}(\vec{r},t) = \frac{ie}{2cV}\sum_{\alpha = 1}^{\infty}k_{\alpha}\omega_{\alpha}\sin{2k_{\alpha}z}\times\\
&\{|i\left[\hat{a}{''}_{\alpha}(t) - \hat{a}{''}^{+}_{\alpha}(t)\right] \mp 
 \left[\hat{a}{}_{\alpha}(t) - \hat{a}{}^{+}_{\alpha}(t)\right]|^2 +\\ &i^2|i\left[\hat{a}{''}_{\alpha}(t) - \hat{a}{''}^{+}_{\alpha}(t)\right] \mp 
 \left[\hat{a}{}_{\alpha}(t) - \hat{a}{}^{+}_{\alpha}(t)\right]|^2\},
\end{split} 
\end{equation}
which is equaled to zero. The same result is obtained in the case of magnetic field operator, determined by (\ref{eq44ab}). For the operator $\hat{j}_{3}^{2,\pm}(\vec{r},t)$ we obtain the relation
\begin{equation}
\label{eq46ab}
\begin{split}
&\hat{j}_{3}^{2,\pm}(\vec{r},t) = Im\hat{j}_{3}^{\pm}(\vec{r},t) = -\frac{2ie}{cV}\sum_{\alpha = 1}^{\infty}k_{\alpha}\omega_{\alpha}\sin{2k_{\alpha}z}\times\\
&\left\{[\hat{a}{}_{\alpha}(t)]^2 + [\hat{a}{}^{+}_{\alpha}(t)]^2 + [\hat{a}{''}_{\alpha}(t)]^2 + [\hat{a}{''}^{+}_{\alpha}(t)]^2\right\}, 
\end{split} 
\end{equation}
which is the same for  magnetic field operator, determined by (\ref{eq44ab}).
Therefore the operator of current density $\hat{j}_{3}^{\pm}(\vec{r},t)$ is independent on sign in expressions for the field operators, based on (\ref{eq34})
and it is independent on  the sequence of $\hat{\vec{H}}^{[i]}(\vec{r},t)$, $i = 1, 2$, in
\begin{equation}
\label{eq47ab}
\hat{\vec{H}}^{ij\pm}(\vec{r},t) = \hat{\vec{H}}^{[i]}(\vec{r},t) \pm i \hat{\vec{H}}^{[j]}(\vec{r},t),
\end{equation}
where $i, j = 1, 2$, $i \neq j$.

It seems to be essential, that EM-field quantization is not binded to Maxwell equations in general case. It means, that the relations (\ref{eq45ab}), (\ref{eq46ab}) are true for more general fields. In the case of
Maxwellian EM-field, using explicit expressions for operator scalar functions given by (\ref{eq30ab}) for 
$\hat{a}{}_{\alpha}(t)$, $\hat{a}{}^{+}_{\alpha}(t)$ and similar relations for $\hat{a}{''}_{\alpha}(t)$, $\hat{a}{''}^{+}_{\alpha}(t)$ the expression (\ref{eq46ab})
has the form
\begin{equation}
\label{eq48ab}
\begin{split}
&\hat{j}_{3}^{2,\pm}(\vec{r},t) = Im\hat{j}_{3}^{\pm}(\vec{r},t) = -\frac{2ie}{cV}\sum_{\alpha = 1}^{\infty}k_{\alpha}\omega_{\alpha}\sin{2k_{\alpha}z}\times\\
&\{[\hat{a}{}_{\alpha}(t = 0)]^2 e^{-i\omega_{\alpha}t} + [\hat{a}{}^{+}_{\alpha}(t = 0)]^2 e^{i\omega_{\alpha}t} + \\ &[\hat{a}{''}_{\alpha}(t = 0)]^2 e^{-i\omega_{\alpha}t} + [\hat{a}{''}^{+}_{\alpha}(t = 0)]^2 e^{i\omega_{\alpha}t}\}. 
\end{split}
\end{equation}
Let us find now the fourth component of 4-vector operator of current density $\hat{j}_{4}^{\pm}(\vec{r},t)$, which determines the charge density. For real part
$\hat{j}_{4}^{1,\pm}(\vec{r},t)$ we have
\begin{equation}
\label{eq49ab}
\begin{split}
&\hat{j}_{4}^{1,\pm}(\vec{r},t) = Re\hat{j}_{4}^{\pm}(\vec{r},t) = \\ &\pm\frac{2e}{c^2V}\sum_{\alpha =1}^{\infty}k_{\alpha}\omega^2_{\alpha} [\{\hat{a}{''}_{\alpha}(t),\hat{a}{}^{+}_{\alpha}(t)\} - \{\hat{a}{}_{\alpha}(t),\hat{a}{''}^{+}_{\alpha}(t)\}],
\end{split}
\end{equation}
where the expressions in braces are anticommutators.
In the case of Maxwellian EM-field there is the connection between $\hat{a}{}_{\alpha}(t)$, $\hat{a}{}^{+}_{\alpha}(t)$ and  $\hat{a}{''}_{\alpha}(t)$, $\hat{a}{''}^{+}_{\alpha}(t)$, since, although they correspond to different particular solutions of Maxwell equations, the solutions are related and the connection between them can be found. It leads to connection between corresponding creation and annihilation operators for two related EM-fields with different $t$-parity.  It can be shown, that the  following relations take place
\begin{equation}
\label{eq50ab}
\begin{split}
&\hat{a}{''}_{\alpha}(t) = \omega^2_{\alpha}\int\limits_{0}^{t}[\int\limits_{0}^{t^{''}}\hat{a}{}_{\alpha}(t')dt']dt^{''}\\
&\hat{a}^{+''}_{\alpha}(t) = \omega^2_{\alpha}\int\limits_{0}^{t}[\int\limits_{0}^{t^{''}}\hat{a}^{+}_{\alpha}(t')dt']dt^{''}
\end{split}
\end{equation}
Then, taking into account the expressions for operator scalar functions, given by (\ref{eq30ab}) for 
$\hat{a}{}_{\alpha}(t)$, $\hat{a}{}^{+}_{\alpha}(t)$ and similar relations for $\hat{a}{''}_{\alpha}(t)$, $\hat{a}{''}^{+}_{\alpha}(t)$, we obtain from (\ref{eq49ab}), that  for Maxwellian EM-field $Re\hat{j}_{4}^{\pm}(\vec{r},t)$ is equal to zero. Therefore we see, that all real part of 4-vector $\hat{j}_{\mu}^{\pm}(\vec{r},t)$ is equal to zero. It corresponds to well known case of nondual single charge electrodynamics. 

For imaginary part $\hat{j}_{4}^{2,\pm}(\vec{r},t)$ we have the relation
\begin{equation}
\label{eq51ab}
\begin{split}
&\hat{j}_{4}^{2,\pm}(\vec{r},t) = Im\hat{j}_{4}^{\pm}(\vec{r},t) = \\ &\frac{2ie}{c^2V}\sum_{\alpha =1}^{\infty}[k_{\alpha}\omega^2_{\alpha} \{[\hat{a}{}^{+}_{\alpha}(t)]^2 - [\hat{a}{}_{\alpha}(t)]^2 + \\ &[\hat{a}{''}^{+}_{\alpha}(t)]^2 - \hat{a}{''}_{\alpha}(t)]^2\}\cos{2k_{\alpha}z} - 2\omega^2_{\alpha}\hat{e}],
\end{split}
\end{equation}
from which the relation for Maxwellian EM-field can be obtained in the manner, analogous to obtaining of expression (\ref{eq48ab}).

Let us verify the implementation of differential conservation law
\begin{equation}
\label{eq52ab}
\frac{\partial \hat{j}^{\pm}_{\mu} (x)}{\partial x_\mu} = 0. 
 \end{equation}

Taking into account (\ref{eq46ab}) and (\ref{eq51ab}) in the case of $Im\hat{j}_{\mu,\pm} (x)$ we have
\begin{equation}
\label{eq53ab}
\begin{split}
&\frac{\partial [Im \hat{j}^{\pm}_{\mu} (x)]}{\partial x_\mu} = \frac{\partial\hat{j}_{3}^{2,\pm}(\vec{r},t)}{\partial x_3} + 
\frac{\partial\hat{j}_{4}^{2,\pm}(\vec{r},t)}{\partial x_4} = \\
&-\frac{4ie}{cV}\sum_{\alpha =1}^{\infty}k^2_{\alpha}\omega_{\alpha}\cos{2k_{\alpha}z} \{[[\hat{a}_{\alpha}(t)]^2 + [\hat{a}^{+}_{\alpha}(t)]^2 + \\
&[\hat{a}{''}_{\alpha}(t)]^2 + [\hat{a}{''}^{+}_{\alpha}(t)]^2] - [[\hat{a}_{\alpha}(t)]^2 + [\hat{a}^{+}_{\alpha}(t)]^2 + \\
&[\hat{a}{''}_{\alpha}(t)]^2 + [\hat{a}{''}^{+}_{\alpha}(t)]^2]\} = 0.
\end{split} 
\end{equation}
Here  the commutation relations 
\begin{equation}
\label{eq54ab}
[\hat{a}_{\alpha}(t), \hat{a}^{+}_{\alpha}(t)] = \hat{e}, \alpha \in N, \end{equation} and the  relations
\begin{equation}
\label{eq55ab}
\frac{d\hat{a}_{\alpha}(t)}{dt} = \frac{1}{i\hbar}[\hat{a}_{\alpha}(t), \hat{\mathcal{H}}_{\alpha}(t)], \alpha \in N,
\end{equation}
where $\hat{\mathcal{H}}_{\alpha}(t)$ is the Hamiltonian, corresponding to cavity $\alpha$-mode, were used. The Hamiltonian $\hat{\mathcal{H}}_{\alpha}(t)$ is given by relation
\begin{equation}
\label{eq56ab}
 \hat{\mathcal{H}}_{\alpha}(t) = \hbar \omega_{\alpha}\left[\hat{a}{}^{+}_{\alpha}(t)\hat{a}{}_{\alpha}(t) + \frac{1}{2}\right]
\end{equation}
 For calculation of derivatives of operators $\hat{a}^{+}_{\alpha}(t)$, $\hat{a}{''}_{\alpha}(t)$, $\hat{a}{''}^{+}_{\alpha}(t)$ the relations,
analogous to (\ref{eq55ab}), were used.
Taking into account (\ref{eq49ab}), (\ref{eq54ab}), (\ref{eq55ab}), (\ref{eq56ab}) in the case of $Re\hat{j}_{\mu,\pm} (x)$ we have
\begin{equation}
\label{eq57ab}
\begin{split}
&\frac{\partial [Re \hat{j}^{\pm}_{\mu} (x)]}{\partial x_\mu} =  
\frac{\partial\hat{j}_{4}^{1,\pm}(\vec{r},t)}{\partial x_4} = \\
&\pm\frac{2e}{c^3V}\sum_{\alpha =1}^{\infty}\omega^3_{\alpha} 
\{[\hat{a}^{+}_{\alpha}(t), \hat{a}{''}_{\alpha}(t)] + [\hat{a}{''}_{\alpha}(t), \hat{a}^{+}_{\alpha}(t)] - \\
&[\hat{a}{''}^{+}_{\alpha}(t), \hat{a}_{\alpha}(t)] - [\hat{a}_{\alpha}(t), \hat{a}{''}^{+}_{\alpha}(t)]\} = 0. 
\end{split} 
\end{equation}
Therefore differential conservation law, given by (\ref{eq52ab}), is fulfilled both for Maxwellian EM-field and in general case. It is seen also, that direct calculation gives really nonzero imaginary part for current densities  for free cavity EM-field. It  can be considered to be direct confirmation of above obtained conclusion on the existence of imaginary component of charge in free EM-field. It gives the possibility to define the set of EM-field functions in general case, that is for  EM-field with and without sources. EM-field functions can be defined to be the components of 4-current density, or in renormalized form, if to divide each component into complex conductivity of the medium $\lambda$. It means in its turn, that the propagation of  EM-field in vacuum, that is free EM-field is also characterized by the value of vacuum conductivity $\lambda_v$, like to dielectric $\epsilon_0$ and magnetic $\mu_0$ vacum permittivities. Therefore, instead unobservable vector and scalar potentials, EM-field can be characterized by, for instance, electric field 4-vector-function with the components $E_\alpha(\vec{r},t) = \{E_x(\vec{r},t), E_y(\vec{r},t), E_z(\vec{r},t), i \frac{c \rho_e(\vec{r},t)}{\lambda}\}$, where  $i {c \rho_e}(\vec{r},t)$ is the $j_4(\vec{r},t)$-component of 4-current density, corresponding to contribution of electric component of EM-field. For the case of EM-field propagation in vacuum $\lambda = \lambda_v$. Alternative characterization by means of magnetic field 4-vector-function $H_\mu(\vec{r},t) = \{H_x(\vec{r},t), H_y(\vec{r},t), H_z(\vec{r},t), i \frac{c rho_m(\vec{r},t)}{\lambda}\}$ seems to be equivalent for free EM-field in vacuum, if to take into account, that for Maxwellian  free EM-field the components of electric field 3-vector-functions and magnetic field 3-vector-functions are bounded up between themselves by Cauchy-Riemann analicity condition.  Here  $i {c \rho_m}(\vec{r},t)$ is the $j_4(\vec{r},t)$-component of 4-current density, corresponding to contribution of magnetic component of EM-field. However the characterization by means of the only single 4-vector-function, for instance by  $H_\mu(\vec{r},t)$ becomes to be nonequivalent in general case, in particular, in the case of single-charge ED, for which $\rho_m(\vec{r},t) = 0$. It means, that in general case both 4-vector-functions $E_\alpha(\vec{r},t)$ and $H_\mu(\vec{r},t)$ have to be used.

 It is remarcable, that the notion of characteristic medium resistance, including characteristic vacuum resistance $Z_0 = \sqrt{\frac{\mu_0}{\epsilon_0}} = 120 \pi$ Ohm is widely used in technique, connected with EM-wave propagation, in particular, in radiospectroscopy technique \cite{Poole}. Therefore, $\lambda_v = \frac{1} {Z_0} = \frac{1} {120\pi} Ohm^{-1}$.

The proof of the existence  for EM-field along with vector force characteristics $\vec{E}(\vec{r},t)$-field strength vector-function and $\vec{H}(\vec{r},t)$-field strength vector-function the additional scalar force characteristic - charge, allows  to be nearing to understanding of, in Dirac characterization, "strange pequliarity of light quantum" \cite{P.Dirac}, consisting in that, that according to Dirac,  light quantum, apparently, discontinues its existence, when it is in one of its stationary states - in zero state - in which its impulse and  energy are equal to zero.  Dirac consideres the absorption process to be jump of  light quantum in zero state and emission process to be jump from given state in the state, where its  existence is physically evident, so it seems that it was recreated. From the absence of restriction on number of light quanta, which can be emerged by given way, Dirac suggested \cite{P.Dirac}, that there is infinitely many of light quanta in zero state, that is, in vacuum state in modern terminology.  

\section{Spin-Charge Separation in Quantized EM-Field Structure}

The conclusion on quaternion structure  of EM-field and the proof of existence of the charge, being to be inherent characteristic of EM-field allow to understand the nature of photons and to explain their two kinds' behavior. On the one hand the photons seem to be charge neutral particles, which is confirmed
 in a number of  observations, including astronomic observations, and experiments on light absorption, transmission, reflection, Rayleigh and Raman scatterings and so on. On the other hand the photons seem to be charged particles, which is confirmed, for example, by observations of zigzag-like light propagation between adjacent rainclouds during storm, since zigzag-like light propagation in the same conditions (that is by the absence of lightning  between the same rainclouds does not takes place) and it is characteristic for charge particles, which is also long ago was confirmed.  It allows to suggest, that EM-field is characterized by spin-charge separation effect, leading to appearance of the photons of two kinds. We have reviewed in details the known mechanisms of spin-charge separation to choose the  mechanism, which seems to be appropriate for description of given effect in EM-field (see the next subsection).

\subsection{Review of Main Mechanisms of Spin-Charge Separation}

\subsubsection{Continuum Limit of 1D Electron Gas Theory and Su, Schrieffer, and Heeger Model of Organic Conductors}

The idea of spin-charge separation
was  explicitly treated  for the first time by Luther and Emery \cite{Luther}
in the context of a continuum limit of the $1D$ electron gas theory, that is by means of field theory.  They have shown, that the Hamiltonian $\hat{H}_{1DEG}$ of the 1D electron gas can
be represented  in the form of
\begin{equation} 
\label{eq60} 
\hat{H}_{1DEG} = \hat{H}_c[\phi_c] + \hat{H}_s[\phi_s] + \hat{H}_{irr}[\phi_c, \phi_s], 
\end{equation} 
where $\hat{H}_c[\phi_c]$ and  $\hat{H}_s[\phi_s]$ are, respectively, the Hamiltonians,
which govern the dynamics of the spin and charge fields,
$\phi_c$ and $\phi_s$, respectively, and $\hat{H}_{irr}[\phi_c, \phi_s]$  (which they did not
treat, explicitly) consists of terms that can be neglected
in the long wave-length limit (irrelevant terms, in the
renormalization group sense). Consequently, concerning the
low energy physics, the spin and charge dynamics become
 completely decoupled from each other. Moreover,
Luther and Emery showed, that any space-time electronic
correlation function can be expressed in the form of the product
of correlators involving only the spin-fields and  only the charge fields correspondingly. In \cite{S_Kivelson} is accentuated the significance of field theory methods at all, since  for the solution of similar tasks they are sufficiently powerful, and  
the work \cite{Luther} is characterised thereupon to be profound and that its significance  has
continued to grow. 

The related model, which describes spin-charge separation is the model of the formation of  solitons
with fractional fermion number.  General idea belongs to  Jackiw and
Rebbi \cite{Jackiw}. They pay attention to the field theories, especially
in one spatial dimension, which lead solitons' formation
with fractional fermion number. However the concrete realization of  given idea in condensed matter physics belongs to  Su, Schrieffer, and Heeger, realized some years later, when with the discovery
of the conducting polymer, trans-polyacetylene (t-PA), by Shirakawa,
MacDiarmid, and Heeger the appropriate subject
really got started. The  model, proposed by Su, Schrieffer, and Heeger (SSH-model) with spin-charge separation
 to be basis phenomenon \cite{SSH}, \cite{SSH_PRB} is the model of conjugated organic 1D-conductors, and it was presented on t-PA example.

 Su, Schrieffer, and Heeger 
 \cite{SSH}, \cite{SSH_PRB} and also Brazovskii \cite{Brazovskii} and Rice \cite{Rice} discovered, that occuring of topological solitons takes place. SSH-model is more general in comparison with the model, proposed by Brazovskii, which can be considered  to be a continuum version
of the SSH-model. The Rice-model \cite{Rice} is also similar to the SSH-model, however the results were obtained by Rice from a somewhat more phenomenological viewpoint. The solitons in cited models
are precisely analogous to those ones of Jackiw and Rebbi.  They  represent themselves
 the  electronic excitations in the model of a set of
noninteracting between themselves electrons coupled to a lattice deformation
(acoustic phonon system), that is, in the
simplest imaginable model, which in fact represents itself the SSH-model. Here, however, the solitons have reversed
charge-spin relations - they can be neutral with
spin 1/2, or if spinless they have charge $\pm e$.
Specifically, what Su, Schrieffer, and Heeger showed, is that when an electron
is added to an  neutral t-PA chain, it
can break up into two pieces, one of which carries the
electron’s charge and the other its spin. Given result
bears a clear family relation with the phenomenon
of spin-charge separation in the 1D electron gas theory of Luther and Emery \cite{Luther},
but it is not identical, how it was remarked in \cite{S_Kivelson}. In the first place, undoped t-PA is a semiconductor, with a moderately large
$(2\Delta\approx 2 eV)$ forbidden gap, and so is nowhere near being quantum
critical. The solitons in  t-PA are not low or zero energy
excitations, the soliton creation energy is approximately \cite{Takayama} $E_s = 2\Delta/\pi$. It leads, for instance, to substantial
attractive interactions between two pieces of electron, that is,  both between the charged and neutral
solitons, which in its turn lead to formation  a bound-state - a polaron \cite{Su}, \cite{Campbell}, \cite{Brasovskii_Kirova}
with a binding energy $E_p = 2E_s  - 2/\pi(2 - \sqrt{2})\Delta$. Given  result indicates, that spin-charge separation does not
occur in the SSH-model in the same precise sense of the 1D electron gas theory. In
other words, spin charge coupling implies in given case, that the lowest
energy excitation of the system with the quantum
numbers of an electron is a quasi-particle with the same
quantum numbers. Then the fractionalization, that is spin-charge separation, 
is, in fact, a high energy feature of the spectrum
of the SSH-model. The lowest energy excitation made by
adding two electrons to the system is a pair of charged
solitons.  Given effect can be understood  qualitatively
from the Hamiltonian in (\ref{eq60}), if to take into account, that under circumstances
in which there is a gap to both spin and charge excitations,
the renormalization group flows carry the system
to a strong coupling fixed point, where the terms in $\hat{H}_{irr}$
are  longer not irrelevant, and in particular can lead to a
short-range attraction between spin and charge solitons. 

It has to be remarked, that  t-PA is a very complicated material, from a
physicist's viewpoint. It is rather disordered, even when
undoped, and doping introduces all sorts of additional
levels of randomness. It is moderately one-dimensional,
with an in-chain bandwidth of order $W \approx 10 eV$ and
an interchain bandwidth of order one or two
tenths of an $eV$. It is not, however, sufficient to permit
truly one-dimensional physics to be manifest at very
long distances, and clearly leads to rather strong soliton
confinement. Nevertheless, many rather spectacular predictions
of the soliton theory, both spectroscopic and dynamical,
were confirmed by experiments performed with
great vigor, determination, and creativity by a large community
of scientists. It means, that the basic features of the soliton model of
t-PA are not only right, in theory, but applicable
to a range of experiments, see review \cite{Heeger_Kivelson}.
However, the real significance of the soliton model of t-PA
 is that it introduced a new paradigm into
the field. It has led to ideas which have  a lasting impact
on condensed matter physics \cite{S_Kivelson}:

 a) That there can exist quasi-particles with fractional
quantum numbers (that is quantum numbers
that are unrelated to those of an integer number
of electrons and holes) that are robust entities, not
just in low energy assymptopia, but in the realworld
realm of materials physics,

 b) That quasi-particles have a topological
character, from which their stability derives,

 c) That fractional quantum numbers are
sharp quantum observables, \cite{Kivelson}, \cite{Bell}, \cite{Jackiw_R}, \cite{Rajaraman}. 

While the character of the
solitons, and the separation of charge and spin are direct
consequences of the broken symmetry (dimerized)
ground state, there are many aspects of the SSH-solution
that are more microscopic, and model dependent. 

In particular, there was a remarkable quantitative prediction
made by Su, Schrieffer, and Heeger, concerning the magnitude of the
dimerization in  t-PA \cite{SSH}, \cite{SSH_PRB}. At the time of
their first work on the subject, the in-chain bandwidth,
$W$, was known within $20$ percent from various quantum chemical
and band-structure calculations \cite{S_Kivelson}. Similarly, the spring
constant $(K \approx 20 eV/{\AA}^2)$ of the $\sigma$- bonds was well known
to depend only on local chemistry, and so was known reliably.
Thus, the only free parameter in the model was
the electron-phonon coupling constant $\alpha$. Su, Schrieffer, and Heeger determined
the value of coupling constant $\alpha$ empirically, by fitting
to the observed optical absorption gap, and were then
able to predict the magnitude of the lattice dimerization,
$u = 0.04 \AA$, on the fitting basis. Quantitative
measurements \cite{Fincher} found $u = 0.03 \AA \pm 0.01 \AA$, thus confirming
the prediction.

There are several additional  results about the
triumph of the SSH-theory, which seem to be also remarkable. They are described in \cite{S_Kivelson} and in \cite{Yearchuck_PL}, \cite{Dovlatova_Yearchuck}. The first  result is the following. Since t-PA is a one
dimensional system, so one might think that mean-field
theory, which was employed, is unreliable. The neglect
of electron-electron interactions seems, at first sight, to
be incorrect, equally, both because these interactions are
known, from quantum chemical studies of small polyenes,
to be strong, and interactions in the
1D electron gas are known to completely destroy free Fermi-gas behavior. However SSH-model is in fact two-component Fermi-gas model, consisting of free electron Fermi-gas component, that is noninteracting between themselves quasiparticles  and the second component - subsystem of also noninteracting between themselves quasiparticles - phonons. It is remarcable, that the interaction between both the components is taken into account in SSH-model. Then it becomes qualitatively to be understandable, that the effect of
electron-electron interactions can be  renormalized by the bringing the effective electron-phonon coupling in, which a priori, has to lead to an  enhancement of electron-phonon coupling constant $\alpha$, that is, the genuine  electron-phonon coupling constant has to be replaced by
the effective electron-phonon coupling constant. The possibility to bring the effective electron-phonon coupling in was proved in \cite{Zimanyi}, \cite{Voit}. What authors of \cite{Zimanyi} and, at
about the same time, J.Voit \cite{Voit} did was to treat
the problem of the 1D electron gas with both electron-electron (instantaneous)
repulsions and electron-phonon induced (retarded)
attractions using standard weak-coupling (one
loop) renormalization group methods. It is argued in given works and in \cite{S_Kivelson} 
 that, even if at the microscopic level the electron-electron
interactions are much stronger than the electron-phonon
interactions, the effective low energy theory is always
dominated by the electron-phonon interactions, since $ W/\hbar\omega_0$ and $W/\Delta$ are sufficiently
large. The
electron-electron interactions have really in given case the effect of renormalizing
the  electron-phonon coupling and it has actually been found, that 
they lead to a strong enhancement of effective value of $\alpha$. Moreover, it was shown, that the renormalizing effect is most pronounced, when
the electron density is commensurate, or nearly so, that is
when there is roughly one electron per site.
 So the remarkable
physical insight of Su, Schrieffer, and Heeger  explains why the empirically determined value of $\alpha$ is large compared
to those found in microscopic, quantum chemical calculations in a natural way -
the empirical $\alpha$ is a renormalized effective coupling. It also
explains the remarkable fact, that when t-PA is
"overdoped", that is, when the electron concentration deviates
by more than about 6 percents from one $\pi$-electron per
carbon atom, it behaves like a nearly non-interacting metal \cite{Kivelson_Heeger}. It is consequence of the strong doping dependence of the effective
electron-phonon coupling, in result the expected Peierls instability
of given quasi-one-dimensional metal is suppressed \cite{Kivelson_Salkola}
to immeasurably small temperatures. 

On the other hand, the
fact that the phonon frequencies, $\hbar\omega_0$, are small on electronic
scales,  $\hbar\omega_0/ \Delta \ll 10^{-1}$, can  be shown to
justify the mean-field theory - indeed, effects of quantum
lattice fluctuations were systematically studied and found
to be quite mild  \cite{Hirsch}, \cite{Fradkin}. The weak-coupling theory applicability
 is guaranteed by the fact that the gap
is small compared to the bandwidth, $\Delta/W \ll 10^{-1}$.
  
The second  result is the following.
The most substantial suggestion in SSH-model is the suggestion, that the only dimerization coordinate $u_n$ of the $n$-th $CH$-group, 
$n = \overline{1,N}$ along chain molecular-symmetry axis $x$ is substantial for determination of main physical properties of the material. The other five degrees of freedom, that is the degrees of freedom, which are relevant to the bonds with the directions not coinciding  with chain molecular-symmetry axis direction, were not taken into consideration. Nevertheless, the model has obtained the magnificent experimental confirmation. It was explained in \cite{Dovlatova_Yearchuck}, that given success is the consequence of some general principle.   Given general principle is really exists and main idea was proposed by Slater at the earliest stage of quantum physics era already in 1924. It is - "Any atom may in fact be supposed to communicate with other atoms all the time it is in stationary state, by means of virtual field of radiation, originating from oscillators having the frequencies of possible quantum transitions..." \cite{Slater}. Given idea will obtain its development, if to clarify  the origin  of virtual field of radiation. It is shown above, that Coulomb field in 1D-systems or 2D-systems can be quantized, that is,  it has the character of radiation field and it can exist without the sources, which have created given field. Consequently, Slater principle can be applied to t-PA. In  t-PA Coulomb field can be considered to be "virtual" field with propagation direction the only along  t-PA chain.  In other words it produces preferential direction in atom  communication the only in one direction (to be consequence of quasionedimensionality), and given direction remains to be preferential by interaction with external EM-field. It explains qualitatively the success of SSH-model in the sense, that degrees of freedom, realized by bonds, which are not coinciding  with chain molecular-symmetry axis direction, can really be not taken into consideration for experiments with the participation of external EM-field and indicating thereby on deep physical insight of Su, Schrieffer, and Heeger in the field.

The third additional  result is the following.
SSH-model along with the physical basis of the existence of solitons, polarons, breathers, formed in $\pi$-electronic subsystem ($\pi$-solitons, $\pi$-polarons, $\pi$-breathers) contains in implicit form  also the basis for the existence of similar quasiparticles in  $\sigma$-electronic subsystem, that is  SSH-model can be developed. It was done in \cite{Yearchuck_PL}, \cite{Dovlatova_Yearchuck}.  The origin of quasiparticles' formation  in  $\sigma$-electronic subsystem is the same two-fold degeneration of ground state  of the whole electronic system, energy of which in ground state has the form of Coleman-Weinberg potential with two minima  at the values of dimerization coordinate $u_0$ and $-u_0$. Really, the appearance of $u_0 \neq 0$ and $-u_0 \neq 0$ indicates on the alternation in interatomic distance. It means, that simultaneously with $\pi$-subsystem, $\sigma$-subsystem will also be  dimerized. Experimental confirmation to given
conclusion is the detection of  SSH-$\sigma$-polarons  in  carbynes \cite{Yearchuck_PL}, at that   the formation of $\sigma$-polaron lattice (PL), which is antiferroelectrically ordered,  has been found.

\subsubsection{Anderson Mechanism of Spin-Charge Separation}

Another way of looking on spin-charge separation phenomenon was proposed by Anderson \cite{Anderson} in 1987, that is substantionally later in comparison with the date of proposal of spin-charge field separation mechanism, made by Luther and Emery, and  with the date of the emergence of the idea of soliton spin-charge  separation mechanism, offered by Jackiw and
Rebbi, which became  a reality in SSH-model. 
To characterize the doped Mott-Hubbard insulator in the metallic regime, two significant ideas were originally
introduced by Anderson - spin-charge separation effect \cite{Anderson}, \cite{Anderson_P_W}, \cite{Anderson_Ren}, which has another origin in comparison with spin-charge separation phenomenon, discussed in  \cite{Jackiw}, \cite{SSH}, \cite{SSH_PRB} and unrenormalizable phase shift effect \cite{Anderson_P}, \cite{Anderson_Cuprates}. 
  Spin-charge separation in Anderson approach means the existence of two independent elementary excitations,
charge-neutral spinon and spinless holon, which carry spin $1/2$ and charge $+e$, respectively. In fact it represents itself the specific realization and generalization for the case of quantum liquids (instead of quantum gases) of general field theory results of the work \cite{Luther} by using of concrete physical models.
  So Anderson spin-charge separation effect  may be mathematically realized in
the so-called slave-particle representation  \cite{Zou_Anderson} of the $t-J$ model
$e_{i\sigma} = h^+_i f_{i\sigma}$, 
where $h^+_i$, $f_{i\sigma}$ are holon  and spinon fields, which can be considered to be analogues of  of the spin and charge fields,
$\phi_c$ and $\phi_s$ in Luther-Emery field model. The occupancy constraint, reflecting the Hubbard gap in its extreme limit, is handled by an equality
 $h^+_i h_i + \sum_{(\sigma)}{f^+_{i\sigma}f_{i\sigma}} = 1$ 
 which commutes with the Hamiltonian. It is seen the close relation of the spin-charge
separation and the constraint condition through the counting of the quantum numbers. But the spin-charge separation  acquires a new meaning here. If those holon 
 and spinon  fields indeed describe elementary excitations,
the hole (electron) is no longer a stable object and must decay into a holon-spinon pair once being injected into the
system. This instability of a hole (electron), being to be free quasiparticles  in solid state physics theory (effective mass method) is referred in the literature, see, for instance \cite{Weng} to be the deconfinement, in
order to distinguish it from the narrow meaning of the Anderson mechanism of spin-charge separation about elementary excitations.  It seems to be be   more strictly to say that injected electron is collectivized (in given process its individual properties are changed). We ought to bear in mind, that decay process is not the procees of genuine decay of  electrons like to nuclear decay. It is the product of model replacement. Since electrons and holes in solid state physics are quasiparticles, that is, they are elementary excitations in effective mass method, when strongly interacting electron sybsysten is replaced by nonintercting gas of given quasiparticles, which have quite other mass in comparison with genuine particles - electrons and positrons. Spinons  and holons are in fact the product of interactions between electrons; or between holes, or mixed electron-hole interaction in the models beyond effective mass model. It is interesting, that the simplest example of spin-charge separation is the formation in result of Coulomb interaction between above indicated quasiparticles - electrons and holes - of usual triplet  excitons, that is the quasiparticles with zero charge and spin $S = 1$. It indicates by the way, that spin-charge separation effect has very long history, starting with prediction by Frenkel of excitons already in 1931. Given simple example clearly indicates, that spin-charge separation effect can not mean the decay into two pieces of electrons in direct sense of word - the replacement by the other collective excitations takes place. However formally the behavior of the system is like to that, quasi the electrons are really decays into two pieces. It 
is especially understandable, if to compare with exciton model, compound element of which - hole - is the absence of electron in valent bond, while the behaviour of given "empty place" in solids is like to some extent to genuine positively charged particles. Different velocity of propagation of spin excitations and charge excitations in the system of strongly interacting electrons is like to some extent to, for example, different velocities of sound and light signals, which accompany the moving aircraft, which especially clearly seen, if the aircraft moves with supersonic velocity. Therefore we come to conclusion, that electrons in the substances with strongly correlated electron system  produce in the objects of solid state physics the lattice and different velocity of propagation of spin and charge excitations along given lattice is like to,  for instance, different propagation velocity of optical and acustic phonons in atomic lattice, that is, Anderson mechanism of spin-charge separation effect is explained qualitatively in a natural way. 

The second Andersons' idea, the so-called "unrenormalizable phase shift", in its applicability to Mott-Hubbard model may be described in the following way. In the presence of
an upper Hubbard band, adding a hole to the lower Hubbard band on-site Coulomb interaction can change the whole Hilbert space. The entire spectrum $\{\vec{k}\}$ of momenta may be shifted through the phase shift effect.
It leads to the orthogonality of a bare doped hole state with its true ground state, which in  its turn leads to equality to zero of  the quasiparticle weight Z, that is  
$Z \equiv 0$. It is the key criterion for a non-Fermi liquid.  In general, it implies
$c_{j\sigma} = e_{j\sigma}\exp{i\theta_{j}}$, 
where $e_{j\sigma}$ is related to elementary excitation fields, for example, $ h^+_j f_{j\sigma}$
  in a spin-charge separation framework. Given
expression means, that in order for a bare hole created by $c_{j\sigma}$ to become low-lying elementary excitations, a many-body
phase shift $\theta_{j}$ must take place in the background. In momentum space the presence of phase shift
changes the Hilbert space (by shifting $\vec{k}$  values). At that $e_{\vec{k},\sigma} = \sum_{(\vec{k}')} h^+_{\vec{k}'} f_{\vec{k}+\vec{k}',\sigma}$,
 where $\vec{k}$ and $\vec{k}'$ belong to the same
set of quantized values. For example, in a 2D square sample with size $L \times L$, the momentum is quantized at $k_q = \frac {2\pi}{L} n$
under the periodic boundary condition with $q = x$, $y$ and $n \in N$. But because of a nontrivial set of phase values $\{\theta_{j}\}$, $c_{\vec{k},\sigma}$  and $e_{\vec{k},\sigma}$
generally cannot be described by the same momentum set  $\{\vec{k}\}$, or, in the same Hilbert space, which thus constitutes an
essential basis for a possible non-Fermi liquid.
The one-dimensional  Hubbard model serves a marvelous example in favor of the decomposition $c_{j\sigma} = e_{j\sigma}\exp{i\theta_{j}}$ over $e_{i\sigma} = h^+_i f_{i\sigma}$. The equality of quasiparticle weight
$Z$ to zero   for a non-Fermi liquid
 becomes to be understandable, if take into account the definition of value $Z$. The quasiparticle weight $Z_{\vec{k}}$ at momentum $\vec{k}$ is given by $Z_{\vec{k}}$ = $|\langle\Psi_G(N_e -1)|c_{\vec{k},\sigma}| \Psi_G(N_e)\rangle|^2$, and
it measures the overlap of a bare hole state at momentum $\vec{k}$, created by operator $c_{\vec{k},\sigma}$ in the ground state $\Psi_G(N_e)$ of $N_e$ electrons, with
the ground state $\Psi_G(N_e -1)$ of $(N_e -1)$ electrons. For a Fermi liquid state, one always has $Z_{\vec{k}_F}$ $\neq 0$ at the Fermi momentum  $\vec{k}_F$.
If $Z_{\vec{k}}$ = $0$ for any $\vec{k}$, then the system is a non-Fermi liquid by definition.

It seems to be interesting, that there are  qualitative differences by description of spin-charge separation effects in 1D systems between SSH-model and in the 1D models,  in which 1D correlated electronic systems are properly
described by the  Luttinger liquid
theory, for which Anderson  mechanism of spin-charge separation takes place. For instance, phonon effects on spin-charge separation in one dimension, realized by Anderson  mechanism, were studied in  \cite{Ning} through the calculation
of one-electron spectral functions in terms of the  cluster perturbation theory
together with an optimized phonon approach. The 1D Holstein-Hubbard model,
which is the simplest model involving both electron-phonon
 and electron-electron  interactions, has
been used. It was found, that the retardation effect, which is the consequence  of the
finiteness of phonon frequency, suppresses the spin-charge separation and eventually makes it invisible
in the spectral function. At the same time  electron-phonon interaction plays the essential role for spin-charge separation presence in SSH-model. 

We are interesting in the studies of the systems with strong 
electron-phonon interaction. Consequently it represents the interest in theoretical study of the possibility of spin-charge separation in 1D systems with strong electron-phonon interaction. Since, according to results of the work  \cite{Ning} phonon effects lead in Anderson model of spin-charge separation and like them, in which 1D correlated electronic systems are properly
described by the  Luttinger liquid,  to vanishing of spin-charge separation effect, we will consider  SSH-model to be basis model for given study. At the same time there is in existing variant of  SSH-model an upper limit on the value of electron-phonon coupling constant. It is consequence of the treatment of electron-phonon coupling to be the linear term in expansion of a hopping integral of tight-binding model about the undimerized state. Given restriction was discussed in \cite{Rice_M} and the maximum allowed value of electron-phonon coupling constant $\alpha$  was evaluated to be $ \approx 1.27$. We will show, that given restriction can be lifted in developed variant of SSH-model.

\subsection{Fermi-Liquid Model of Spin-Charge Separation in Quantized EM-Field Structure}

It seems to be substantial the result, obtained by Dirac, that dynamical system, which consists of the ensemble of identical bosons is equivalent to dynamical system, which consists of the ensemble of oscillators. In other words quantized EM-field can be represented instead oscillator system by equivalent many-particle
system of bosons, interacting with each other.  It seems to be evident, that each single boson will possess by spin with S = 1. The simplest analogue in the physics of condensed matter of the system of interacting S = 1 bosons is carbon. So we come to the model of linearly polarized EM-field to be the chain of bosons, which is like in its mathematical description to the chain of carbon atoms in trans-polyacetylene (t-PA), at that both in "atomic" and "electronic" structure. One-dimensionality of the task can be argued in the following way. It is shown above, that for description of EM-field instead unobservable vector and scalar potentials the 4-vector of electrical and/or magnetic field strength can be used. Consequently, to describe linearly polarized EM-field in Euclidian space $R_3$ it is sufficient to specify the propagation direction, that is vector $\vec k$  and to define $\vec E$. Given vectors determine the plane, in which a  frame of reference with $z$-axis along  the propagation direction and orthogonal to it $x$-axis can be setted. Taking into account the homogeneity of Minkowski space $R_4$ and homogeneity of free EM-field in it, free EM-field can be modelled by the set of noninteracting (or weak interacting) between themselves "boson-atomic" chains, similarly to many carbon-based, that is, also spin-1 boson-based, polymer structures with equidistant interatomic spacing (by the degeneracy absence) and located along propagation direction. What concerned the "atomic" structure, we have to include the contribution of vacuum fluctuations, which presents in oscillator task and which are absent in the case of boson set \cite{Scully}. The presence of charge, being to be scalar characteristic of 
EM-field gives the possibility to  model "electronic" structure of equivalent boson chain like to t-PA  electronic structure, that is consisting of "$\sigma$-subsystem" and "$\pi$-subsystem". It becomes to be understandable, if to take into account, that charge space distribution is directly connected with  $\vec E$ space distribution. In other words the presence of $E_z$-component will determine the appearance of EM-field charge "$\sigma$-subsystem", while $E_x$-component will determine the appearance of EM-field charge "$\pi$-subsystem", at that, like to  t-PA, its distribution in space $R_3$  will be twice degenerated. Given conclusion can be argued on the basis above established quaternion structure of EM-field in the following way. $E_x$-polar component by EM-field propagation every other half-period alters its sign, at the same time $E_x$-axial component does not alters its sign, which is equivalent to appearance of alternating single-double interbosonic  "$\pi$-bonds" in EM-field charge "$\pi$-subsystem", at that two configuration - single-double and double-single are topologically equivalent. Consequently we come to conclusion, that the interaction between equivalent to oscilators "bosonic atoms" can be described in the frames of Fermi gas model in zero-th order approximation or in the frames of Fermi liquid model in the first order approximation. Mathematical description in zero-th order approximation will be similar to well known SSH-model with some corrections, concerning two branch of quasiparticles, given in \cite{Yearchuck_Dovlatova}. Fermi  liquid model  will represent the generalization of  SSH-model and will be further considered in details  in applicability to EM-field description. 

We will start from Hamiltonian
\begin{equation}
\label{Eq1m}
\hat{\mathcal{H}}(u) = \hat{\mathcal{H}}_{0}(u) + \hat{\mathcal{H}}_{\pi,t}(u) + \hat{\mathcal{H}}_{\pi,u}(u).
\end{equation}
Like to works  \cite{SSH}, \cite{SSH_PRB} we will consider Born-Oppenheimer approximation.
 The first term in 
(\ref{Eq1m}) is 
\begin{equation}
\label{Eq2m}
\begin{split}
\hat{\mathcal{H}}_{0}(u) = \sum_{m}\sum_{s}(\frac{\hat{P}_m^2}{2M^*}\hat{a}^+_{m,s} \hat{a}_{m,s} + K u_m^2 \hat{a}^+_{m,s} \hat{a}_{m,s})
\end{split}
\end{equation}
is the  operator of kinetic energy of "boson atom" motion (the first term), and the  operator of the $\sigma$-bonding energy
{the second term}, $K$ is effective $\sigma$-bonds spring constant, $M^*$ is effective  mass  of "boson atom", $u_m$ is configuration coordinate for $m$-th "boson atom",  which corresponds to translation of $m$-th "boson atom"  along the symmetry axis $z$ of the chain, $m = \overline{1,N}$, $N$ is number of "boson atom"  in the chain, $\hat{P}_m$ is operator of impulse, conjugated to configuration coordinate $u_m$, $m = \overline{1,N}$, $\hat{a}^+_{m,s}$, $\hat{a}_{m,s}$ are creation and annihilation operators of creation or annihilation of quasipartile with spin projection $s$ on the $m$-th chain site in  "$\sigma$-subsystem". 

The second term in 
(\ref{Eq1m}) is 
\begin{equation}\begin{split}
\label{Eq3m}
\hat{\mathcal{H}}_{\pi,t}(u) = \sum_{m}\sum_{s}[t_0(\hat{c}^+_{m+1,s} \hat{c}_{m,s} + (\hat{c}^+_{m,s} \hat{c}_{m+1,s})].
\end{split}
\end{equation}
It is the resonance interaction (hopping interaction in tight-binding model approximation) of
 quasiparticles in "$\pi$-subsystem" of all charge system, which is cosidered to be Fermi liquid, and
in which the only constant term in Taylor series expansion of resonance integral about the dimerized state is taking into account. Here $\hat{a}^+_{m,s}$, $\hat{a}_{m,s}$ are creation and annihilation operators of creation or annihilation of quasipartile with spin projection $s$ on the $m$-th chain site in  "$\pi$-subsystem".

The third term in 
(\ref{Eq1m}) is 
\begin{equation}\begin{split}
\label{Eq4m}
&\hat{\mathcal{H}}_{\pi,u}(u) = \sum_{m}\sum_{s}[(-1)^m 2 \alpha_1 u (\hat{c}^+_{m+1,s} \hat{c}_{m,s}  + \\
&\hat{c}^+_{m,s} \hat{c}_{m+1,s})  +  
(-1)^m 2 \alpha_2 u \hat{c}^+_{m,s}\hat{c}^+_{m+1,s} \hat{c}_{m+1,s}\hat{c}_{m,s}].
\end{split}
\end{equation}
It represents correspondingly the terms, which are proportional to linear terms  in Taylor series expansion about the dimerized state of the resonance interaction  of
 quasiparticles in "$\pi$-subsystem" of  charge system and  potential energy of the pairwise interaction of quasiparticles in "$\pi$-subsystem" between themselves. It is taken into account, that in  Born-Oppenheimer approximation in perfectly dimerized chain the coordinates  $\{u_m\}$, $m = \overline{1,N}$, can be represented in the form  $\{u_m\} = \{(-1)^m u\}$, where $u$ is displacement amplitude, corresponding to minimum of ground state energy being to be a function of displacement value \cite{SSH}. It seems to be evident, taking into account the symmetry, that there are two values of $u$ with opposite signs minimizing ground state energy, and indicating on its two-fold degeneration. Physically the third part of Hamiltonian describes electron-phonon interaction with total constant $\alpha =  \alpha_1 + \alpha_2$, which a priori seems to be extending the range of applicability of SSH-model to essentially more strong values of electron-phonon interaction because electron-electron interactions (see above)  are known (from quantum chemical studies of small polyenes)
to be strong.

 Operator $\hat{\mathcal{H}}(u)$ is invariant under spatial translations with period $2a$, where $a$ is projection of spacing between two adjacent  "boson atoms" in undimerized lattice on chain axis direction. It means, that all various wave vectors $\vec{k}$  in $\vec{k}$-space will be in reduced zone with module of $\vec{k}$ in the range $-\frac{\pi}{2a} \leq k \leq \frac{\pi}{2a}$. It can be considered like to usual semiconductors to be consisting of two subzones - conduction ($c$) band and valence ($v$) band. Then it seems to be convenient to represent the operators $\hat{c}^+_{m,s}$,   $\hat{c}_{m,s}$,
 $m = \overline{1,N}$ in the form
\begin{equation}
\begin{split}
\label{Eq5m}
&\hat{c}_{m,s} = \hat{c}^{(c)}_{m,s} + \hat{c}^{(v)}_{m,s},\\
&\hat{c}^+_{m,s} = \hat{c}^{+(c)}_{m,s} + \hat{c}^{+(v)}_{m,s},
\end{split}
\end{equation} related to $\pi-c$- and $\pi-v$-band correspondingly,
and to define $\vec{k}$-space operators
\begin{equation}
\begin{split}
\label{Eq6m}
&\hat{c}^{(c)}_{k,s} = \frac{i}{\sqrt{N}}\sum_{m}\sum_{s}(-1)^{m+1}\exp(-ikma)\hat{c}^{(c)}_{m,s},\\
&\hat{c}^{(v)}_{k,s} = \frac{1}{\sqrt{N}}\sum_{m}\sum_{s}\exp(-ikma)\hat{c}^{(v)}_{m,s},
\end{split}
\end{equation}
$m = \overline{1,N}$. The principle, like to  MO LCAO is used in fact to build the operators $\hat{c}^{(c)}_{k,s}$ and $\hat{c}^{(v)}_{k,s}$, at that the antibonding character of $c$-band orbitals is taken into account by means of factor $i(-1)^{m+1}$.  Inverse to (\ref{Eq6m}) transform is
\begin{equation}
\begin{split}
\label{Eq7m}
&\hat{c}^{(c)}_{m,s} = \frac{1}{\sqrt{N}}\sum_{k}\exp{i[m(ka + \pi) - \frac{\pi}{2}]}\hat{c}^{(c)}_{k,s},\\
&\hat{c}^{(v)}_{m,s} = \frac{1}{\sqrt{N}}\sum_{k}\exp(ikma)\hat{c}^{(v)}_{k,s},
\end{split}
\end{equation}
$m = \overline{1,N}$. 

The $\sigma$-operators $\hat{a}^+_{m,s}$ and $\hat{a}_{m,s}$, $m = \overline{1,N}$ can also be represented in the form like to (\ref{Eq5m}) for $\pi$-operators and analogous to (\ref{Eq6m}), (\ref{Eq7m}) transforms can be defined. Then the expression for the operator $\hat{\mathcal{H}}_{0}(u)$ can be rewritten
\begin{equation}
\begin{split}
\label{Eq8m}
&\hat{\mathcal{H}}_{0}(u) = \hat{\mathcal{H}}^{\sigma,c}_{0}(u) + \hat{\mathcal{H}}^{\sigma,v}_{0}(u) =
\sum_{m}\sum_{s}(\frac{\hat{P}_m^2}{2M^*} + K u_m^2) \times \\ 
&\frac{1}{N}\sum_{k}(\hat{a}^{+\sigma,c}_{k,s} \hat{a}^{\sigma,c}_{k,s} + \hat{a}^{+\sigma,v}_{k,s} \hat{a}^{\sigma,v}_{k,s}),
\end{split}
\end{equation}
where $\hat{a}^{+\sigma,c}_{k,s}$,  $\hat{a}^{\sigma,c}_{k,s}$ and  $\hat{a}^{+\sigma,v}_{k,s}$, $\hat{a}^{\sigma,v}_{k,s}$ are $\sigma$-operators of creation and annihilation, related to $\sigma$-$c$-band and to $\sigma$-$v$-band correspondingly. The independence of $|u_m|$ on $m$, $m = \overline{1,N}$ means, that the expression  
$(\frac{\hat{P}_m^2}{2M^*} + K u_m^2)$ is independent on $m$. Then we obtain
\begin{equation}
\begin{split}
\label{Eq9m}
\hat{\mathcal{H}}_{0}(u) = \sum_{k}\sum_{s}(\frac{\hat{P}^2}{2M^*} + K u^2)(\hat{n}^{\sigma,c}_{k,s} + 
\hat{n}^{\sigma,v}_{k,s}),
\end{split}
\end{equation}
where
$\hat{n}^{\sigma,c}_{k,s}$ and  
$\hat{n}^{\sigma,v}_{k,s}$ are operators of number of  $\sigma$-quasiparticles in $\sigma$-$c$-band and  $\sigma$-$v$-band correspondingly.

The expression for $\hat{\mathcal{H}}_{\pi,t}(u)$ in terms of $\{\hat{c}^{(c)}_{k,s}\}$ and $\{\hat{c}^{(v)}_{k,s}\}$ is conciding in its mathematical form with known corresponding expression in \cite{SSH}, \cite{SSH_PRB} and it is
\begin{equation}
\begin{split}
\label{Eq10m}
\hat{\mathcal{H}}_{\pi,t}(u) =  \sum_{k}\sum_{s} 2t_0 \cos ka (\hat{c}^{+(c)}_{k,s}\hat{c}^{(c)}_{k,s} - \hat{c}^{+(v)}_{k,s}\hat{c}^{(v)}_{k,s})
\end{split}
\end{equation}
The expression for the first part of operator $\hat{\mathcal{H}}_{\pi,u}(u)$ in terms of $\{\hat{c}^{(c)}_{k,s}\}$ and $\{\hat{c}^{(v)}_{k,s}\}$ is also conciding in its form with known corresponding expression in \cite{SSH}, \cite{SSH_PRB} and it is given by
\begin{equation}
\begin{split}
\label{Eq10ma}
\hat{\mathcal{H}}_{\pi,u,\alpha_1}(u) =  \sum_{k}\sum_{s} 4 \alpha_1 u \sin ka (\hat{c}^{+(v)}_{k,s}\hat{c}^{(c)}_{k,s} + \hat{c}^{+(c)}_{k,s}\hat{c}^{(v)}_{k,s}), 
\end{split}
\end{equation}
where subscript $\alpha_1$ in Hamiltonian designation indicates on the taking into account the part of charge quantum-phonon interaction, connected with resonance interaction  (hopping) processes.
The expression for the second part of operator $\hat{\mathcal{H}}_{\pi,u}(u)$, which describes the part of charge quantum-phonon interaction, determined by interaction  between quasiparticles in Fermi liquid state of "$\pi$-subsystem" in terms of $\{\hat{c}^{(c)}_{k,s}\}$ and $\{\hat{c}^{(v)}_{k,s}\}$ is the following
\begin{equation}
\begin{split}
\label{Eq11m}
\hat{\mathcal{H}}_{\pi,u,\alpha_2}(u) = \sum_{k}\sum_{k'}\sum_{s}\alpha_2(k, k',s) \hat{c}^{+(c)}_{k',s}\hat{c}^{+(v)}_{k',s}  \hat{c}^{(v)}_{k,s}\hat{c}^{(c)}_{k,s}.
\end{split}
\end{equation}
Here the consideration is  restricted  by the taking into account the contribution of the term, corresponding to interaction  between the quasiparticles in different bands, which seems to be the most essential. 
Physically the identification of linear on displacement $u$ part for both resonance interaction (hopping) and for the pairwise interaction of quasiparticles in "$\pi$-subsystem" between themselves  with charge quantum-phonon interaction   is understandable, if to take into account, that by boson-"atomic" displacements the phonons  are generated, which in its turn can by release of, for instance, the m-place on,  to deliver the energy and impulse, which are necessary for transfer of the quasiparticle with charge quantum  ( which can be associated with electron with corresponding effective mass) from adjacent $(m-1)$- or $(m+1)$-position in chain in the case of  resonance interaction (hopping). For the case the pairwise interaction of given quasiparticles, it means, that its linear on displacement $u$ part is realized by means of phonon field,  which transfers the energy and impulse from one quasiparticle to another (which can be not inevitable adjacent). Mathematically it can be proved in the following way. The processes of interaction in $c$ ($v$) band can be considered to be independent on each other. It means, that transition probability from the $\langle k_{l,s}|$-state to $\langle k_{j,s}|$-state in $c$-band and from $\langle k'_{l,s}|$-state to  $\langle k'_{j,s}|$-state  in $v$-band, which is proportional to coefficient $\alpha_2(k, k',s)$, can be expressed in the form of product of real parts of corresponding matrix elements, that is in the form
\begin{equation}
\begin{split}
\label{Eq12m}
&\alpha_2(k, k',s) \sim  Re\langle k_{l,s}|\hat{V}^{(c)}|k_{j,s}\rangle Re\langle k'_{l,s}|\hat{V}^{(v)}|k'_{j,s}\rangle = \\
&\sum_{k_{ph}}Re\langle k_{l,s}|\hat{V}^{(c)}|k_{ph}\rangle \langle k_{ph}|k_{j,s}\rangle \times \\
&\sum_{k_{ph}}Re\langle k'_{r,s}|\hat{V}^{(v)}|k_{ph}\rangle\langle k_{ph}|k'_{n,s}\rangle,  
\end{split}
\end{equation} 
where $\hat{V}^{(v)}$ = $V_{0(v)}\hat{e}$ ($\hat{e}$ is unit operator) is the first term in Taylor expansion of pairwise interaction of quasiparticles, for instance with wave vectors  $k'_{r}$,  $k'_{n}$ and spin projection $s$ in $v$-band, that is, in ground state, $\hat{V}^{(c)}$, = $V_{1(c)} u \hat{e}$ is the second term in Taylor expansion of pairwise interaction in excited state (in c-band), that is, it is   product of configuration coordinate $u$ and coordinate derivative at $u = 0$  of operator of pairwise interaction of quasiparticles with for example, wave vectors $k_{l}$, $k_{j}$ and spin projection $s$   in $c$-band. At that, since the linear density of pairwise interaction is independent on $k$, which seems to be consequence of translation invariance of the chain, $V_{0(v)}$, $V_{1(c)}$ are constants. Therefore, if pairwise interaction is accompanying by process of phonon generation in $c$-band, then we have  $\hat{V}^{(c)} = {V}_{0(c)}u \hat{e}$, $\hat{V}^{(v)} = V_{0(v)} \hat{e}$. It means, that  phonon, which was generated by the displacemnt formation leding to its accompanying change of magnitude of pairwise interaction in $c$-band  between, for example, quasiparticle in $l$- and $j$-positions, delivers the energy and impulse to the quasiparticle in $v$-band, for instance in $n$-position,  without displacement excitation, since it is in  binding with related quasiparticle in $r$-position. A number of variants are possible along with process of phonon generation in $c$-band above described.  The result will be quite similar, if to change the place of excitation, that is, if to interchange the role of $c$ and $v$ bands for given process. There seem to be possible the realization of both the stages (that is phonon generation and absorption) in single $c$ or $v$ band and simultaneous realization both the stages in both the bands. Mathematical description will for all possible variants be similar and for distinctness we will consider only the first variant. For the simplicity we consider also the processes, in which the spin projection is keeping to be the same. It is evident also,  that in $z$-direction the impulse distribution is quasi-continuous (the chain has  the macroscopic length $L = N a$).
Consequently, standard way $\sum_{k_{ph}}\rightarrow \frac{L}{2\pi}\int_{k_{ph}}$ can be used. Further, phonon states can be described by wave functions $\langle k_{ph}| = v_0 exp(ik_{ph} z)$, where $z \in [0,L]$, $k_{ph} \in [-\frac{\pi}{2a}$, $\frac{\pi}{2a}$], $v_0$ is constant. Therefore, from (\ref{Eq12m}) we have the expression
\begin{equation}
\begin{split}
\label{Eq13m}
&\alpha_2(k, k',s) = b |v_{0v}|^2 |v_{0c}|^2  V_{0(c)} u V_{0(v)} |\phi_{0cs}|^2 |\phi_{0vs}|^2 \times\\ 
&\frac{N}{2\pi(q_l - q_j)(q_r - q_n)} Re\{\exp[{i(k_l m_l - k_j m_j)a}] \exp{ika}\} \times \\
&Re\{\exp[{i(k'_r m_r - k'_n m_n)a}] \exp {ik'a}\},
\end{split}
\end{equation}
where $|\phi_{0cs}|^2$, $|\phi_{0vs}|^2$ are squares of the modules of the wave functions $|k_{j,s}\rangle$ and $|k'_{j,s}\rangle$ respectively, $k = k_{ph}(q_l - q_j)$, $k' = k'_{ph}(q_r - q_n)$ $q_l, q_j, q_r, q_n \in N$ with additional conditions $(q_l - q_j)a \leq L$, $(q_r - q_n)a  \leq L$, $b$ - is aspect ratio, which in principle can be determined by comparison with experiment. Here
the values $(q_l - q_j)$, $(q_r - q_n)$ determine the steps  in pairwise interaction with phonon participation and they are considered to be fixed. We will consider the processes for which $k = k'$, consequently, $(q_r - q_n)$ =  $(q_l - q_j)$. 

The relation (\ref{Eq13m})
 by $k_{l }m_l = k_{j} m_j$ and by $k_r m_r = k_n m_n$ transforms into
\begin{equation}
\begin{split}
\label{Eq14m}
&\alpha_2(k, k',s) = b |v_{0v}|^2 |v_{0c}|^2  V_{0(c)} u V_{0(v)} |\phi_{0cs}|^2 |\phi_{0vs}|^2 \times \\
&\frac{N}{2\pi[(q_l - q_j)]^2} \sin ka  \sin k'a,
\end{split}
\end{equation}
where $|\phi_{0cs}|^2$, $|\phi_{0vs}|^2$ are squares of the modules of the wave functions $|k_{j,s}\rangle$ and $|k'_{j,s}\rangle$ respectively, $k = k_{ph}(q_l - q_j)$, $k' = k'_{ph}(q_r - q_n)$, $q_l, q_j, q_r, q_n \in N$ with additional conditions $(q_l - q_j)a \leq L$, $(q_r - q_n)a  \leq L$. Here
the values $(q_l - q_j)$, $(q_r - q_n)$ determine the steps  in pairwise interaction with phonon participation and they are considered to be fixed. We will consider the processes for which $k = k'$, consequently, $(q_r - q_n)$ =  $(q_l - q_j)$. Let us designate 
\begin{equation}
\begin{split}
\label{Eq15m}
&b |v_{0v}|^2 |v_{0c}|^2  V_{0(c)}  V_{0(v)} |\phi_{0cs}|^2 |\phi_{0vs}|^2 \times \\
&\frac{N}{2\pi[(q_l - q_j)]^2} = 4 \alpha_2(s)
\end{split}
\end{equation}
Then, taking into account that spin projection $s$ is fixed, the dependence on $s$ can be omitted,  consequently  $\alpha_2(s) =  \alpha_2$. So we have
\begin{equation}
\begin{split}
\label{Eq16m}
&\hat{\mathcal{H}}_{\pi,u,\alpha_2}(u) = \\ 
&\sum_{k}\sum_{k'}\sum_{s}4 \alpha_2 u \sin ka  \sin k'a \hat{c}^{+(c)}_{k',s}\hat{c}^{+(v)}_{k',s}  \hat{c}^{(v)}_{k,s}\hat{c}^{(c)}_{k,s}.
\end{split}
\end{equation}

Something otherwise will be treated the participation of the phonons in linear on $u$ part of   pairwise interaction,
if phonon generation is accompanying process of quasiparticle transition from $v$-band into $c$-band. It is the case of strongly doped chain, when Peierls transition is supressed and vorbidden gap is vanished. In given case
the expression for density of the  charge quantum carrier-(in particular, electron)-phonon coupling parameter $\alpha^d_2(k, k',s)$, which is related to the part of  charge quantum carrier-phonon interaction, determined by interaction  between quasiparticles in Fermi liquid, which produces "$\pi$-subsystem", is the following
\begin{equation}
\begin{split}
\label{Eq17m}
&\alpha_2(k, k',s) \sim  Re\langle k_{l,s}|\hat{V}^{}|k'_{j,s}\rangle = |v_{0v}|^2 |v_{0c}|^2   u V_{1} |\phi_{0s}|^2  \times\\
&\frac{N^2}{[2\pi]^2} 
\int_{k_{ph}}\exp[i(k_{ph} q a - k_{l} m_l a)]  \times\\
&\{\int_{k'_{ph}}\exp[i(k'_{ph} - k_{ph})q'a]  \times\\
&\exp[-i(k'_{ph} q' a - k'_{j} m_n a)]dk'_{ph}\}dk_{ph},
\end{split}
\end{equation}
where $q = m_j - m_l$, $q' = m_r - m_n$ are integers, satisfying foregoing relations, subscrips in left part are omitted, since fixed step is considered.
Then,  taking into account, that in  continuous limit by integration  the modules $k$ and $k'$ of  wave vektors $\vec{k}$ and $\vec{k'}$ are running all the $k$- and $k'$-values in $k$- and $k'$-spaces, we can  designate $(k_{ph} q a - k_{l}m_l a) = ka$, $(k'_{ph} q' a - k'_{j} m_j a) = k'a$ omitting the subscrips. Thus, we obtain
\begin{equation}
\begin{split}
\label{Eq18m}
&\alpha_2(k, k',s) \sim  Re\langle k_{l,s}|\hat{V}^{}|k'_{j,s}\rangle = |v_{0v}|^2 |v_{0c}|^2   u V_{1} |\phi_{0s}|^2  \times\\
&\frac{N^2}{[2\pi]^2} (\sin ka  \sin k'a + \cos ka  \cos k'a). \end{split}
\end{equation}
It was taken into account, that by $v$-band $\rightarrow$ $c$-band  transition of  quasiparticle, the impulse of emitted phonon in $v$-band is equal to the impulse of absorbed phonon in  $c$-band.
System of operators $\hat{c}^{+(c)}_{k',s}$, $\hat{c}^{+(v)}_{k',s}$,  $\hat{c}^{(v)}_{k,s}$, $\hat{c}^{(c)}_{k,s}$ corresponds to noninteracting  quasiparticles, and it is understandable, that in the case of  interacting  quasiparticles their linear combination has to be used 
\begin{equation}
\begin{split}
\label{Eq19m}
\left[\begin{array} {*{20}c}  \hat{a}^{(v)}_{k,s} \\  \hat{a}^{(c)}_{k,s} \end{array}\right] = \left[\begin{array} {*{20}c} \alpha_{k,s} & -\beta_{k,s}  \\  \beta_{k,s} & \alpha_{k,s} \end{array}\right] \left[\begin{array} {*{20}c}  \hat{c}^{(v)}_{k,s} \\  \hat{c}^{(c)}_{k,s}  \end{array}\right], 
\end{split}
\end{equation}
where matrix of transformation coefficients $||A||$ is
\begin{equation}
\begin{split}
\label{Eq20m}
||A|| = \left[\begin{array} {*{20}c} \alpha_{k,s} & -\beta_{k,s} \\  \beta_{k,s} & \alpha_{k,s} \end{array}\right]  
\end{split}
\end{equation}
is unimodular matrix with determinant $det||A||= \alpha^2_{k,s} + \beta^2_{k,s} = 1$.  Since $det||A|| \neq 0$, it means, that inverse 
transformation exists and it is given by the matrix
\begin{equation}
\begin{split}
\label{Eq21m}
||A||^{-1} = \left[\begin{array} {*{20}c} \alpha_{k,s} & \beta_{k,s}  \\ - \beta_{k,s} & \alpha_{k,s} \end{array}\right].  
\end{split}
\end{equation}
Then we obtain for the Hamiltonian $\hat{\mathcal{H}}_{\pi,u,\alpha_1}(u)$, which corresponds to SSH one-electron treatment of electron-phonon coupling, the following relation
\begin{equation}
\begin{split}
\label{Eq22m}
&\hat{\mathcal{H}}_{\pi,u,\alpha_1}(u) = \\
&\sum_{k}\sum_{s}\Delta_k [\alpha^2_{k,s} \hat{a}^{+(v)}_{k,s} \hat{a}^{(c)}_{k,s} - 
\alpha_{k,s} \beta_{k,s} \hat{a}^{+(v)}_{k,s}\hat{a}^{(v)}_{k,s} \\
&+ \beta_{k,s} \alpha_{k,s} \hat{a}^{+(c)}_{k,s}\hat{a}^{(c)}_{k,s} - 
\beta^2_{k,s} \hat{a}^{+(c)}_{k,s} \hat{a}^{(v)}_{k,s} 
+ \alpha^2_{k,s} \hat{a}^{+(c)}_{k,s} \hat{a}^{(v)}_{k,s} \\
&+ \alpha_{k,s} \beta_{k,s} \hat{a}^{+(c)}_{k,s} \hat{a}^{(c)}_{k,s}  -
\beta_{k,s} \alpha_{k,s} \hat{a}^{+(v)}_{k,s}\hat{a}^{(v)}_{k,s} - \beta^2_{k,s} \hat{a}^{+(v)}_{k,s} \hat{a}^{(c)}_{k,s}], 
\end{split}
\end{equation}
where $\Delta_k = 4 \alpha_1 u \sin ka$.
The diagonal part $\hat{\mathcal{H}}^d_{\pi,u,\alpha_1}(u)$ of operator $\hat{\mathcal{H}}_{\pi,u,\alpha_1}(u)$  is
\begin{equation}
\begin{split}
\label{Eq23m}
&\hat{\mathcal{H}}^d_{\pi,u,\alpha_1}(u) = \sum_{k}\sum_{s}2 \Delta_k \alpha_{k,s} \beta_{k,s} (\hat{n}^{(c)}_{k,s} - \hat{n}^{(v)}_{k,s}), 
\end{split}
\end{equation}
where $\hat{n}^{(c)}_{k,s}$ is density of operator of quasiparticles' number in $c$-band,  $\hat{n}^{(v)}_{k,s}$ is density of operator of quasiparticles' number in $v$-band.
 The part of pairwise interaction, which is  linear in displacement coordinate $u$ and leads to participation of the phonons,  is given by 
the Hamiltonian 
\begin{equation}
\begin{split}
\label{Eq24m}
&\hat{\mathcal{H}}_{\pi,u,\alpha_2}(u) = \sum_{k}\sum_{k'}\sum_{s} 4 \alpha_2 u \sin ka  \sin k'a \times\\ &(\alpha^2_{k',s} \hat{a}^{+(c)}_{k',s}\hat{a}^{(v)}_{k',s} - \beta^2_{k',s} \hat{a}^{(v)}_{k',s} \hat{a}^{+(c)}_{k',s}\\
&+ \alpha_{k',s} \beta_{k',s} \hat{a}^{(c)}_{k',s} \hat{a}^{+(c)}_{k',s} -
\beta_{k',s} \alpha_{k',s} \hat{a}^{(v)}_{k',s} \hat{a}^{+(v)}_{k',s}) \\
&\times (\alpha^2_{k,s} \hat{a}^{+(c)}_{k,s} \hat{a}^{(v)}_{k,s} - \beta^2_{k,s} \hat{a}^{+(v)}_{k,s} \hat{a}^{(c)}_{k,s}\\
&+ \alpha_{k,s} \beta_{k,s} \hat{a}^{+(c)}_{k,s} \hat{a}^{(c)}_{k,s} - \beta_{k,s} \alpha_{k,s} \hat{a}^{+(v)}_{k,s} \hat{a}^{(v)}_{k,s}).
\end{split}
\end{equation}
The diagonal part $\hat{\mathcal{H}}^d_{\pi,u,\alpha_2}(u)$ of operator $\hat{\mathcal{H}}_{\pi,u,\alpha_2}(u)$  is
\begin{equation}
\begin{split}
\label{Eq25m}
&\hat{\mathcal{H}}^d_{\pi,u,\alpha_2}(u) = 4 \alpha_2 u \sum_{k}\sum_{k'}\sum_{s} \alpha_{k'} \beta_{k'} (\hat{n}^{(v)}_{k',s} - \hat{n}^{(c)}_{k',s}) \\
&\times \alpha_{k,s} \beta_{k,s} (\hat{n}^{(v)}_{k,s} - \hat{n}^{(c)}_{k,s}) \sin k'a \sin ka 
\end{split}
\end{equation}

The Hamiltonian $\hat{\mathcal{H}}_{\pi,t}(u)$ in terms of operator variables $\hat{a}^{(c)}_{k,s}$ $\hat{a}^{(v)}_{k,s}$ is
\begin{equation}
\begin{split}
\label{Eq26m}
&\hat{\mathcal{H}}_{\pi,t}(u) =  \sum_{k}\sum_{s} 2t_0 \cos ka [(\alpha^2_{k,s}  - \beta^2_{k,s}) (\hat{a}^{+(c)}_{k,s} \hat{a}^{(c)}_{k,s} - \\
&\hat{a}^{+(v)}_{k,s} \hat{a}^{(v)}_{k,s}) - 2 \alpha_{k,s} \beta_{k,s} (\hat{a}^{+(v)}_{k,s} \hat{a}^{(c)}_{k,s} + \hat{a}^{+(c)}_{k,s} \hat{a}^{(v)}_{k,s})]
\end{split}
\end{equation}
The diagonal part $\hat{\mathcal{H}}^d_{\pi,t}(u)$ of operator $\hat{\mathcal{H}}_{\pi,t}(u)$  is given by the relation
\begin{equation}
\begin{split}
\label{Eq27m}
\hat{\mathcal{H}}^d_{\pi,t}(u) = \sum_{k}\sum_{s} \epsilon_k (\alpha^2_{k,s}  - \beta^2_{k,s}) (\hat{n}^{(c)}_{k,s} -
\hat{n}^{(v)}_{k,s}), 
\end{split}
\end{equation}
where $\epsilon_k = 2t_0 \cos ka$. 

The operator transformation for the $\sigma$-subsystem, analogous to (\ref{Eq19m}) shows, that the Hamiltonian $\hat{\mathcal{H}}_{0}(u)$ is invariant under given transformation, that is, it can be represented in the form, given by (\ref{Eq9m}).

To find the values of elements of matrices $||A||$ and  $||A||^{-1}$, the Hamiltonian $\hat{\mathcal{H}}_{}(u)$
has to be tested for conditional extremum on the variables $\alpha_{k}$, $\beta_{k}$ with condition $\alpha^2_{k,s}  - \beta^2_{k,s} = 1$. The corresponding Lagrange function $\hat{\mathfrak{E}}^L_{}(u)$  
is
\begin{equation}
\begin{split}
\label{Eq28m}
&\hat{\mathfrak{E}}^L_{}(u) = \sum_{k}\sum_{s}(\frac{\hat{P}^2}{2M^*} + K u^2)(\hat{n}^{\sigma,c}_{k,s} + 
\hat{n}^{\sigma,v}_{k,s}) \\
&+ \sum_{k}\sum_{s} \epsilon_k (\alpha^2_{k,s}  - \beta^2_{k,s}) (\hat{n}^{(c)}_{k,s} -
\hat{n}^{(v)}_{k,s}) \\
&+ \sum_{k}\sum_{s} 2 \Delta_k \alpha_{k,s} \beta_{k,s} (\hat{n}^{(c)}_{k,s} - \hat{n}^{(v)}_{k,s}) \\
&+ 4 \alpha_2 u \sum_{k}\sum_{k'}\sum_{s} \alpha_{k',s} \beta_{k',s} (\hat{n}^{(c)}_{k',s} - \hat{n}^{(v)}_{k',s}) \alpha_{k,s} \beta_{k,s}\\
&\times(\hat{n}^{(c)}_{k,s} - \hat{n}^{(v)}_{k,s}) \sin k'a \sin ka  + \lambda_{k,s} (\alpha^2_{k,s}  - 1 + \beta^2_{k,s})
\end{split}
\end{equation}
Then, the necessary condition for extremum is determined by  Lagrange equations
\begin{equation}
\begin{split}
\label{Eq29m}
&\frac{\partial{\hat{\mathfrak{E}}^L_{}(u)}}{\partial\alpha_{k}} = 2 \alpha_{k,s}\epsilon_k (\hat{n}^{(c)}_{k,s} - \hat{n}^{(v)}_{k,s}) + 2 \Delta_k  \beta_{k,s} (\hat{n}^{(c)}_{k,s} - \hat{n}^{(v)}_{k,s}) \\
&\times [1 + \frac{\alpha_2}{\alpha_1} \sum_{k'}\sum_{s} \alpha_{k',s} \beta_{k',s} \sin k'a (\hat{n}^{(c)}_{k',s} - \hat{n}^{(v)}_{k',s})] \\
&+ 2 \lambda_{k,s} \alpha_{k,s} = 0,
\end{split}
\end{equation}
\begin{equation}
\begin{split}
\label{Eq30m}
&\frac{\partial{\hat{\mathfrak{E}}^L_{}(u)}}{\partial\beta_{k,s}} = 2 \beta_{k,s}\epsilon_k (\hat{n}^{(v)}_{k,s} - \hat{n}^{(c)}_{k,s}) + 2  \Delta_k  \alpha_{k,s} (\hat{n}^{(c)}_{k,s} - \hat{n}^{(v)}_{k,s}) \\
&\times [1 + \frac{\alpha_2}{\alpha_1} \sum_{k'}\sum_{s} \alpha_{k',s} \beta_{k',s} \sin k'a (\hat{n}^{(c)}_{k',s} - \hat{n}^{(v)}_{k',s})] \\
&+ 2 \lambda_{k,s} \beta_{k,s} = 0
\end{split}
\end{equation}
and
\begin{equation}
\begin{split}
\label{Eq31m}
\frac{\partial{\hat{\mathfrak{E}}^L_{}(u)}}{\partial\lambda_{k,s}} = \alpha^2_{k,s}  - 1 + \beta^2_{k,s} = 0.
\end{split}
\end{equation}
Let us designate
\begin{equation}
\begin{split}
\label{Eq32m}
[1 + \frac{\alpha_2}{\alpha_1} \sum_{k'}\sum_{s} \alpha_{k',s} \beta_{k',s} \sin k'a (\hat{n}^{(c)}_{k',s} - \hat{n}^{(v)}_{k',s})] = \hat{\mathcal{Q}},
\end{split}
\end{equation}
then, passing on to observables in the Lagrange equations (\ref{Eq29m}) - (\ref{Eq31m}), we obtain for $\beta^2_{k,s}$, $\alpha^2_{k,s}$ and for product $\alpha_{k,s} \beta_{k,s}$ the relations
\begin{equation}
\begin{split}
\label{Eq33m}
\beta^2_{k,s} = \frac{1}{2}(1 \pm \frac{\epsilon_k}{\sqrt{\epsilon^2_k + \mathcal{Q}^2 \Delta^2_k}}), 
\end{split}
\end{equation}
\begin{equation}
\begin{split}
\label{Eq34m}
\alpha^2_{k,s} = \frac{1}{2}(1 \mp \frac{\epsilon_k}{\sqrt{\epsilon^2_k + \mathcal{Q}^2 \Delta^2_k}}), 
\end{split}
\end{equation}
\begin{equation}
\begin{split}
\label{Eq35m}
\alpha_{k,s} \beta_{k,s} = \frac{1}{2}\frac{\mathcal{Q} \Delta_k}{\sqrt{\epsilon^2_k + \mathcal{Q}^2  \Delta^2_k}}, 
\end{split}
\end{equation}
where $\mathcal{Q}$ is eigenvalue of operator $\hat{\mathcal{Q}}$.
The equation for factor $\mathcal{Q}$ is
\begin{equation}
\begin{split}
\label{Eq36m}
[1 + \frac{\alpha_2}{2\alpha_1} \sum_{k}\sum_{s}\frac{\mathcal{Q} \Delta_k \sin ka }{\sqrt{\epsilon^2_k + \mathcal{Q}^2 \Delta^2_k}} ({n}^{(c)}_{k,s} - {n}^{(v)}_{k,s})] = \mathcal{Q},
\end{split}
\end{equation}
where superscript {'} is omitted and  ${n}^{(c)}_{k,s}$ is eigenvalue of density  operator of quasiparticles' number in $c$-band,  ${n}^{(v)}_{k,s}$ is eigenvalue of density operator of quasiparticles' number in $v$-band.
It is evident, that at $Q = 1$ in (\ref{Eq33m}) - (\ref{Eq35m}) we have the case of SSH-model. It corresponds to the case, if $\frac{\alpha_2}{\alpha_1} \sum_{k}\sum_{s} \frac{1}{2}\frac{\Delta_k}{\sqrt{\epsilon^2_k +  \Delta^2_k}} \sin ka ({n}^{(c)}_{k,s} - {n}^{(v)}_{k,s})] \rightarrow 0$, which is realized, if $\alpha_2  \rightarrow 0$. Consequently, it seems to be interesting to consider the opposite case, when $|\frac{\alpha_2}{\alpha_1} \sum_{k}\sum_{s} \frac{1}{2}\frac{\Delta_k}{\sqrt{\epsilon^2_k + \Delta^2_k}}\sin ka ({n}^{(c)}_{k,s} - {n}^{(v)}_{k,s})]| \gg 1$. Passing on to continuum limit, in which $\sum_{k}\sum_{s}
\rightarrow 2 \frac{Na}{\pi} \int\limits_0^{\frac{\pi}{2a}}$, and assuming ${n}^{(v)}_{k,s} = 1$, ${n}^{(c)}_{k,s} = 0$, we have integral equation
\begin{equation}
\begin{split}
\label{Eq37m}
\frac{2 N u a \alpha_2}{\alpha_1 \pi t_0} \int\limits_0^{\frac{\pi}{2a}}\frac{\sin^2 ka}{\sqrt{1 - \sin^2 ka [1-(\frac{2 u \mathcal{Q}}{t_0})^2] }} dk = 1.
\end{split}
\end{equation}
In the case $|\frac{2 u \mathcal{Q}}{t_0}| < 1$ the relation (\ref{Eq37m}) can be rewritten in the form
\begin{equation}
\begin{split}
\label{Eq38m} 
&K\left\{\sqrt{1-\left(\frac{2\alpha_1 u \mathcal{Q}}{t_0}\right)^2}  \right\} - E\left\{\sqrt{1-\left(\frac{2\alpha_1 u \mathcal{Q}}{t_0}\right)^2} \right\} = \\
&\frac{\pi [t^2_0 - (2 u \mathcal{Q})^2]}{2 N u \alpha_2},
\end{split}
\end{equation}
where $K\left\{\sqrt{1-(\frac{2\alpha_1 u \mathcal{Q}}{t_0})^2}  \right\}$ and $E\left\{\sqrt{1-(\frac{2 u \mathcal{Q}}{t_0})^2} \right\}$ are complete elliptic integrals of the first and  the second kind, respectively. Expanding given  integrals into the series and restricting by the terms of the second-order of smallness, we obtain
\begin{equation}
\begin{split}
\label{Eq39m} \mathcal{Q} \approx \frac{t_0}{6 u } \sqrt{25 - 32 \frac{t_0 \alpha_1}{N u \alpha_2}}.
\end{split}
\end{equation}
It is evident, that the condition $|\frac{2 u \mathcal{Q}}{t_0}| < 1$  is held true by $\frac{1}{3}\sqrt{25 - 32 \frac{t_0 \alpha_1}{N u \alpha_2}} < 1$.

In the case $|\frac{2 u \mathcal{Q}}{t_0}| > 1$ the relation (\ref{Eq37m}) can be represented in the form
\begin{equation}
\begin{split}
\label{Eq40m}
\int\limits_0^{\frac{\pi}{2}}\frac{\cos^2 y} {\sqrt{1 - \sin^2 y [1-(\frac{t_0}{2 u \mathcal{Q}})^2] }} dy = - \frac{\pi \mathcal{Q} \alpha_1}{\alpha_2 N},
\end{split}
\end{equation}
where $y = \frac{\pi}{2} - ka$. It is equivalent to the equation
\begin{equation}
\begin{split}
\label{Eq41m}
&\left(\frac{t_0}{2 u \mathcal{Q}}\right) F\left\{\frac{\pi}{2},\sqrt{1-\left(\frac{t_0}{2 u \mathcal{Q}}\right)^2}  \right\}\\
&- E\left\{\frac{\pi}{2},\sqrt{1-\left(\frac{t_0}{2 u \mathcal{Q}}\right)^2}  \right\} = \\
&\frac{\pi \mathcal{Q} \alpha_1}{\alpha_2 N}\left[1 - \left(\frac{t_0}{2 u \mathcal{Q}}\right)^2\right],
\end{split}
\end{equation}
where $F\left\{\frac{\pi}{2},\sqrt{1-\left(\frac{t_0}{2 u \mathcal{Q}}\right)^2}  \right\}$ is the complete elliptic integral of the first kind.
The  approximation of elliptic integrals, like to  the approximation, given by  (\ref{Eq39m}), leads to the relation
\begin{equation}
\begin{split}
\label{Eq42m} \mathcal{Q} \approx \frac{-3 \alpha_2 N}{16} \left[ 1 \pm \sqrt{1 + \frac{80 \alpha_1 t_0}{9 N u \alpha_2  }}\right].
\end{split}
\end{equation}

In the case $\frac{2 u \mathcal{Q}}{t_0} = 1$ the relation (\ref{Eq37m}) is 
\begin{equation}
\begin{split}
\label{Eq43m}
\int\limits_0^{\frac{\pi}{2}}\cos^2 y dy = - \frac{\pi \alpha_1 \mathcal{Q}}{\alpha_2 N},
\end{split}
\end{equation}
where $y = \frac{\pi}{2} - ka$.
It is seen, that  in given case the value of parameter $Q$ is calculated exactly and it is
\begin{equation}
\begin{split}
\label{Eq44m}
\mathcal{Q} = \frac{\alpha_2 N}{ 4\alpha_1} 
\end{split}
\end{equation} 
The values of energy of $\pi$-quasiparticles in $c$-band $E_k^{(c)}(u)$ and in  $v$-band $E_k^{(v)}(u)$ can be obtained in the following way
\begin{equation}
\begin{split}
\label{Eq45m}
E_k^{(c)}(u) = \frac{\partial{\mathfrak{E}^L_{}(u)}}{\partial{n^{(c)}_{k,s}}},
E_k^{(v)}(u) = \frac{\partial{\mathfrak{E}^L_{}(u)}}{\partial{n^{(v)}_{k,s}}},
\end{split}
\end{equation}
 where $\mathfrak{E}^L_{}(u)$ is the energy of "$\pi$-subsystem" of chain, which is obtained from Lagrange function operator, (\ref{Eq28m}), by passing on to observables. Therefore, we have
\begin{equation}
\begin{split}
\label{Eq46m}
&E_k^{(c)}(u) = \epsilon_k (\alpha^2_{k,s}  - \beta^2_{k,s}) + 2 \Delta_k \alpha_{k,s} \beta_{k,s} 
+ 8 \alpha_2 u \sin ka \\
&\times \sum_{k'}\sum_{s} \alpha_{k',s} \beta_{k',s} (\hat{n}^{(c)}_{k',s} - \hat{n}^{(v)}_{k',s})  \sin k'a  \alpha_{k,s} \beta_{k,s} \\
&= \epsilon_k (\alpha^2_{k,s}  - \beta^2_{k,s}) + 2 \Delta_k \alpha_{k,s} \beta_{k,s} \mathcal{Q}
\end{split}
\end{equation}
and 
\begin{equation}
\begin{split}
\label{Eq47m}
&E_k^{(v)}(u) = - \epsilon_k (\alpha^2_{k,s}  - \beta^2_{k,s}) - 2 \Delta_k \alpha_{k,s} \beta_{k,s} 
 - 8 \frac{\alpha_2} u \sin ka \\
&\times \sum_{k'}\sum_{s} \alpha_{k',s} \beta_{k',s} (\hat{n}^{(c)}_{k',s} - \hat{n}^{(v)}_{k',s})  \sin k'a  \alpha_{k,s} \beta_{k,s} \\
&=- \epsilon_k (\alpha^2_{k,s}  - \beta^2_{k,s}) - 2 \Delta_k \alpha_{k,s} \beta_{k,s} \mathcal{Q}.
\end{split}
\end{equation}
It is seen from (\ref{Eq46m}) and (\ref{Eq47m}), that $E_k^{(v)}(u) = - E_k^{(c)}(u)$. Taking into account the relations (\ref{Eq33m}) - (\ref{Eq35m}), we obtain
\begin{equation}
\begin{split}
\label{Eq48m}
E_k^{(c)}(u) =  \mp \frac{\epsilon^2_k}{\sqrt{\epsilon^2_k + \mathcal{Q}^2 \Delta^2_k}} + \frac{\mathcal{Q}^2 \Delta^2_k}{\sqrt{\epsilon^2_k + \mathcal{Q}^2  \Delta^2_k}}, 
\end{split}
\end{equation}

\begin{equation}
\begin{split}
\label{Eq49m}
E_k^{(v)}(u) =  \pm \frac{\epsilon^2_k}{\sqrt{\epsilon^2_k + \mathcal{Q}^2 \Delta^2_k}} - \frac{\mathcal{Q}^2 \Delta^2_k}{\sqrt{\epsilon^2_k + \mathcal{Q}^2  \Delta^2_k}}. 
\end{split}
\end{equation}
Therefore, we have two values for the enegy of quasiparticles, indicating on the possibility of formation of the quasiparticles of two kinds. Upper sign in the first terms in (\ref{Eq48m}),   (\ref{Eq49m}) corresponds to the quasiparticles with the energy
\begin{equation}
\begin{split}
\label{Eq50m}
&E_k^{(c)}(u) =   \frac{\mathcal{Q}^2 \Delta^2_k  - \epsilon^2_k}{\sqrt{\epsilon^2_k + \mathcal{Q}^2  \Delta^2_k}},\\ 
&E_k^{(v)}(u) =   \frac{\epsilon^2_k - \mathcal{Q}^2 \Delta^2_k}{\sqrt{\epsilon^2_k + \mathcal{Q}^2  \Delta^2_k}}
\end{split}
\end{equation}
in $c$-band and $v$-band respectively. Lower sign in the first terms in (\ref{Eq48m}),   (\ref{Eq49m}) corresponds to  the quasiparticles with the energy
\begin{equation}
\begin{split}
\label{Eq51m}
&E_k^{(c)}(u) =   \sqrt{\epsilon^2_k + \mathcal{Q}^2  \Delta^2_k},\\ 
&E_k^{(v)}(u) =   - \sqrt{\epsilon^2_k + \mathcal{Q}^2  \Delta^2_k}
\end{split}
\end{equation}
in $c$-band and $v$-band respectively.
The quasiparticles of the second kind  at  $\mathcal{Q} = 1$ are quite similar quasiparticles in its mathematical description, that those ones, which were obtained in \cite {SSH}, \cite{SSH_PRB}. 

We have used the only necessary condition for extremum of the functions $E(\alpha_{k,s} \beta_{k,s})$. It was shown in \cite{Yearchuck_Dovlatova}, that for the  SSH-model  
the  sufficient conditions for the minimum are substantial, they change the role of both solutions.  Sufficient conditions for the minimum of the functions $E(\alpha_{k,s} \beta_{k,s})$
 can be  obtained by standard way, which was used in \cite{Yearchuck_Dovlatova}. It consist in that, that the second differential of  the energy, being to be the function of  three variables  ${\alpha}_{k,s}$,  ${\beta}_{k,s}$ and  
 $\lambda_{k,s}$,  has to be positively defined quadratic form. From the condition of positiveness of three principal minors of quadratic form coefficients we obtain  the following three sufficient conditions for the energy minimum

\subsubsection{The First Condition}

The first condition is
\begin{equation}
\label{Eq52m}
\begin{split}
&\{ \epsilon_k (1 - \frac{\epsilon_k}{\sqrt{\epsilon^2_k + \mathcal{Q}^2  \Delta^2_k}}) < \frac{(\mathcal{Q}\Delta)^2_k}{\sqrt{\epsilon^2_k + \mathcal{Q}^2  \Delta^2_k}} | 
({n}^c_{k,s} - {n}^v_{k,s}) < 0\}, \\
&\{\epsilon_k (1 - \frac{\epsilon_k}{\sqrt{\epsilon^2_k + \mathcal{Q}^2  \Delta^2_k}} > \frac{(\mathcal{Q}\Delta)^2_k}{\sqrt{\epsilon^2_k + \mathcal{Q}^2  \Delta^2_k}} | ({n}^c_{k,s} - {n}^v_{k,s}) > 0 \}
\end{split}
\end{equation}
for the solution, which coincides in its form with SSH-solution at $\mathcal{Q} = 1$ (SSH-like solution) and
\begin{equation}
\label{Eq53m}
\begin{split}
&\{\epsilon_k (1 + \frac{\epsilon_k}{\sqrt{\epsilon^2_k + \mathcal{Q}^2  \Delta^2_k}} < \frac{{\mathcal{Q}\Delta}^2_k}{\sqrt{\epsilon^2_k + \mathcal{Q}^2  \Delta^2_k}} | ({n}^c_{ks} - {n}^v_{ks}) < 0 \},\\
&\{\epsilon_k (1 + \frac{\epsilon_k}{E_k} > \frac{(\mathcal{Q}\Delta)^2_k}{\sqrt{\epsilon^2_k + \mathcal{Q}^2  \Delta^2_k}} | ({n}^c_{ks} - {n}^v_{ks}) > 0 \}
\end{split}
\end{equation}
 for the additional solution. It is seen, that the first condition is realizable for the quasiparticles of both the kinds, at that, for both the kinds of states - near equilibrium  state $({n}^c_{ks} - {n}^v_{ks} < 0)$  and  strongly nonequilibrium state $(n^c_{ks} - {n}^v_{ks} > 0$.  

\subsubsection{The Second Condition}

The second condition is the same for both the solutions and it is
\begin{equation}
\label{Eq54m}
 (\frac{\epsilon^2_k}{\sqrt{\epsilon^2_k + \mathcal{Q}^2  \Delta^2_k}} - 2\frac{{\mathcal{Q}\Delta}^2_k}{\sqrt{\epsilon^2_k + \mathcal{Q}^2  \Delta^2_k}})^2 - \epsilon^2_k   + \frac{1}{4} (\mathcal{Q}\Delta)^2_k > 0 
\end{equation}
It  is realizable for the quasiparticles of both the kinds.

\subsubsection{The Third Condition}

For the SSH-like solution we have
\begin{equation}\label{Eq55m}
(3\frac{(\mathcal{Q}\Delta)^2_k}{\sqrt{\epsilon^2_k + \mathcal{Q}^2  \Delta^2_k}} + 4\frac{\epsilon^2_k}{\sqrt{\epsilon^2_k + \mathcal{Q}^2  \Delta^2_k}})({n}^c_{ks} - {n}^v_{ks}) > 0. 
\end{equation}
It  means,  that   SSH-like solution is unapplicable for the description of standard processes, passing near equilibrium state by any parameters. The quasiparticles, described by   SSH-like solution, can be created the only in strongly nonequilibrium state with inverse                                                                                                   
population of the levels in $c$- and $v$-bands. At the same time  the solution, which corresponds to upper signs in (\ref{Eq48m}), has to satisfy  to the following condition
\begin{equation}
\label{Eq56m}
(3\frac{(\mathcal{Q}\Delta)^2_k}{\sqrt{\epsilon^2_k + \mathcal{Q}^2  \Delta^2_k}} - 4\frac{\epsilon^2_k}{\sqrt{\epsilon^2_k + \mathcal{Q}^2  \Delta^2_k}})({n}^c_{ks} - {n}^v_{ks}) > 0, 
\end{equation}
which can be realized  both in near equilibrium and in strongly nonequilibrium states of the "$\pi$-subsystem" of boson-"atomic" chain, which is considered to be quantum Fermi liquid.

\subsubsection{Ground State of Boson-"Atomic" Chain}

The continuum limit for the ground state energy of boson-"atomic"  chain with SSH-like quasiparticles will coincide in its mathematical form with known solution \cite{SSH_PRB}, if to  replace $\Delta_k \mathcal{Q} \rightarrow \Delta_k$.  Let us calculate  the ground state energy $E^{[u]}_0(u)$ of the boson-"atomic"  chain with  quasiparticles' branch, which is stable near equilibrium. Taking into account, that in ground state ${n}^c_{k,s} = 0$, ${n}^v_{k,s} = 1$, in the continuum limit we have
\begin{equation}
\label{Eq57m}
E^{[u]}_0(u) = - \frac{2N a}{\pi}\int\limits_0^{\frac{\pi}{2a}} \frac{(\mathcal{Q}\Delta_k)^2 - 
\epsilon^2_k}{\sqrt{\mathcal{Q}\Delta_k)^2 + 
\epsilon^2_k}}dk + 2NKu^2,
\end{equation}
then, calculating the integral, using the complete elliptic integral of the first kind $F(\frac{\pi}{2}, 1 - z^2)$ and the complete elliptic integral of the second kind
$E(\frac{\pi}{2}, 1 - z^2)$,  we obtain
\begin{equation}
\label{Eq58m}
\begin{split}
&E^{[u]}_0(u) =  \frac{4Nt_0}{\pi}\{F(\frac{\pi}{2}, 1 - z^2) + \\
&\frac{1 + z^2}{1 - z^2}[E(\frac{\pi}{2}, 1 - z^2) - F(\frac{\pi}{2}, 1 - z^2)]\} + 2NKu^2, 
\end{split}
\end{equation}
where
$z^2 = \frac{2\mathcal{Q}\alpha_1 u}{t_0}$.
Approximation of ({\ref{Eq58m}}) at $z \ll 1$ gives
\begin{equation}
\label{Eq59m}
\begin{split}
&E^{[u]}_0(u) = N \{\frac{4t_0}{\pi} - \frac{6}{\pi}\ln\frac{2t_0}{\mathcal{Q}\alpha_1 u} \frac{4 (\mathcal{Q}\alpha_1)^2 u^2}{t_0} + \\
&\frac{28 (\mathcal{Q}\alpha_1)^2 u^2}{\pi t_0} + ...\} + 2NKu^2.
\end{split}
\end{equation}
It is seen from (\ref{Eq59m}), that the energy of quasiparticles, described by   solution, which corresponds to upper signs in (\ref{Eq48m}) has the form of Coleman-Weinberg potential with two minima at the values of dimerization coordinate $u_0$ and $-u_0$ like to the energy of quasiparticles, described by   SSH-solution  for t-PA \cite{SSH_PRB}. 

Therefore, all qualitative conclusions of the model proposed in \cite{SSH_PRB} are holding in Fermi-liquid consideration of "$\pi$-subsystem" of the chain (instead Fermi-gas consideration) for the quasiparticles, corresponding to the second-branch-solution. It seems to be also substantial, that Fermi-liquid treatment of electron-phonon interaction extends the applicability limits of SSH-model to 1D conjugated conductors, allowing its use  in the case of strong electron-phonon interaction. It is evident, that the mechanism of the phenomenon
of spin-charge separation in Fermi-liquid is soliton mechanism, analogous to proposed by Jackiw and Rebbi on the basis of field theory positions and its applicability to physical fields, in particular to EM-field, seems to be natural. It means like to SSH-Fermi-gas model, that when elementary spin and charge carrier, for instance an electron,
is added to an  neutral  chain, it
can break up into two pieces, one of which carries the
electron’s charge and the other its spin. Given result
bears a clear family relation with the phenomenon
of antecedent spin-charge separation in the 1D electron gas theory of Luther and Emery \cite{Luther},
but it is quite different from Anderson spinon-holon mechanism. The results obtained  make more exact and correct  prevalent viewpoint, that spin-charge separation effect is indication on non-Fermi-liquid behavior of electronic systems and that it can be  reasonably described the only in the frames of Luttinger liquid theory. Given 
viewpoint is true the only for Anderson mechanism. In the other hand  given results allow to propose the reasonable explanation of the existence in the fields with charges of chargeless particles - solitons with nonzero spin value, which in the case of EM-field seems to be equal $\frac{1}/{2}$ instead prevalent viewpoint,
that photons possess by spin $S = 1$. Therefore, the photons in quantized EM-field are main excitations  in  oscillator structure, which is equivalent to spin S = 1 "boson-atomic" structure, like  matematically to well known spin S = 1 boson matter  structure - carbon atomic backbone structure  in many conjugated polymer chains. The  photons have two kind nature.  The photons of the first kind represent themselves neutral EM-solitons of  SSH-soliton family. They are  main excitations in so-called "undoped " structure of EM-field, including free EM-field in vacuum. Naturally, they have nonzero size, that is  they cannot be considered to be point objects. It seems to be evident, that like to Fermi-gas SSH-model, the main excitations in "doped" "boson-atomic" structure of EM-field will be charged spinless EM-solitons, which also can be referred  to  SSH-soliton family. It seems to be reasonable to suggest, that "doping" can be effective in the medium like to rain-clouds, although detailed mechanism has  to be additionally studied.

\section{Conclusions}

It is shown on the basis of complex number theory,  that any quantumphysical quantity is complex quantity.

It has been established the partition of
linear space $\left\langle\mathcal F,+,\cdot \right\rangle$ over the ring of scalars and pseudoscalar set, the vectors in which are sets of contravariant (or covariant) EM-field tensors and pseudotensors $\left\{{F}^{\mu\nu}\right\}$, $\left\{\tilde{F}^{\mu\nu}\right\}$, (or $\left\{{F}_{\mu\nu}\right\}$, $\left\{\tilde{F}_{\mu\nu}\right\}$), into 4 subspaces $\left\langle\mathcal F^{(i)},+,\cdot \right\rangle$, $i = \overline {1,4}$. 
In subspaces  $\left\langle \mathcal F^{(1)},+,\cdot \right\rangle$,  $\left\langle \mathcal F^{(2)},+,\cdot \right\rangle$  vector $\vec {E}$ is polar vector, and vector $\vec {H}$ is axial.  At the same time,  the components of vector $\vec {E}$ in the second subspace correspond to pure space components of  field tensor, and the components of  vector $\vec {H}$ correspond  to time-space mixed components of given field tensor. Arbitrary element of subspace  $\left\langle \mathcal F^{(3)},+,\cdot \right\rangle$  is 4-\textit{pseudo}tensor. Its space components are in fact the components of antisymmetric 3-\textit{pseudo}tensor, which determines polar magnetic field vector $\vec {H}$, (that is vector, which is dual to given tensor), while time-space mixed components are the components of axial 3-vector $\vec {E}$ of electric field.   
Therefore, the symmetry properties of the components of the vectors $\vec {E}$ and $\vec {H}$ under improper rotations in the third subspace  will be opposite to those ones in the first subspace. The symmetry properties of the components of $\vec {E}$ and $\vec {H}$ under improper rotations in the fourth subspace  will be opposite to those ones in the second subspace.
  
 The analysis of the symmetry properties of 4-vectors of EM-field potentials  $A_{\mu}$, performed on the basis of general Lorentz group representations,  leads to the same result, that is to existence of four kinds of 4-vectors of EM-field potentials, which transform according to direct product of representations of proper Lorentz group and uneven representations of two subgroups - space inversion subgroup and time reversal subgroup  of general Lorentz group. 
 
The algebraic properties of the set of functionals, determined on the space $\left\langle F,+,\cdot \right\rangle$ were established. In particular, the statement, indicating, that the set of  functionals 
 $\left\{\Phi [\tilde {F}^{\mu \nu}(x)]\right\}$, $\left\{\tilde{\Phi}[\tilde {F}^{\mu \nu}(x)]\right\}$, preassigned on the space $\left\langle F,+,\cdot \right\rangle$, produces  linear space $\left\langle \Phi',+,\cdot \right\rangle$ over a field of scalars $P$, which is dual to the space $\left\langle F,+,\cdot \right\rangle$, however it is nonselfdual,
 is established.
Given result is in fact the consequence of the existence of two kinds of independent tensor and scalar characteristics  of EM-field, that is genuine tensor and scalar functions on the one hand  and pseudotensor and pseudoscalar functions on the other hand.
The practical significance of given result consists in the 
necessity, if some physical phenomenon with participation of EM-field is studied, to take into consideration always both the spaces, that is   $\left\langle F,+,\cdot \right\rangle$ and  $\left\langle \Phi,+,\cdot \right\rangle$. More strictly, known Gelfand triple, which includes together with spaces $\left\langle F,+,\cdot \right\rangle$ and  $\left\langle \Phi,+,\cdot \right\rangle$ the Hilbert space with topology, determined in the proper way, has to be taken into consideration.

Additional gauge invariance of complex relativistic fields was studied. It has been found, that conserving quantity, corresponding to invariance of generalized relativistic equations under the operations of additional gauge symmetry group - multiplicative group $\mathfrak R$ of all real numbers (without zero) - is purely imaginary charge. So, it was shown, that complex fields are characterized by complex charges. It gives key for correct generalization of field equations, in particular for electrodynamics.  
 
 Additional hyperbolic dual symmetry of Maxwell equations is established, which includes Lorentz-invariance to be its particular case. The essence of additional hyperbolic dual symmetry of Maxwell equations is that, that Maxwell equations along with dual transformation symmetry, established by Rainich, given by (\ref{eq1b}) - (\ref{eq1c}), are symmetric relatively the dual transformations of another kind.  Hyperbolic dual transformations for electric and magnetic field strengh vector functions are
\begin{equation}
\label{eq1bbca}
 \left[\begin{array} {*{20}c}  \vec {E{''}} \\ \vec {H{''}} \end{array}\right] = \left[\begin{array} {*{20}c} \cosh\vartheta& i\sinh\vartheta  \\ -i\sinh\vartheta&\cosh\vartheta \end{array}\right]\left[\begin{array} {*{20}c}  \vec {E} \\ \vec {H} \end{array}\right],
\end{equation}
where  $\vartheta$ is arbitrary continuous parameter,
$\vartheta \in [0,2\pi]$. 

Generalized Maxwell equations are obtained on the basis of both dual and hyperbolic dual symmetries of  EM-field. It is shown, that in general case both scalar and vector quantities, entering  equations, are quaternion quantities, four components of which have different parities under improper rotations.  

Invariants for  EM-field, consisting of  dually symmetric parts, for both the cases of dual symmetry and hyperbolic dual symmetry are found. 
It is concluded, that Maxwell equations with all quaternion vector and scalar variables give concrete connection between dual and gauge symmetries of EM-field.

 It is shown, that there exists physical conserving
quantity, which is simultaneously invariant under  both Rainich dual and additional hyperbolic dual symmetry transformation of Maxwell
equations. It is spin in general case and spirality in the geometry, when vector $\vec{E}$ is directed along absciss axis, $\vec{H}$ is directed along ordinate axis in  $(\vec{E}, \vec{H})$ functional space. It is additional proof for quaternion four component 
structure of EM-field to be a single whole.Spin takes on special leading significance among the physical characteristics of EM-field, since the only spin (spirality  in the geometry considered) combine two subsystems of photon fields,  which have definite $P$-parity  (even and uneven) with the subsystem of two fields, which have definite $t$-parity (also even and uneven) into one system. It is considered to be the proof for four component structure of EM-field to be a single whole, that is, it is the confirmation along with the possibility of the representation of EM-field in four component quaternion form, given by (\ref{eq5abcce}), (\ref{eq6bcdde}), (\ref{eq7abccde}), (\ref{eq8abccde}),
the necessity of given representation. It
extends the overview on the nature of EM-field itself. It seems to be remarkable, that given result on the special leading significance of spin is in agreement with result in \cite{D_Yearchuck_A_Dovlatova}, where was shown,  that spin is quaternion vector of the state in Hilbert space, defined under ring of quaternions, of any quantum system (in the
frame of the chain model considered) interacting with EM-field

The  connection between symmetry of dynamical systems, in particular, between gauge symmetry and analytical properties of quantities, which are invariant under corresponding symmetry qroup has been established. For example, the
 analicity of complex-valued function $Q(z)$ = $Q_1(z) + iQ_2(z)$ of complex variable $z = \vec{r} + ict$, representing itself the complex charge of EM-field   
is proved.

 Canonical Dirac quantization method is developed in
two aspects. The first aspect is its application the only to observable quantities.

New principle of EM-field  quantization is proposed. It is development of canonical Dirac quantization method, which  is realized in two aspects. The first aspect consist in  choosing of field functions, which are immediately  observable quantities - 4-vector-functions  $E_\mu(\vec{r},t)$ and $H_\mu(\vec{r},t)$, the first three components of which are the components of 3-vector-functions $\vec{E}(\vec{r},t)$ and $\vec{H}(\vec{r},t)$ of electric and magnetic field strengths correspondingly, the fourth component is $\frac{i c \rho_e(\vec{r},t)}{\lambda}$, $\frac{i c \rho_m(\vec{r},t)}{\lambda}$, where $\rho_e(\vec{r},t)$ is electric charge density,  $\rho_m(\vec{r},t)$ is magnetic charge density, which is equaled to zero in the case of single-charge electrodynamics, $\lambda$ is medium conductivity, which in the case of free field in vacuum is $\lambda_v = \frac{1} {Z_0} = \frac{1} {120\pi} Ohm^{-1}$. The second  aspect is the realization along with well known time-local quantization of space-local quantization and space-time-local quantization, which allow to establish,  correspondingly, the time of photon creation (annihilation), the space coordinate of photon creation (annihilation) and the the space and time coordinates simultaneously of photon creation (annihilation). 

It is shown, that Coulomb field can be quantized in 1D- and 2D-systems, that is, it is radiation field in given low-dimensional systems.

New model of photons is proposed. The photons in quantized EM-field are main excitations  in oscillator structure of EM-field, which is equivalent to spin S = 1 "boson-atomic" structure,  like  matematically to well known spin S = 1 boson matter  structure - carbon atomic backbone chain structure in many conjugated polymers. They have two kind nature.  The photons of the first kind  represent themselves neutral chargeless EM-solitons of  SSH-soliton family. The photons of the second kind  represent themselves charged spinless EM-solitons, which also can be referred  to  SSH-soliton family.

\section{Aknowledgement} Authors are grateful to Y.Yerchack for discussions and the help in the work.

\end{document}